\documentclass[twocolumn,floatfix,groupedaddress,superscriptaddress,prb,showpacs]{revtex4}

\usepackage{graphicx}
\usepackage{amsmath}
\usepackage{amssymb}
\usepackage{amsfonts}
\usepackage{dsfont}
\usepackage{longtable}
\usepackage{dcolumn}
\usepackage{bbm}
\usepackage{float}

\begin{document}
\title{Derivation of the $t$-$J$ model for finite doping}

\author{Simone A. Hamerla}
\email{hamerla@fkt.physik.tu-dortmund.de}
\affiliation{Lehrstuhl f\"{u}r Theoretische Physik I, 
Technische Universit\"{a}t Dortmund,
 Otto-Hahn Stra\ss{}e 4, 44221 Dortmund, Germany}

\author{Sebastian Duffe}
%\email{}
\affiliation{Lehrstuhl f\"{u}r Theoretische Physik I, 
Technische Universit\"{a}t Dortmund,
 Otto-Hahn Stra\ss{}e 4, 44221 Dortmund, Germany}

\author{G\"otz S. Uhrig}
\email{goetz.uhrig@tu-dortmund.de}
\affiliation{Lehrstuhl f\"{u}r Theoretische Physik I, 
Technische Universit\"{a}t Dortmund,
 Otto-Hahn Stra\ss{}e 4, 44221 Dortmund, Germany}

\date{\textrm{\today}}

\begin{abstract} 
Mapping complex problems to simpler effective models is a key tool
in theoretical physics. One important example in the realm of
strongly correlated fermionic systems is the mapping of the Hubbard model
to a $t$-$J$ model which is appropriate for the treatment
of doped Mott insulators. Charge fluctuations across the
charge gap are eliminated. So far the derivation of the
$t$-$J$ model is only known at half-filling or in its 
immediate vicinity. Here we present the necessary conceptual
advancement to treat finite doping. The results for the
ensuing coupling constants are presented. Technically, the extended
derivation relies on self-similar continuous unitary transformations
(sCUT) and normal-ordering relative to a doped reference ensemble.
The range of applicability of the derivation of $t$-$J$ model
is determined as function of the doping $\delta$ and 
the ratio bandwidth $W$ over  interaction $U$.
\end{abstract}

%71.10.Fd Lattice fermion models (Hubbard model, etc.) 
%75.10.Jm Quantized spin models 
%71.27.+a Strongly correlated electron systems; heavy fermions 
%71.30.+h Metal-insulator transitions and other electronic transitions 
%71.28.+d Narrow-band systems; intermediate-valence solids

\maketitle

%%%%%%%%%%%%%%%%%%%%%%%%%%%%%%%%%%%%%%%
%%%%%%%%%%%%%%%%%%%%%%%%%%%%%%%%%%%%%%%

\section{Introduction}

\label{chap:intro}

The Hubbard model \cite{gutzw63,hubba63,kanam63} 
is one of the most common models for the description of 
strongly correlated electron systems on lattices. Because it contains the 
motion of the electrons as well as the interaction between two electrons at 
the same site it is capable to describe  charge degrees of 
freedom as well as magnetic degrees of freedom.
Due to the rich physical behavior 
of the Hubbard model an analytic solution is not possible 
except in one dimension \cite{essle05}. 

One common route to simplify the model for large 
repulsion $U$ is to derive an effective model 
which does no longer contain charge fluctuations across
the charge gap. Processes which  change the number of doubly occupied sites
(double occupancies, DOs)
are eliminated. For large  enough repulsion $U$ and at half-filling the 
electrons  are fixed on their lattice  sites. 
In this Mott-insulating phase the model can be mapped onto a Heisenberg model 
describing only the energetically low-lying 
spin degrees of freedom. In the
immediate vicinity of half-filling the motion and the interaction
of doped holes is described by the extension
of the Heisenberg model to the $t$-$J$ model 
\cite{harri67,takah77,macdo88,stein97,reisc04}.
The metallic behavior for small values of
the repulsion is beyond the applicability of this mapping \cite{milli91}.

In the present work the mapping of the Hubbard model
to the $t$-$J$ model is extended to finite macroscopic doping concentration
$\delta$. The influence of the doping $\delta$  on the resulting
parameters of the  $t$-$J$ model 
is studied for sufficiently large repulsion. Our approach 
provides a systematic and controlled derivation of the effective coupling 
constants as  function of the doping concentration. Thereby, an important
gap between the applicability of the derivation of the $t$-$J$
model and its actual applications is closed. 
 
First, we consider the half-filled case. The elimination of the charge 
fluctuations across the charge gap is performed by a self-similar
 continuous unitary transformation with various types of generators. 
Besides the magnetic exchange couplings, the resulting $t$-$J$ model 
contains the motion and the  interaction of holes and doubly occupied sites. 
Since the mapping starts from a 
reference ensemble comprising the two spin states with equal weight and without
any correlations between the spins on neighboring sites the 
spin state in the effective model remains unspecified.

\begin{figure}[htb]
\begin{center}
     \includegraphics[width=1\columnwidth]{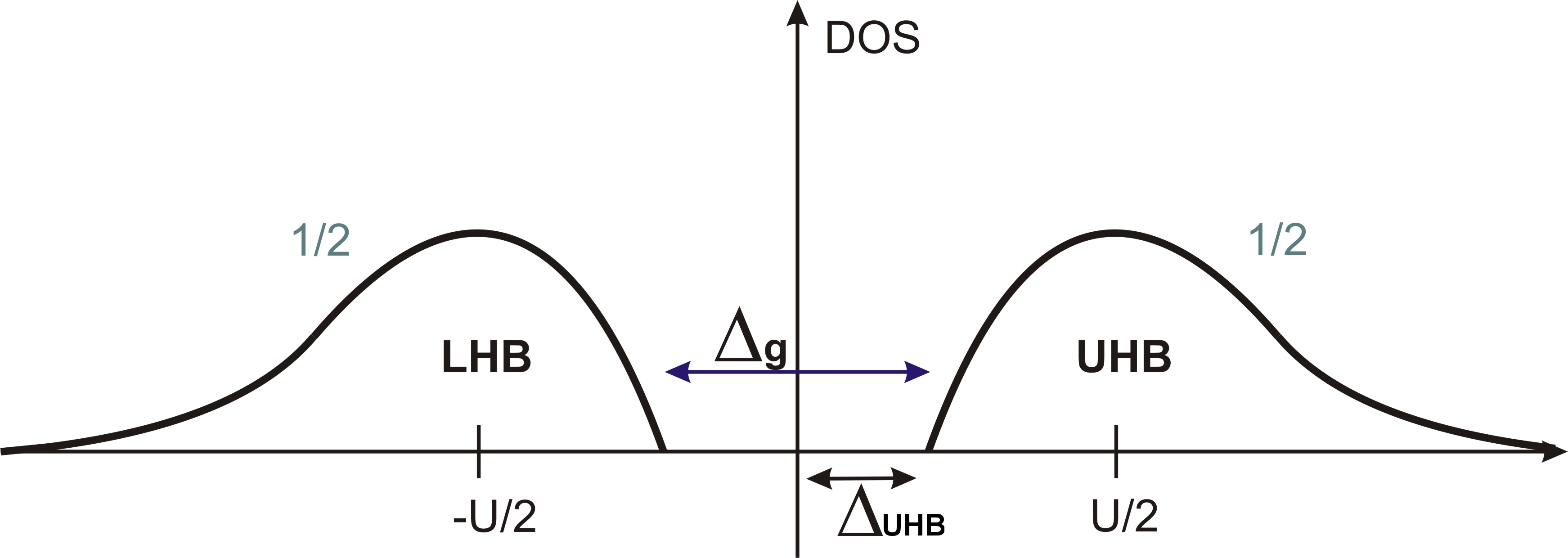}
\end{center}
\caption{(Color online) Density of states for the half-filled Hubbard model 
with large repulsion $U$ \cite{gutzw63,hubba63,kanam63}. 
The density of states exhibits two distinct, 
equally weighted bands, the lower Hubbard band (LHB) and the upper Hubbard 
band (UHB) \cite{hubba63,hubba64a,hubba64b,georg96}.}
\label{fig_DOS_half}
\end{figure}

The mapping relies on the elimination of processes changing the number of 
holes and doubly occupied sites.
Note that an empty site represents a double occupancy of two holes.
In the half-filled case the density of states exhibits two distinct bands for 
large $U$ (see Fig.\  \ref{fig_DOS_half}). The bands display equal weight $1/2$ 
and  they are well separated for large $U$ 
\cite{hubba63,hubba64a,hubba64b,georg96}. 
Thus the states without holes or doubly occupied sites are energetically 
well separated from the ones with one or more holes or  doubly occupied sites.
If $U$ is decreased the bands approach each other. As soon as they 
touch the insulating phase  is no longer the
appropriate phase and metallic behavior occurs resulting in the breakdown
of the  mapping to the $t$-$J$ model.\footnote{There may be hysteresis
in the sense that the insulating phase becomes metastable before
it is really unstable on  decreasing $U$ \cite{georg96}.}

\begin{figure}[htb]
\begin{center}
     \includegraphics[width=1\columnwidth]{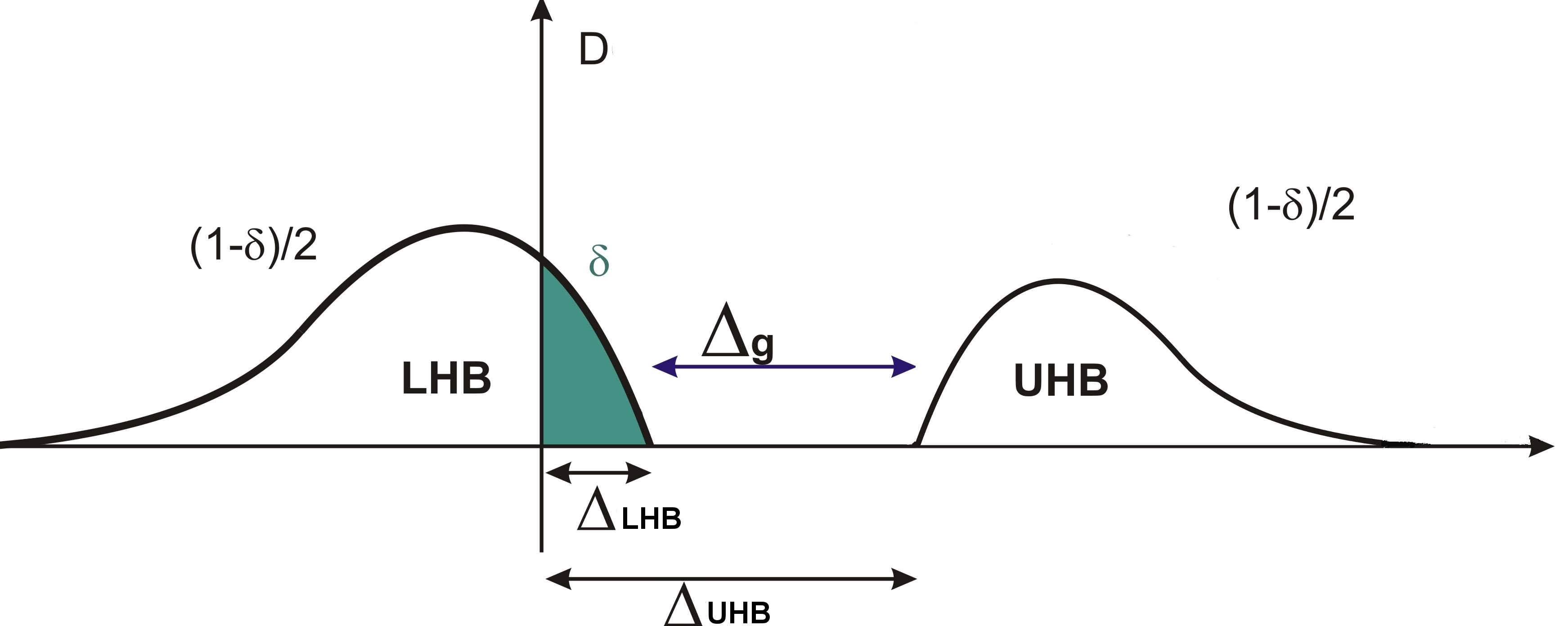}
\end{center}
\caption{(Color online) Density of states for the case of hole doping with a 
doping concentration $\delta$. The weight of the hole state is given by 
$\delta$. The two half-filled states both carry the weight 
$\frac{1-\delta}{2}$. Note that the weight of the lower band is larger than the
 one of the upper Hubbard band \cite{jarre95}. }
\label{fig_DOS_dop}
\end{figure}
The generic density of states obtained in the doped case is depicted in Fig.\ 
\ref{fig_DOS_dop}. The effect of the doping on the density of states consists 
in shifting the Fermi energy into the
lower Hubbard band for hole doping and redistributing the weight of the bands. 
Electron doping is completely analogous in shifting the 
Fermi energy into the upper Hubbard band. Since we focus here
on particle-hole symmetric, bipartite lattices we will
consider hole doping without loss of generality.

It is obvious from the comparison of Fig.\  \ref{fig_DOS_half} with Fig.\ 
\ref{fig_DOS_dop} that the energetic separation
of the Hubbard bands is more subtle in the doped case
than in the half-filled case. 
The bands are shifted depending on $\delta$ \emph{and}
spectral weight is transferred as well. The shift of spectral weight
is a smoking gun evidence for strongly correlated fermionic systems.

Simple counting arguments in the limit $U\rightarrow \infty$ tell us
the distribution of weight. Adding
an $\uparrow$ electron to a site succeeds  with probability $p=\delta
+(1-\delta)/2$. The first term results from the fraction of empty sites; 
the corresponding weight is found at low energy because no doubly occupied 
site has to be created. The second term results from the fraction of sites 
occupied
by $\downarrow$ electrons; the corresponding weight is found at about
$\omega\approx U$ because a doubly occupied site is created. Together with
the sum rule the weights shown in  Fig.\ \ref{fig_DOS_dop} result.

We continue to use
the number of doubly occupied sites (DOs) as criterion to distinguish
different sectors of the Hilbert space. In order to have a
quantitative measure for the energy separation of the 
 sectors with differing number of DOs
the apparent charge gap  $\Delta_g$ is introduced which
 measures the energy separation of 
subspaces. It does not measure the energy gap between two pure states
which is the reason why we call this separation of energy scales
``apparent''. While the apparent charge gap is not an energy gap
in a rigorous sense it is experimentally significant: It quantifies
the energy needed in an Mott insulator  to create a charge
excitation irrespective of the spin state of the system.
For instance, the system can be at a temperature which implies
a disordered paramagnetic spin state while it preserves the
insulating properties.

In the half-filled case the apparent charge gap measures the minimal energy of 
a doubly occupied site moving in an arbitrary spin background. 
The apparent charge gap is reduced for increasing values of the 
band width $W$. 
If the gap vanishes the system is no longer stable against charge fluctuations 
and  the mapping fails \cite{milli91}. 

Obviously, the physical properties of strongly interacting
fermionic systems depend considerably on the doping level  \cite{damas03}. 
This fact leads automatically to the question how the validity of the 
mapping from the Hubbard model to a generalized $t$-$J$ model 
is influenced by doping.
Thus the apparent charge gap has to be determined in dependence on the doping 
$\delta$. Keeping track of the apparent charge  gap we determine the 
parameter range in which the mapping is still justified.
This range of applicability may not be misinterpreted as
phase diagram although it has some similarities \cite{milli91}. 
For instance, for large repulsion $U$ a doped $t$-$J$ model is still 
perfectly well defined while it displays metallic behavior. Of course, 
it is expected that the  applicability of a $t$-$J$ model
decreases upon increasing doping \cite{milli91}.

In view of the above, it is one of our central objectives to
derive  a diagram showing the range of applicability in dependence on the 
doping which has, to our knowledge, not been done before.
Our findings provide access to the limitations of the use of 
$t$-$J$ models in the context of planar cuprates to the extent that
they can be described by a single-band Hubbard model.

The approach used is based on two conceptual ingredients. The first
is a systematically controlled change of basis by means
of continuous unitary transformations \cite{wegne94,stein97,reisc04}.
The second is the choice of a doped reference ensemble without
spin or charge order. 
Within the range of applicability the effective $t$-$J$ model is derived. 
The doping dependence of the effective coupling constants is studied.
 The results are given in dependence on the ratio
$W/U$ and on the dopant concentration 
$\delta$.  The method implemented here uses a self-similar truncation scheme 
to reduce the amount of proliferating terms in the running
Hamiltonian. The truncation is performed according
to the range of the processes, that means rather local
processes are kept while ones of longer range are neglected. 
Hence the local processes acquire a non-perturbative dependence on the 
bare, initial coupling constants of the system.

Furthermore, recently introduced modified generators of the CUT are 
implemented to cope with the vast amount of terms arising 
during the transformation \cite{fisch10a}. Their results
are very close to the previously used particle-conserving generator
\cite{mielk98,knett00a,reisc04}, but they significantly 
facilitate the calculation in terms of required memory and CPU time.

After this introduction,  the model and the method (see Sects.\ 
\ref{chap:model}, \ref{chap:method}) are introduced. In Sect. \ref{chap:gap} 
results for the apparent charge gap are presented and Sect.\ 
\ref{chap:results_coup} provides exemplary results for the doping 
dependence of the coupling constants.
Sect.\ \ref{chap:summ} concludes the article.

%%%%%%%%%%%%%%%%%%%%%%%%%%%%%%%%%%%%%%%%%%%%%%%%%%%%%%%%%%%%%%%%%%%%%%%%%%%%%%%
%%%%%%%%%%%%%%%%%%%%%%%%%%%%%%%%%%%%
\section{Hubbard Model}
\label{chap:model}
%%%%%%%%%%%%%%%%%%%%%%%%%%%%%%%%%%%%%%%%%%%%%%%%%%%%%%%%%%%%%%%%%%%%%%%%%%%%%%%
%%%%%%%%%%%%%%%%%%%%%%%%%%%%%%%%%%%%%

We consider the  fermionic Hubbard model \cite{hubba63,kanam63,gutzw63}. 
It describes electrons with spin $\sigma$ on a lattice site 
$i$ by  their creation operator $\hat{c}_{i,\sigma}^\dagger$
and their annihilation operator $\hat{c}_{i,\sigma}^{\phantom\dagger}$.
The Hamiltonian consists of two terms describing the
single-fermion kinetics ($H_t$) and their interaction ($H_U$)
\begin{subequations}
\begin{align}
\label{hu}
  H &= H_t+ H_U\\
 H_t &= t\sum_{<i,j>}( \hat{c}_{i\sigma}^\dagger 
\hat{c}_{j\sigma}^{\phantom\dagger} + \text{h.c.}) \\
 H_U &= U\sum_i \left(\hat{n}_{i,\uparrow}-
\frac{1}{2}\right)\left(\hat{n}_{i,\downarrow}-\frac{1}{2}\right).
\end{align}
\end{subequations}
The kinetic part consists of the hopping of an electron with spin $\sigma$ 
from site $i$ to site $j$ and vice versa. For this process to take place $i$ 
and $j$ have to be nearest neighbors as indicated by the bracket under the sum.
The corresponding matrix element is denoted by $t$. The 
band width $W$ of the model is given by $W = 2zt$ with the coordination number 
$z$ (number of nearest neighbors).  
In this work the lattice studied is the two dimensional 
square lattice with coordination number $z=4$ so that $W=8t$. 

The second part of the Hamiltonian determines the interaction of the electrons.
 This term constitutes a pure on-site interaction. In $H_U$ the operator 
$\hat{n}_{i,\sigma} = \hat{c}_{i,\sigma}^\dagger
\hat{c}_{i,\sigma}^{\phantom\dagger}$ represents the number operator for the 
electrons. This indicates that putting two electrons on the same site costs 
the additional energy $U$.

In the Hubbard model there are four possible states per site. The site may be 
singly occupied by one electron with spin up or spin down 
$|\uparrow\rangle$,$|\downarrow\rangle$, doubly occupied by two electrons with 
opposite spin $|\downarrow\uparrow\rangle$ or completely
empty $|0\rangle$. The last two 
configurations correspond to charge fluctuations and are referred to as double 
occupancies (DO) in this context.

The interplay of  motion and interaction of electrons in 
the Hubbard model provides a description of the metal-insulator transition 
\cite{imada98}. Another important field of application of the single-band
Hubbard model is the physics of high-$T_\text{C}$ cuprates
\cite{bedno86,lee06}.

For a large Hubbard repulsion $U$ in the half-filled case  
the density of states exhibits two separate bands, see Fig.\ 
\ref{fig_DOS_half}, the so-called lower (LHB) and the upper Hubbard band (UHB).
For infinite $U$ each site of the lattice is occupied by one electron which is 
energetically fixed to its site. If $U$ is finite the electron can move and
virtually hop to adjacent site. Thereby,  DOs are created but the the physics 
remains rather local, that means, the charge correlation length stays
small.
 
Based on the locality of the important processes one may map the Hubbard model 
onto an effective $t$-$J$ model. The generalized $t$-$J$ model  conserves the 
number of DOs. It
comprises a part which describes the magnetic degrees of freedom, which 
is a generalized Heisenberg model, and
a part which describes the motion and interaction of DOs reflecting the
charge degrees of freedom.
In order to obtain a model conserving the number of DOs, processes which
create or annihilate DOs have to be  eliminated. One
systematic way to achieve this objective is the application
of continuous unitary transformations to the Hamiltonian.

For smaller values of $U$ the local picture used here is no 
longer appropriate and the derivation of the $t$-$J$ model is not
justified.

%%%%%%%%%%%%%%%%%%%%%%%%%%%%%%%%%%%%%%%%%%%%%%%%%%%%%%%%%%%%%%%%%%%%%%%%%%%%%%
%%%%%%%%%%%%%%%%%%%%%%%%%%%%%%%%%%%%%%%%%%%%%%%%%%%%%%%%%%%%%%%%%%%%%%%%%%%%%%
%%%%%%%%%%%%%%%%%%%%%%%%%%%%%%%%%%%%%%
%
\section{Continuous Unitary Transformations}
\label{chap:method}
%%%%%%%%%%%%%%%%%%%%%%%%%%%%%%%%%%%%%%%%%%%%%%%%%%%%%%%%%%%%%%%%%%%%%%%%%%%%%%
%%%%%%%%%%%%%%%%%%%%%%%%%%%%%%%%%%%%%%

\subsection{General Framework}

The effective $t$-$J$ model is derived from the Hubbard model by eliminating 
processes which change the number of DOs. This elimination is performed using 
continuous unitary transformations (CUT) 
\cite{wegne94,stein97,mielk98,knett00a,reisc04,fisch10a}.
The elimination is based on a systematic change of the basis
\begin{align}
H(\ell) = \hat{U}^{\phantom\dagger} (\ell)H^{\phantom\dagger}
\hat{U}^\dagger(\ell)
\end{align}  
with a unitary operator $\hat{U}$ and a continuous auxiliary variable 
$\ell$ referred to as the flow parameter.
The transformation is determined by the flow equation
\begin{align}
\frac{d}{d\ell} H(\ell) = \left[\eta(\ell), H(\ell)\right]
\label{flow}
\end{align}
where $\eta(\ell)$ denotes an antihermitian infinitesimal generator.
At $\ell=0$ the transformation starts with the initial Hamiltonian $H$. 
The unitary transformation can be stopped at any arbitrary value of the 
flow parameter $\ell$. Usually, the effective  Hamiltonian is reached 
for $\ell=\infty$. 
Due to the continuity of the transformation it is readjusted to the 
flowing Hamiltonian for every value of $\ell$.

The transformation stops automatically when the commutator 
$[H(\ell),\eta(\ell)]$ vanishes which is generically the case for 
$\ell\rightarrow \infty$, i.e., for convergence for $\ell\rightarrow \infty$.
The structure of the effective Hamiltonian is determined by the 
choice of the generator $\eta(\ell)$.
We first choose the generator which leads to an
effective model conserving the number of DOs.
To this end, we introduce the operator 
\begin{align}
\hat{D} := \sum_i [\hat{n}_{i,\uparrow}\hat{n}_{i,\downarrow} + 
\left(1-\hat{n}_{i,\uparrow}\right)\left(1-\hat{n}_{i,\downarrow}\right)]
\end{align}
counting the number of DOs. 

By the use of $\hat D$ the repulsive part of the Hamiltonian can be 
written as
\begin{align}
\hat{H}_U = \frac{U}{2}\left(\hat{D} - \frac{N}{2}\right)
\end{align}
with $N$ denoting the number of sites.
The kinetic part is split into three parts according to their 
effect on the number of DOs
\begin{align}
\hat{H}_t &= \hat{T}_0 + \hat{T}_{+2} + \hat{T}_{-2}\label{eq:H_2}
\end{align}
where $\hat T_i$ creates $i$ DOs. The terms are given by
\begin{subequations}
\begin{align}
\hat{T}_0 = t_0\sum_{<i,j>, \sigma} \left[\left(1-\hat{n}_{i,\sigma}\right)
\hat{c}_{i,\bar\sigma}^\dagger \hat{c}_{j,\bar\sigma}^{\phantom\dagger}
\left(1-\hat{n}_{j, \sigma}\right)+\right.
\nonumber\\
\left.
\hat{n}_{i,\sigma}\hat{c}_{i,\bar\sigma}^\dagger 
\hat{c}_{j,\bar\sigma}^{\phantom\dagger} \hat{n}_{j,\sigma}+ 
\text{h.c.}\right]
\label{TO}\\
\hat{T}_{+2} = t_{+2}\sum_{<i,j>, \sigma} \left[\hat{n}_{i,\sigma}
\hat{c}_{i,\bar\sigma}^\dagger \hat{c}_{j,\bar\sigma}^{\phantom\dagger}
\left(1-\hat{n}_{j, \sigma}\right)+\right.
\nonumber\\
\left.\hat{n}_{j,\sigma}\hat{c}_{j,\bar\sigma}^\dagger 
\hat{c}_{i,\bar\sigma}^{\phantom\dagger} 
\left(1-\hat{n}_{i,\sigma}\right)\right]
\\
\hat{T}_{-2} = t_{-2}\sum_{<i,j>, \sigma} 
\left[\left(1-\hat{n}_{i,\sigma}\right)
\hat{c}_{i,\bar\sigma}^\dagger 
\hat{c}_{j,\bar\sigma}^{\phantom\dagger}\hat{n}_{j, \sigma}+\right.
\nonumber\\
\left.\left(1-\hat{n}_{j,\sigma}\right)\hat{c}_{j,\bar\sigma}^\dagger 
\hat{c}_{i,\bar\sigma}^{\phantom\dagger}\hat{n}_{i, \sigma}\right]\,.
\end{align}
\end{subequations}
with $\bar{\sigma} := -\sigma$.

The terms contained in $T_0$ have no effect on the number of DOs while
the terms in $T_{+2}$($T_{-2}$) increase (decrease) the number of DOs by two. 
These terms  are the ones that we intend to  eliminate by the transformation.
In the initial Hamiltonian the prefactors $t_{+2}$,$t_{-2}$ and $t_0$ are 
equal, but they evolve differently under the CUT. 

\begin{figure}[htb]
\begin{center}
     \includegraphics[width=0.35\columnwidth]{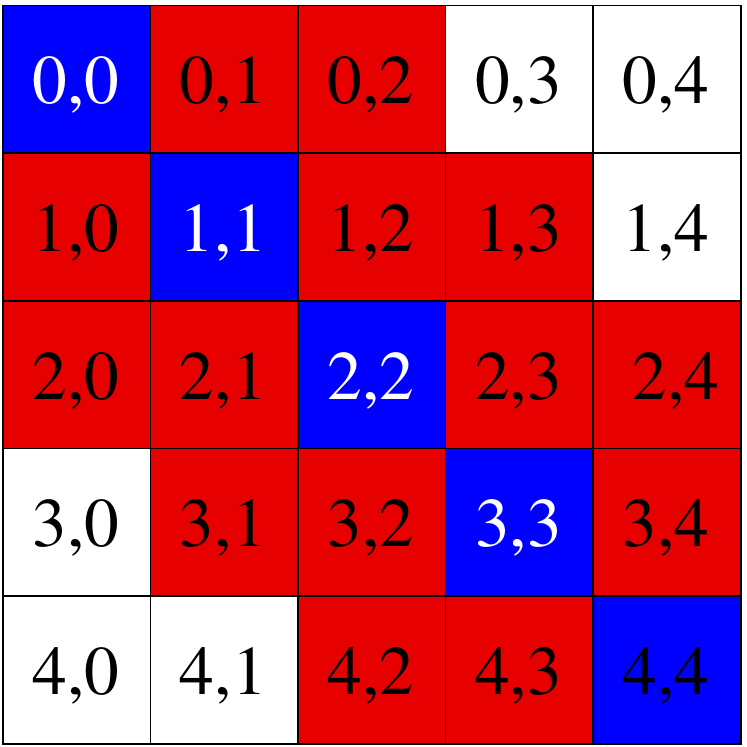}
\end{center}
\caption{(Color online) Schematic diagram of the Hamiltonian with the terms 
contained in the $\eta_\text{pc}$ generator highlighted in red (dark grey). 
These terms will be  eliminated. The terms in the uncoloured squares 
are zero initially and stay zero in the flow 
induced by $\eta_{\text{pc}}$.}
\label{fig_MKU}
\end{figure}

The first generator we use, the so-called \emph{quasiparticle conserving} 
generator $\eta_\text{pc}$, can be expressed by the commutator
\begin{align}
\eta_{\text{pc}}(\ell) = \left[\hat{D}, \hat{H}(\ell)\right]\,.
\label{pc}
\end{align}
This generator corresponds to the generator defined in Refs.\ 
\onlinecite{mielk98,knett00a,reisc04,fisch10a} 
except for a global factor of two, 
which just  implies a multiplicative renormalization of the flow parameter.
The terms comprised by this generator are sketched in Fig.\ \ref{fig_MKU}.
The terms of the Hamiltonian are classified according to their number of 
quasiparticle creation and annihilation operators. The $\left(j,l\right)$ 
block consists of terms with $j$ creation and $l$ annihilation operators. 
Note that such a term requires at least $l$ excitations to be present
in order to become active. But it is also active if more than $l$
excitations are present in the system. In this respect, the 
scheme in Fig.\ \ref{fig_MKU} may not be mistaken to be a matrix.
For a comprehensive presentation we refer the reader to
Ref.\ \onlinecite{fisch10a}.

The quasiparticle conserving generator $\eta_{\text{pc}}$ comprises all terms 
of  the off-diagonal blocks. Due to the structure of the generator 
the block-band structure of the Hamiltonian is preserved during the flow 
\cite{mielk98,knett00a,dusue04a}. 
During the whole flow there will only be terms 
created which change the number of DOs by $0,+2$ or $-2$.

With the definition (\ref{pc}) of the generator the flow equation (\ref{flow}) 
can be calculated. Comparing the contributions on both sides of 
Eq.\ \ref{flow} a set of differential equations for the prefactors of 
the monomials in the creation and annihilation operators  is obtained.
These differential equations are first order in $\ell$ and they
are bilinear in the prefactors entering on the right hand side.

The equations do not form a closed set because in infinite systems
new terms continue to arise on each application of the 
commutator in \eqref{flow}. For $\ell=0$ these terms carry the prefactor zero
because they are not part of the initial Hamiltonian. If we kept all these 
new terms in the remaining calculations we would obtain exact results for the 
effective model. But the number of arising terms is rising exponentially so 
that we have to limit them in number.
For this purpose a truncation scheme is introduced
which specifies the relevance of all term. Less important terms are 
neglected, leading to a closed set of differential equations which can be 
solved numerically. 

In many previous applications a small parameter is used to classify
the arising terms so that a  perturbative treatment results.
In contrast, we are adopting here a truncation scheme  
which classifies the terms according to their 
structure. Such a CUT scheme is usually called self-similar. It
resembles more conventional renormalizations. Effects of infinite order are 
present in the prefactors of the kept terms.

The truncation scheme used in this work keeps or
neglects terms according to their locality, i.e., according
to the range of the represented physical process.
This approach is well justified if the model under study is governed by 
a small correlation length. This is exactly the case for
a Hubbard model at large $U$ where the propagation of charge
degrees of freedom is suppressed by the high energetic cost
of creating a DO.

\subsection{Reference Ensemble and Normal Order}

Before we discuss how we  measure the degree of locality we
 have to find a unique representation for the operators to be
 sure to treat similar terms in the same way. To this end, the monomials 
are expressed as normal-ordered products of local operators.
The normal-ordering we are using is not the standard one
known for the fermionic or bosonic algebra because the
creation or annihilation of a DO can not be represented by
interaction free fermions or bosons. Instead, we
use a reference ensemble. Non-trivial operators are only those
which create or annihilate fluctuations away from the reference ensemble.
 For a given doping  concentration $\delta$ the reference ensemble is 
defined by the statistical operator
\begin{align}
\hat{\rho}_\delta = \Pi_i\left\{\frac{1-\delta}{2}\Big[| 
\uparrow\rangle_i \langle \uparrow|_i + | \downarrow\rangle_i \langle \downarrow|_i \Big]+
\delta | 0\rangle_i \langle 0|_i \right\}
\label{ro}
\end{align}
where the product extends over all lattice sites $i$.  In the half-filled case 
($\delta =0$) the reference ensemble is paramagnetic and the magnetic
degrees of freedom are totally disordered. Each site is
equally probably occupied by an $\uparrow$ or by a $\downarrow$
electron; no direction is singled out, no correlation between
neighboring sites exists.
Charge fluctuations from this reference ensemble are
the empty $|0\rangle$ and the doubly occupied site $|\downarrow\uparrow\rangle$.
Magnetic fluctuations are induced by the application by 
spin operators, see below.

Considering doping we focus on hole doping only because the
model at hand is particle-hole symmetric so that electron doping
leads exactly to the same results.
Hence we include  the empty state $|0\rangle$ in the reference ensemble
\eqref{ro} besides 
the half-filled states with a probability given by the doping level
$\delta$. The remaining weight is again equally distributed over the two
spin states. Note that this extension to the doped case
does not introduce any bias. There is no correlation between
sites nor on each site. Hence the reference ensemble \eqref{ro}
is the mixture with the maximum entropy at given level of doping.

Based on the reference ensemble we define a term as normal-ordered if the 
expectation value of each of its factors of local operators vanishes with 
respect to this ensemble. Thus a normal-ordered local operator fulfills
\begin{align}
\langle A_i\rangle_{\text{ref}} &= 
\delta \langle 0|_i A_i | 0\rangle_i + 
\frac{1-\delta}{2}\left(\langle \uparrow|_i A_i | \uparrow\rangle_i +
 \langle \downarrow|_i A_i | \downarrow\rangle_i\right)
\nonumber\\&=0.
\label{normal-condition}
\end{align}
Based on this condition we define a basis of local normal-ordered 
operators (see Tab.\ \ref{table:list}). Of course, the
identity does not fulfull the condition \eqref{normal-condition}.
But the identity is the trivial action of an operator and obviously
does not create or annihilate any fluctuation away from the
reference ensemble. Without the identity the list of operators
would not be complete.
Any monomial occuring during the flow 
is expressed in the operator basis in Tab.\ \ref{table:list}. 
\begin{table}
	\centering
		\begin{tabular}{|c|c|}
		%\label{table:list}
		\hline
		bosonic & fermionic\\
		\hline
		\hline
			$\mathds{1}$ & $\left(1-\hat{n}_\downarrow\right)
\hat{c}_\uparrow^{\phantom\dagger}$
\\
			$\sigma^z = \hat{n}_\uparrow-\hat{n}_\downarrow$ & 
$\left(1-\hat{n}_\uparrow\right)\hat{c}_\downarrow^{\phantom\dagger}$
\\
		 $\hat{c}_\uparrow^\dagger\hat{c}_\downarrow^{\phantom\dagger}$
 & $\hat{n}_\downarrow\hat{c}_\uparrow$
\\
			$\hat{c}_\downarrow^\dagger
\hat{c}_\uparrow^{\phantom\dagger}$ & $\hat{n}_\uparrow\hat{c}_\downarrow$
\\
				$\hat{c}_\downarrow\hat{c}_\uparrow$ & 
$\hat{n}_\uparrow\hat{c}_\downarrow^\dagger$
\\
				$\hat{c}_\uparrow^\dagger
\hat{c}_\downarrow^\dagger$& $\hat{n}_\downarrow\hat{c}_\uparrow^\dagger$
\\
				$\bar{n}_\delta = \hat{n}_\uparrow+ 
\hat{n}_\downarrow-1 + \delta\mathds{1}$ & $\left(1-\hat{n}_\downarrow\right)
\hat{c}_\uparrow^{\dagger}$
\\
				$\hat{D}_\delta = 
2\hat{n}_\uparrow\hat{n}_\downarrow - \bar{n}_\delta$ & 
$\left(1-\hat{n}_\uparrow\right)\hat{c}_\downarrow^{\dagger}$
\\
				%	\label{table:list}
				\hline
		\end{tabular}
		\caption{Basis of normal-ordered local operators}
		\label{table:list}
\end{table}\noindent

Among the normal-ordered operator basis the operator 
$\bar{n}_\delta = \hat{n}_\uparrow+ \hat{n}_\downarrow-1 + \delta\mathds{1}$ 
occurs. This operator counts the number of electrons on one site relative to 
the mean value of the filling $1-\delta$. In the half-filled case ($\delta=0$) 
the mean value of the filling is $1$. Thus $\bar{n}_0$ applied to an empty site
 yields $-1$. Applied to a doubly occupied site yields $+1$ and a singly 
occupied site leads to $0$. 
In the doped case the counting operator $\hat{D}_\delta$ can be determined 
from the operator for the half-filled case through
\begin{align}
\hat{D}_\delta = \hat{D}_0 +\delta\sum_i\mathds{1}_i\,.
\end{align}

A unique representation for a possible operator occurring in the
Hamiltonian or in the generator is given by the appropriate linear
combination of monomials of the basis operators.
The monomial is the product of local operators 
acting on different sites. Hence the 
expectation value of each monomial also vanishes.

\subsection{Implementation: Truncation Schemes}

\begin{figure}[htb]
\begin{center}
     \includegraphics[width=0.3\columnwidth]{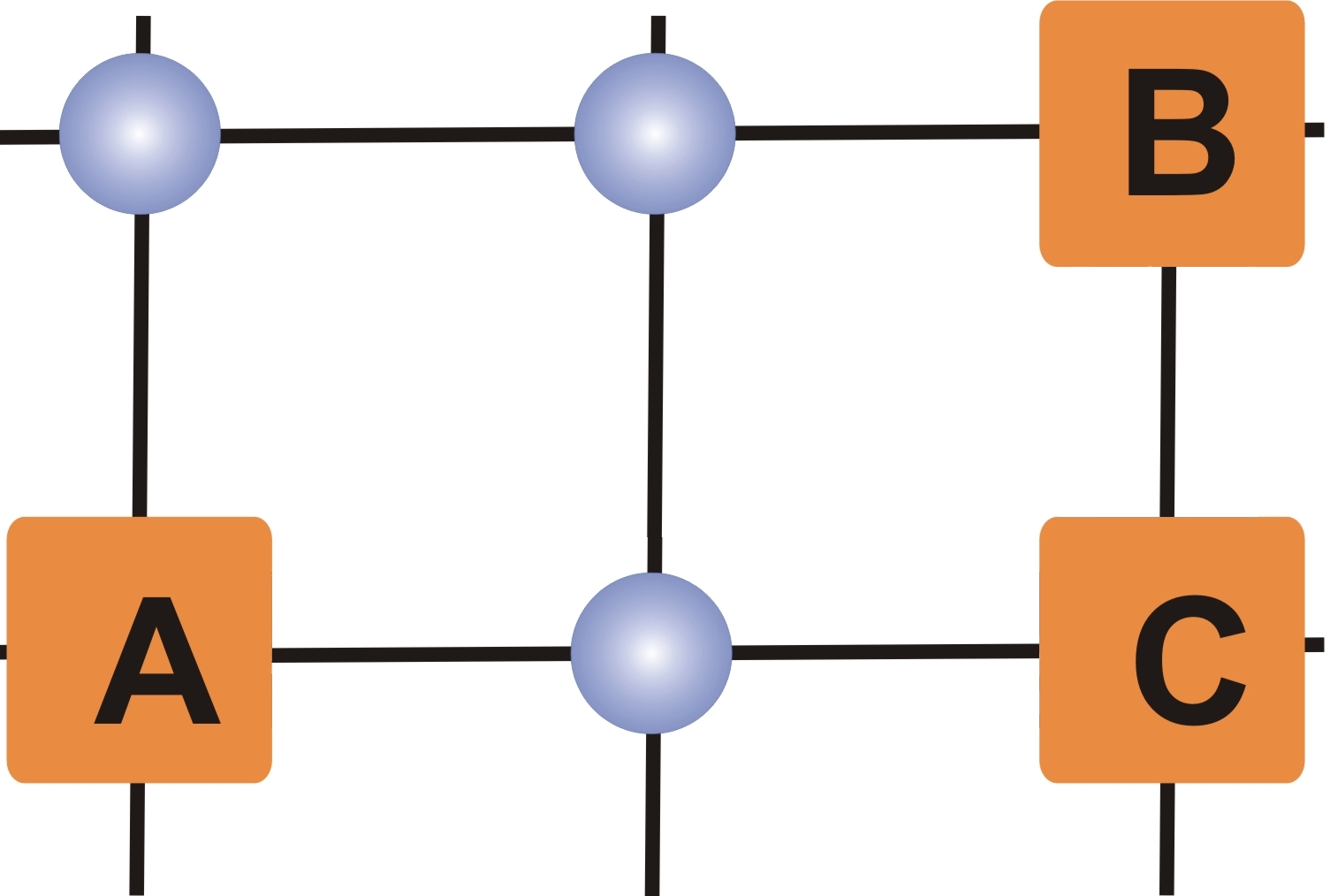}
\end{center}
\caption{(Color online) Cluster of a term with an extension of 3. 
The term consists of the operators A,B and C. 
The taxi cab distance between the outermost operators A 
and C determines its extension.}
\label{fig:ext3}
\end{figure}

The cluster of sites of a monomial is the set of sites on
which the monomial has a non-trivial action. Here `non-trivial'
simply means no to be the identity. Based on the clusters 
a truncation scheme is defined by measuring its locality
by the extension of its cluster.  The extension is defined as the 
maximum taxi cab distance between the outermost cluster sites. 
Thus the extension in $x$ and in $y$ direction have to be summed up.
An exemplary term with an extension of 3 is shown in Fig.\ \ref{fig:ext3}. 
With  the normal-ordered operators given in Tab.\
\ref{table:list}, a monomial with the cluster
shown in Fig.\ \ref{fig:ext3} can be expressed 
as product of 3 local operators. These operators  act on 
the lattice sites (0,0), (2,0) and (2,1). The extension of this cluster in 
x-direction is 2 and its extension in y-direction is 1. 

\begin{figure}[htb]
\begin{center}
     \includegraphics[width=0.35\columnwidth]{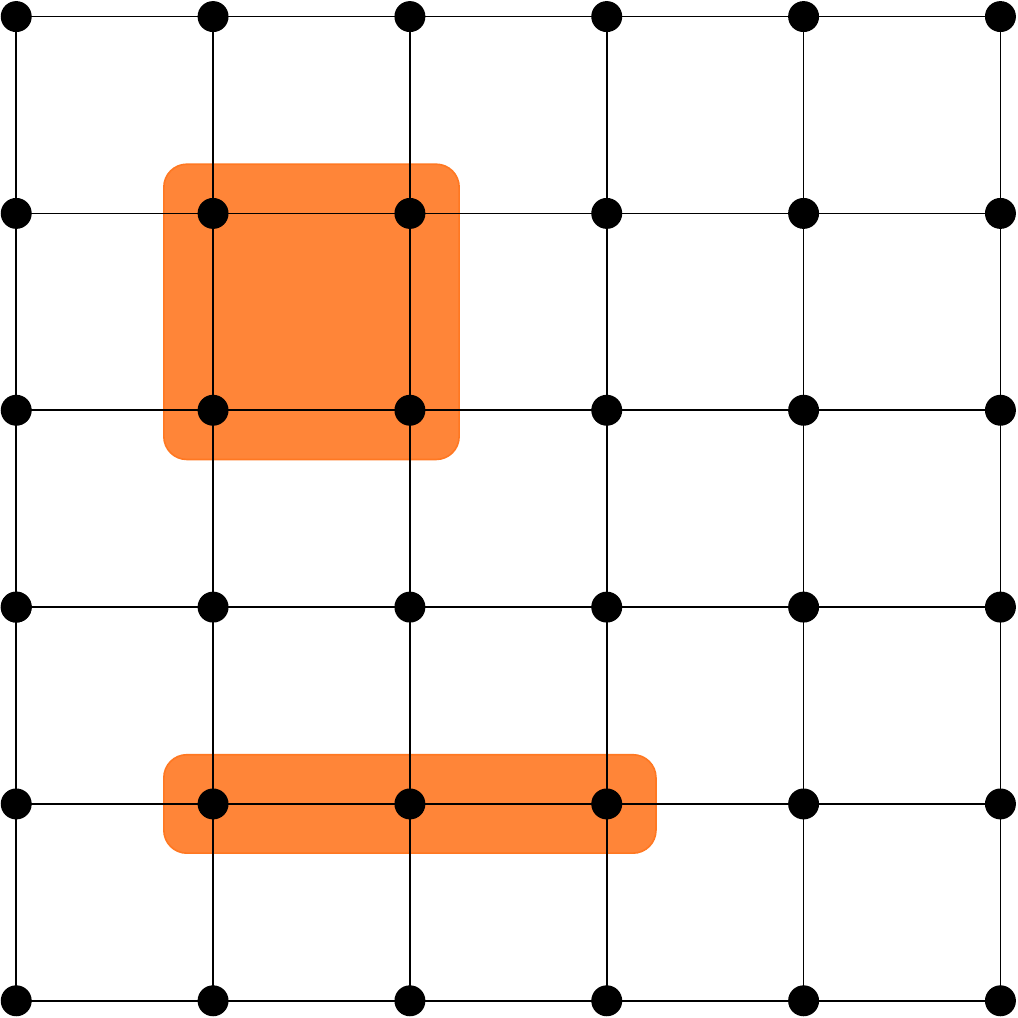}
\end{center}
\caption{(Color online) The maximum clusters occurring in a
calculation with a maximum extension 2. In the following
we will call the calculation based on this truncation
the \emph{plaquette} calculation.}
\label{fig:plaquette}
\end{figure}

To limit the number of generated terms in the course of the flow
we define a maximum extension. For each normal-ordered term generated by
 the commutator the extension is 
determined. A term with an extension higher than the defined maximum
extension is neglected.
A CUT truncated to a maximum extension of two only considers terms 
whose clusters have an extension two  or less.
(see Fig.\ \ref{fig:plaquette}).

\begin{figure}[htb]
\begin{center}
     \includegraphics[width=0.95\columnwidth]{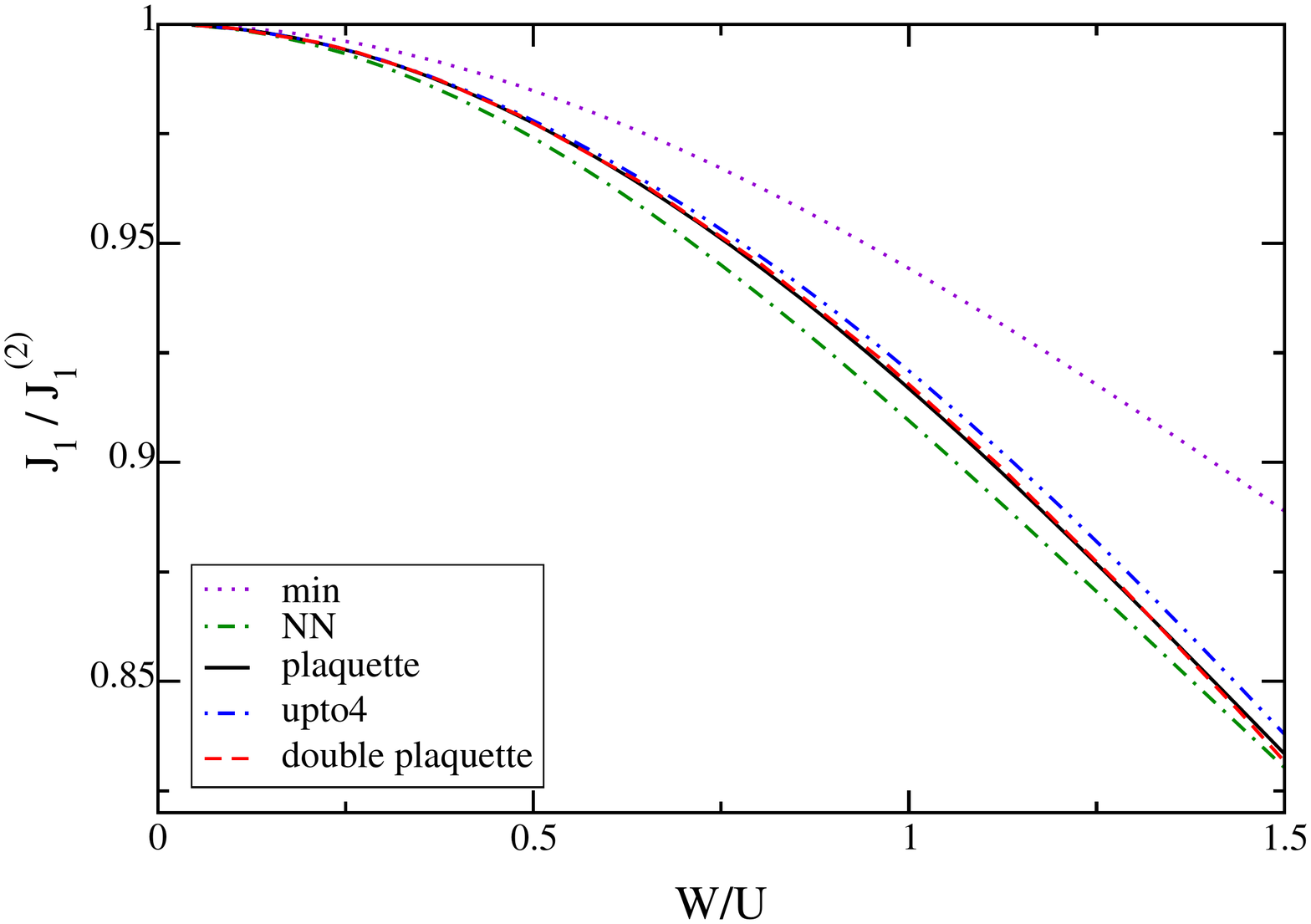}
\end{center}
\caption{(Color online) Effective nearest-neighbor Heisenberg exchange 
$J_1$ obtained in various truncation schemes relative to the 
leading perturbative result in $t/U$.}
\label{fig:trunc_J1}
\end{figure}

In this way more and more extended truncation schemes are used until the 
numerical results do not change noticeably anymore. Then the calculation
is sloppily said to be `converged'. 
To illustrate how the couplings change under the 
influence of different truncation schemes we consider the nearest-neighbor 
magnetic exchange constant $J_1$ 
\begin{align}
H_{\text{Heisenberg}} = 
J_1 \sum_{\langle i,j\rangle} \vec{S}_i\vec{S}_j .
\label{J_1}
\end{align}
In leading perturbation order one obtains $J_1^{(2)} = \frac{4t^2}{U}$ 
\cite{harri67,takah77,macdo88}. 
The results for this coupling constant obtained by CUT are shown in Fig.\ 
\ref{fig:trunc_J1}.

The results are shown for various truncation schemes, where `min' denotes the 
minimal model in which \emph{only} the Heisenberg exchange term is kept
in addition to the terms present in the initial Hamiltonian.
The NN truncation represents a nearest-neighbor calculation 
defined by the maximum extension 1. This calculation 
reproduces the second order perturbative  result for $J_1$. The 
plaquette calculation contains all terms which fit on the clusters shown in 
Fig.\ \ref{fig:plaquette}. This truncation corresponds to a maximum extension 
2. It reproduces $J_1$ up to fourth order in $t/U$. 

A maximum extension of 3 corresponds to the so-called `double plaquette' 
calculation. This truncation scheme is 
sufficient to describe 6th order processes of $J_1$. Since the double 
plaquette calculation results in a large number of terms 
an additional truncation scheme  is introduced, the `upto4' truncation. 
In this calculation a subset of processes with extension 3 are considered
which consist at most of  4 non-trivial local operators on different sites. 

From the results for $J_1$ we deduce
that $J_1$ already converged for a maximum extension 2. Thus the terms 
contained in a plaquette calculation are sufficient to describe the
nearest-neighbor exchange appropriately. 
Considering even more extended terms 
does not change the result significantly.

We emphasize that the calculation with a truncation according to the extension
is \emph{not} equivalent to a finite-size cluster calculation.
The latter acts on a finite cluster with finite Hilbert space
only. The former only restricts the maximum range
of physical processes but remains a calculation on the
thermodynamic, infinitely large system. The latter computes
quantities on finite clusters and a second approximation, for instance
finite-size scaling, is needed
to extend these finite-cluster results to the infinite system.

\subsection{Implementation: Flow Equations}

The choice of a truncation scheme implies that the set of
differential equations describing the flow is finite.
Hence, two tasks have to be accomplished, both of which
are implemented on computers. First, the flow equations
have to be set up. Even though the large number of running
coupling constants makes the use of computer aid indispensable,
this step is an essentially analytic calculation. Second, the flow equations
are integrated numerically which results in the effective 
coupling constants determining the effective model. In the effective model the 
most important subspaces, subspaces of different number of DOs,
 are decoupled from the rest of the Hilbert space. Thus important 
observables can be calculated with less effort.

The derivation of the flow equation is realized by a program implemented in 
C++. This program performs the calculation of the commutators and collects all 
contributions to the same term.
Due to the vast amount of terms it is advantageous to use symmetries to 
increase the efficiency. If a particular term can be generated from another 
term by applying symmetry transformations both terms have the same prefactor. 
The model under study displays the SU(2) spin rotation symmetry and the
point group symmetry of the square lattice. This group contains rotation 
symmetries about $\pi/2$, $\pi$ and $3/2\pi$, reflection symmetries about 
$x$, $y$ and the diagonal. In the half-filled case the particle-hole symmetry 
may be used additionally.
Of the spin rotation symmetry we only exploited the
spin flip symmetry, i.e., the U(1) symmetry of rotations around
$S_z$. In addition, we used that the Hamiltonian is hermitian conjugate
so that adjoint terms also must have the same prefactor.

 By applying the above symmetries up to 64 terms are created
 out of a single term. Since they all carry the same prefactor it is 
sufficient to treat one representative instead of all 64 terms separately. 
By this technique,  the number of terms is reduced from more 
than 1.6 million to 26251 in the double plaquette calculation.
Yet the double plaquette calculation remains costly. It requires
14.7 weeks of CPU  time and more than 20 GB RAM memory.

Compared to the derivation of the flow equations the solution of them 
is straightforward. We start at $\ell=0$ with the initial 
Hamiltonian and integrate the differential equations. At $\ell = \infty$ the 
effective model is reached. Since the integration is performed numerically, 
this limit can not be reached and we stop before at large enough 
values of $\ell$.

In order to have a measure to which extent the CUT is accomplished we 
introduce  the \emph{residual off-diagonality} (ROD) \cite{fisch10a}.
The name is motivated by the idea that the CUT eliminates
the off-diagonal terms. As we will see in the next subsection
the precise choice which terms are eliminated and which are not
depends on the choice of the generator $\eta$. Hence in practice 
the ROD is a measure of the norm of the generator. The ROD is calculated by 
squaring the (real) prefactors of the terms of the generator, 
summing them and finally taking the square  root of this sum.

The ROD measures to what extent the terms in the generator are eliminated 
at the current value of the flow parameter $\ell$. When the ROD vanishes, the 
generator vanishes and consequently the transformation is finished. 
When the ROD is decreased to some small value, for instance $10^{-15}$,
 the calculation can be stopped at 
$\ell < \infty$. The contributions of the remaining off-diagonal terms 
are  negligible so that we consider the model obtained to be
the wanted effective model.

%%%%%%%%%%%%%%%%%%%%%%%%%%%%%%%%%%%%%%%%%%%%%%%%%%%%%%%%%%%%%%%%%%%%%%%%%%%%%%%
%%%%%%%%%%%%%%%%%%%%%%%%%%%%%%%%%%%%%
\subsection{Various Choices of the Generator}
%%%%%%%%%%%%%%%%%%%%%%%%%%%%%%%%%%%%%%%%%%%%%%%%%%%%%%%%%%%%%%%%%%%%%%%%%%%%%%%
%%%%%%%%%%%%%%%%%%%%%%%%%%%%%%%%%%%%%

Because the number of generated terms during the flow 
leads to computational costly  calculations, we consider
various choices of generators for simplification \cite{fisch10a}.
The basic idea of the modified generators is that the
most relevant physics requires only a very small number of DOs.
Hence it may be sufficient to separate subspaces with zero or one DO from
the remaining Hilbert space instead of applying $\eta_\text{pc}$
which eliminates all terms changing the number of DOs.

\begin{figure}[htb] 
\begin{minipage}[hbt]{0.5\columnwidth}
	\centering
	\includegraphics[width=0.65\columnwidth]{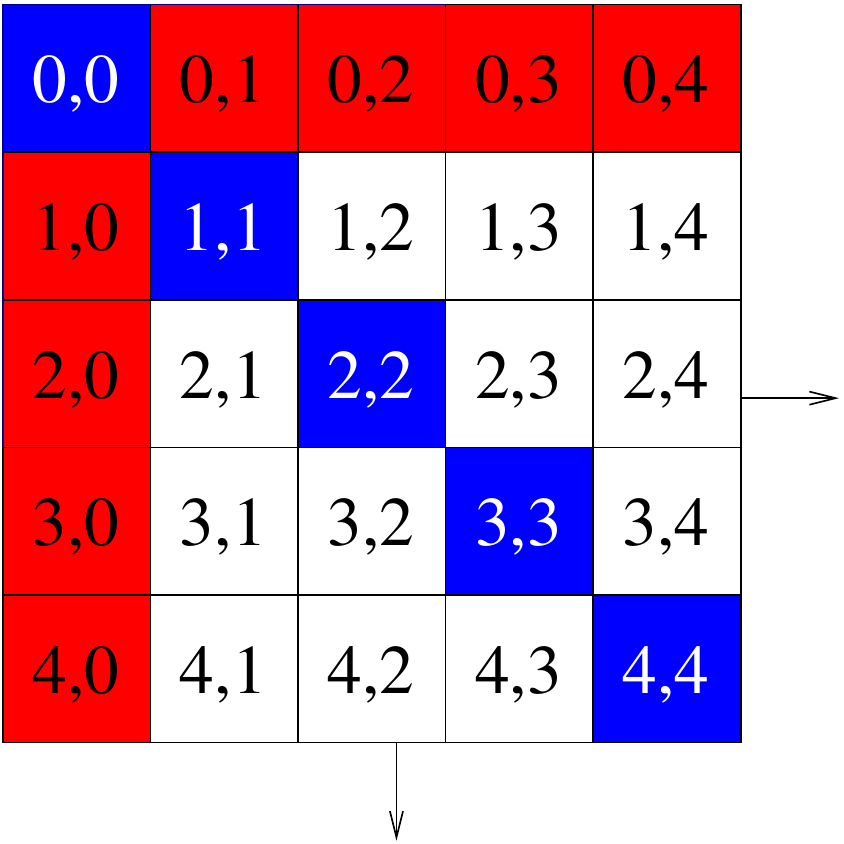}
\end{minipage}
\hspace*{-0.5cm}
$\Rightarrow$
\hspace*{-0.5cm}
\begin{minipage}[hbt]{0.5\columnwidth}
	\centering
	\includegraphics[width=0.65\columnwidth]{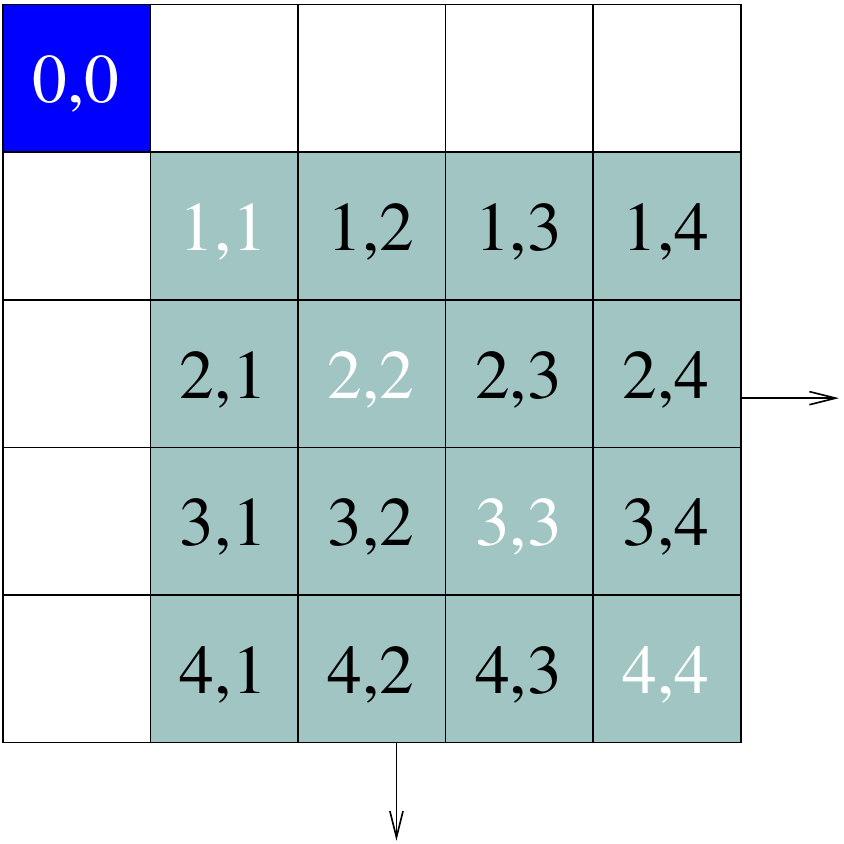}
\end{minipage}
\caption{(Color online) Effect of the gs-generator. The terms of the 
Hamiltonian are labelled according to the number of creation and annihilation 
operators they contain. Thus a term in the  $\left\{i,j\right\}$ block creates 
$i$ DOs after annihilating $j$ DOs.}
\label{fig:gs}
\end{figure}

An obvious example is the derivation of the Heisenberg model
describing the magnetic degrees of freedom without any charges.
Here the separation of the subspace without any DO from 
the remaining Hilbert space is completely sufficient. 
Thus we consider the generator $\eta_{\text{gs}}$ \cite{fisch10a}
\begin{align}
\eta_{\text{gs}}(\ell) = 2\sum_{i>0}^N\left(\hat{H}_0^i(\ell)-
\hat{H}_i^0(\ell)\right)
\label{gs}
\end{align}
where $N$ denotes the number of quasiparticles. The operator 
$\hat{H}_j^i(\ell)$ represents all terms which contain $j$
annihilation operators of DOs and $i$ creation operators of DOs.
This generator contains all 
terms which couple to the subspace without DOs. Note that this subspace is a 
high-dimensional subspace and not a single ground state 
for the model under study in contrast to the situation considered
by Fischer et al.\ \cite{fisch10a}. But the other
conceptual points, e.g., concerning the formulation in second quantization 
and the differences to a matrix formulation \cite{dawso08} are the same.
The Hamiltonian and its evolution under the CUT induced by the
gs-generator \eqref{gs} is graphically represented in Fig.\ \ref{fig:gs}.

\begin{figure}[htb] 
\begin{minipage}[hbt]{0.5\columnwidth}
	\centering
	\includegraphics[width=0.65\columnwidth]{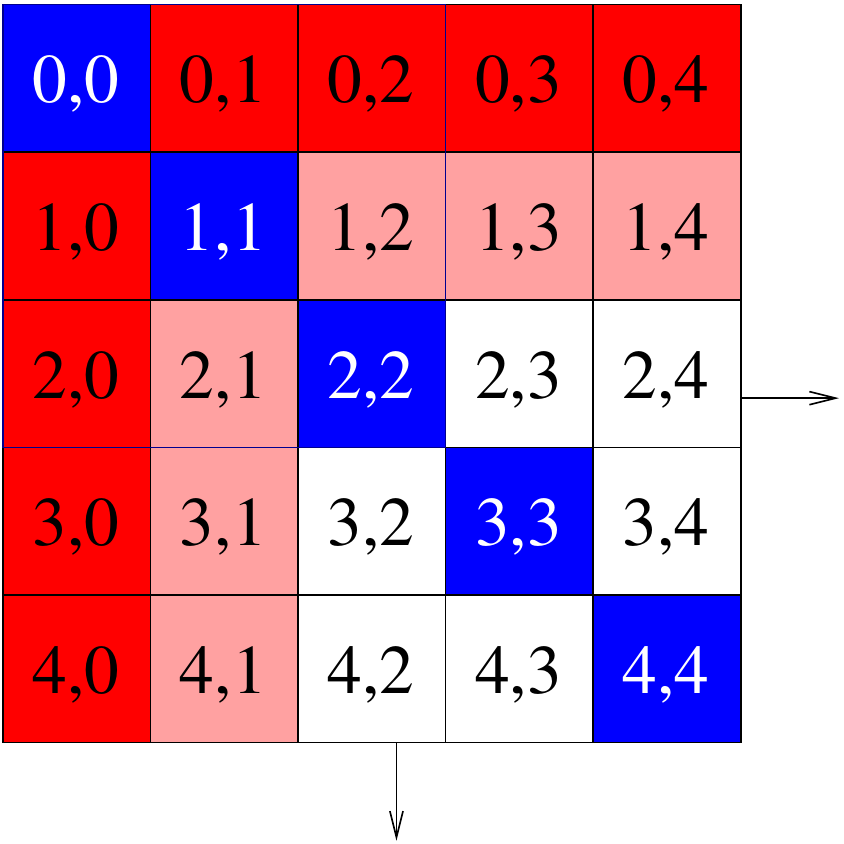}
\end{minipage}
\hspace*{-0.5cm}
$\Rightarrow$
\hspace*{-0.5cm}
\begin{minipage}[hbt]{0.5\columnwidth}
	\centering
	\includegraphics[width=0.65\columnwidth]{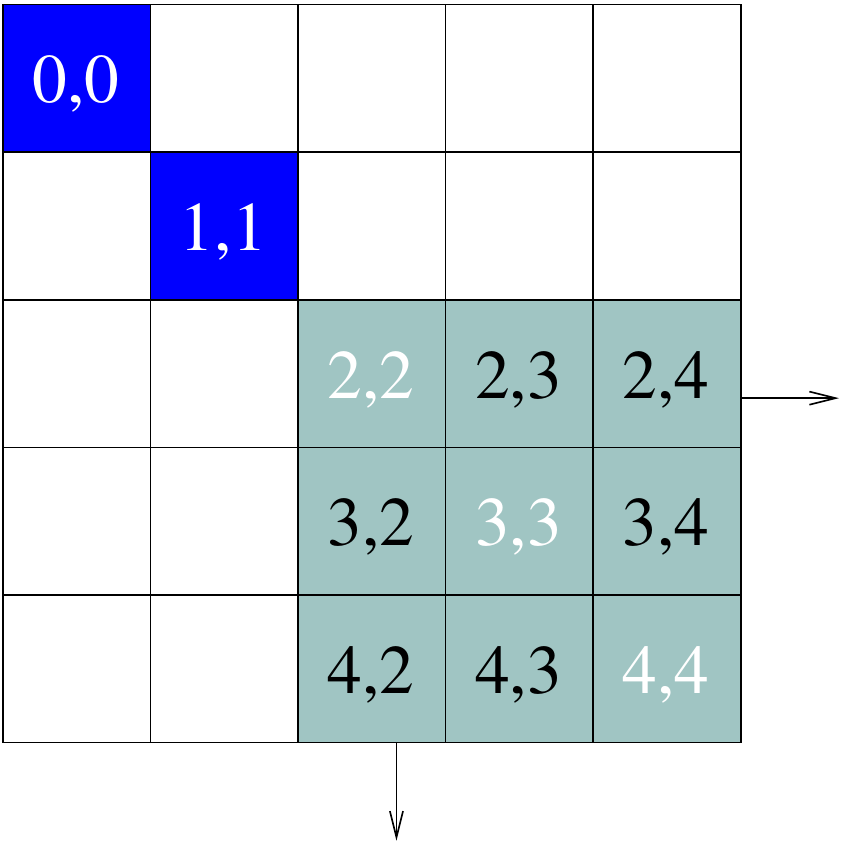}
\end{minipage}
\caption{(Color online)  Effect of the gs,1p-generator. The terms of the 
Hamiltonian are labelled according to the number of creation and annihilation 
operators they contain. Thus a term in the  $\left\{i,j\right\}$ block creates 
$i$ DOs after annihilating $j$ DOs.}
\label{fig:0n1n}
\end{figure}

If in addition we aim at an explicit description of the motion of a single DO 
the generator $\eta_{gs,1p}$ has to be used
\begin{align}
\label{gs1p}
& \eta_{\text{gs,1p}}(\ell) =
\\
\nonumber
& 2\sum_{i>0}^N\left(\hat{H}_0^i(\ell)-\hat{H}_i0(\ell)\right)+
2\sum_{i>0}^N\left(\hat{H}_1^i(\ell)-\hat{H}_i1(\ell)\right).
\end{align}
In this generator the idea of decoupling some subspaces from the
remainder of the Hilbert space is extended to the subspace with one DO. 
Thus also terms coupling to this 
subspace are included as depicted in Fig.\ 
\ref{fig:0n1n}. Whereas the subspaces with zero and with one DO are decoupled 
at the end of the transformation, the other subspaces are still coupled. To 
compute eigenvalues in the subspaces of zero or one DO \emph{only}
these subspaces need to be taken into account.
In contrast, eigenstates involving two or more DOs still require the
diagonalization of the full Hilbert space.

The CPU time needed for a double plaquette calculation using various
generators are given in Tab.\ \ref{table:CPU}.
In the case of a CUT based on the gs,1p-generator the use of
symmetries is even more  efficient. 
Using all of them except for the particle-hole symmetry reduces 
the number of terms from 5 million to 55049. The reader may be
surprised that these numbers are larger than those for
the particle-conserving $\eta_\text{pc}$ although
terms linking subspaces with higher number of DOs are not
decoupled. The explanation is that $\eta_\text{pc}$
has the additional feature that it preserves the
block-banded structure of the Hamiltonian while
the other generators do not \cite{mielk98,knett00a,fisch10a}.
Yet the modified generators induce a simpler and faster
CUT as shown by the numbers in Tab.\ \ref{table:CPU}.
\begin{table}
	\centering
		\begin{tabular}{|c|c|c|}
			\hline
		pc-generator & gs,1p-generator & gs-generator\\
		\hline
		\hline
			102 days & 51 days & less than 10 days\\
				\hline
		\end{tabular}
		\caption{Comparison of CPU time needed for a double plaquette 
calculation using all symmetries in the half-filled case with different 
generators.}
		\label{table:CPU}
\end{table}

%%%%%%%%%%%%%%%%%%%%%%%%%%%%%%%%%%%%%%%%%%%%%%%%%%%%%%%%%%%%%%%%%%%%%%%%%%%%%%
%%%%%%%%%%%%%%%%%%%%%%%%%%%%%%%%%%%%%%%%%%%%%%%%%%%%%%%%%%%%%%%%%%%%%%%%%%%%%%%
%%%%%%%%%%%%%%%%%%%%%%%%%%%%%%%%%%%%%
\section{Minimal and Nearest Neighbor Model}
\label{chap:NN}
%%%%%%%%%%%%%%%%%%%%%%%%%%%%%%%%%%%%%%%%%%%%%%%%%%%%%%%%%%%%%%%%%%%%%%%%%%%%%%
%%%%%%%%%%%%%%%%%%%%%%%%%%%%%%%%%%%%%%

In this section we present analytic solutions of the 
flow equations which are possible for the two simplemost truncation schemes.
Due to the simplicity of the truncation schemes
no difference between the different choices of the generator are found.
For concreteness, we consider the particle-conserving $\eta_\text{pc}$
here.

\subsection{Minimal Model}

The calculation of the minimal model starts by studying all processes on 
between adjacent sites. 
This nearest neighbor (NN) calculation is equivalent to a maximal extension of 
$e=1$. To arrive at the minimal model all terms not present in the initial 
Hamiltonian except the NN Heisenberg exchange are neglected.

The initial generator takes the form 
\begin{align}
\eta(\ell) = \left[\hat{D},\hat{H}(\ell)\right] = 2\hat{T}_{+2}-2\hat{T}_{-2}
\end{align}  
with the flow parameter $\ell$. 
Inserting this definition in the flow equation \eqref{flow} we calculate
\begin{align}
\frac{d}{d\ell} H(\ell) = \left[\eta(\ell),\hat{H}_U(\ell) + \hat{H}_t(\ell)
\right].
\end{align}
From the commutator new terms arise \cite{hamer10}. 
In the minimal model all terms except the NN Heisenberg exchange 
\begin{subequations}
\begin{align}
H_{\text{Heisenb.}} = J_1(\ell) \sum_{\langle i, j \rangle} \vec{S}_i\vec{S}_j
\\
\vec{S_i}\vec{S_j} = \frac{1}{2}\left(\sigma_i^+\sigma_j^- + 
\sigma_i^-\sigma_j^+\right) + \frac{1}{4}\sigma_i^z\sigma_j^z
\end{align}
\end{subequations}
are omitted. 
The coupling constant $J_1(\ell)$ starts at $J_1(0)=0$ because it is not 
part of the initial Hamiltonian $H(0)$. For $\ell \neq 0$ it evolves according 
to the flow equation. 

In this simple case the flow equation can be solved analytically
for a general lattice with coordination number $z$ \cite{hamer10}. For 
$\ell = \infty$ the effective model is reached. The effective coupling 
constants take the form
\begin{subequations}
\begin{align}
t_{0,\text{eff}} &= t_0
\\
t_{+2,\text{eff}} &= 0 
\label{t2}\\
U_{\text{eff}} &= \frac{1}{2}\sqrt{4U_0^2+16 z t_0^2}
\\
J_{1,\text{eff}} 
&= \frac{1}{z}\sqrt{4U_0^2+16 z t_0^2}-\frac{2}{z}U_0.
\label{j1-min}
\end{align}
\end{subequations}
To obtain the equations for the square lattice $z=4$ must be inserted.
The variables $t_0$ and $U_0$ 
represent the initial, unrenormalized values of the hopping and the 
Hubbard repulsion. For simplicity we will omit the subscript 0 and label the 
unrenormalized values by $t$ and $U$ henceforth.
Since the terms $T_{+2}$ and $T_{-2}$ are hermitian conjugates and we assume 
their coefficients to be real $t_{+2} = t_{-2}$ holds. 
From (\ref{t2}) we see that the terms contained in $T_{+2}$ and $T_{-2}$ are 
eliminated as it should be because they change the number of DOs. 
The effective model is eventually given by
\begin{align}
\hat{H}_{\text{eff}} = U_{\text{\text{eff}}}\frac{1}{2}\hat{D}+ \hat{T}_0 + 
J_{1,\text{eff}}\sum_{<i,j>}\vec{S_i}\vec{S_j}\,.
\end{align}

%%%%%%%%%%%%%%%%%%%%%%%%%%%%%%%%%%%%%%%%%%%%%%%%%%%%%%%%%%%%%%%%%%%%%%%%%%%%%%%
%%%%%%%%%%%%%%%%%%%%%%%%%%%%%%%%%%%%%
\subsection{Nearest Neighbor Model}
%%%%%%%%%%%%%%%%%%%%%%%%%%%%%%%%%%%%%%%%%%%%%%%%%%%%%%%%%%%%%%%%%%%%%%%%%%%%%%%
%%%%%%%%%%%%%%%%%%%%%%%%%%%%%%%%%%%%%

In the nearest neighbor model all terms arising from a nearest neighbor 
calculation are included in the effective model as well as in the generators. 
This is the full calculation with extension 1. It can still be solved 
analytically \cite{hamer10}. 
In this truncation scheme the Heisenberg exchange coupling is 
given by
\begin{align}
 J_{1,\text{eff}} = \frac{2}{3+z}\left(\sqrt{U_0^2+4(3+z)t_0^2}-U_0\right).
\end{align}
Note the differences to the result of the minimal model \eqref{j1-min}.
Of course, this differences arises only beyond leading order.

Besides the Heisenberg exchange the calculation contains the term $\hat{H}_V$ 
describing the interaction of two DOs
\begin{align}
\hat{H}_V(\ell) = V(\ell) \sum_{<i,j>} \bar{n}_i\bar{n}_j.
\end{align}
The operator $\bar{n}$ counts the amount of electrons 
compared to the mean value of the filling. For the half-filled case the mean 
value is $1$.The third term created during the flow is 
\begin{align}
\hat{H}_p(\ell) = V_p(\ell) \sum_{<i,j>} \left(\hat{c}_{i\uparrow}^\dagger 
\hat{c}_{i\downarrow}^\dagger \hat{c}_{j\downarrow}^{\phantom\dagger}
\hat{c}_{j\uparrow}^{\phantom\dagger}+\text{h.c.}\right)\,.
\label{eq:V_p}
\end{align}
describing pair hopping processes of DOs. 
One of the processes contained in this 
term is the hopping of two electrons from site $j$ to an empty site $i$. As 
the empty state as well as the doubly occupied state represents a DO this 
process does not change the number of DOs.
 In the effective model $V$ and $V_p$ take the values
\begin{subequations}
\begin{align}
V_{\text{eff}} &= -\frac{2}{3+4z}\sqrt{U2+4(3+z)t2}+\frac{2}{3+4z}U_0\\
V_{p,\text{eff}} &= \frac{4}{3+4z}\sqrt{U2+4(3+z)t2}+\frac{4}{3+4z} U_0.
\end{align}
\end{subequations}

In the case of doping another contribution to the flow equation also arises. 
It reads
\begin{align}
\hat{H}_\mu &= \mu \sum_i \bar{n}_{i,\delta}
\label{eq:mu}
\end{align}
and determines the chemical potential $\mu$. In the effective model this 
constant takes the value
\begin{align}
\mu_{\text{eff}} &= \frac{\delta z}{2(3+z)} U\sqrt{1+4(3+z)\frac{t^2}{U^2}} - 
\frac{z}{2(3+z)}U\delta\,.
\end{align}
with the coordination number $z$.
In leading order in $\frac{t}{U}$ this yields a chemical potential which 
depends linearly on the doping constant $\delta$ and on the coordination 
number of the lattice
\begin{align}
\mu^{(2)} = \delta z \frac{t^2}{U}\,.
\end{align}

%%%%%%%%%%%%%%%%%%%%%%%%%%%%%%%%%%%%%%%%%%%%%%%%%%%%%%%%%%%%%%%%%%%%%%%%%%%%%%
\section{Influence of the Choice of Generator}
\label{chap:generators}
%%%%%%%%%%%%%%%%%%%%%%%%%%%%%%%%%%%%%%%%%%%%%%%%%%%%%%%%%%%%%%%%%%%%%%%%%%%%%%%
%%%%%%%%%%%%%%%%%%%%%%%%%%%%%%%%%%%%%

\begin{figure}[htb]
\begin{center}
       \includegraphics[width=0.98\columnwidth]{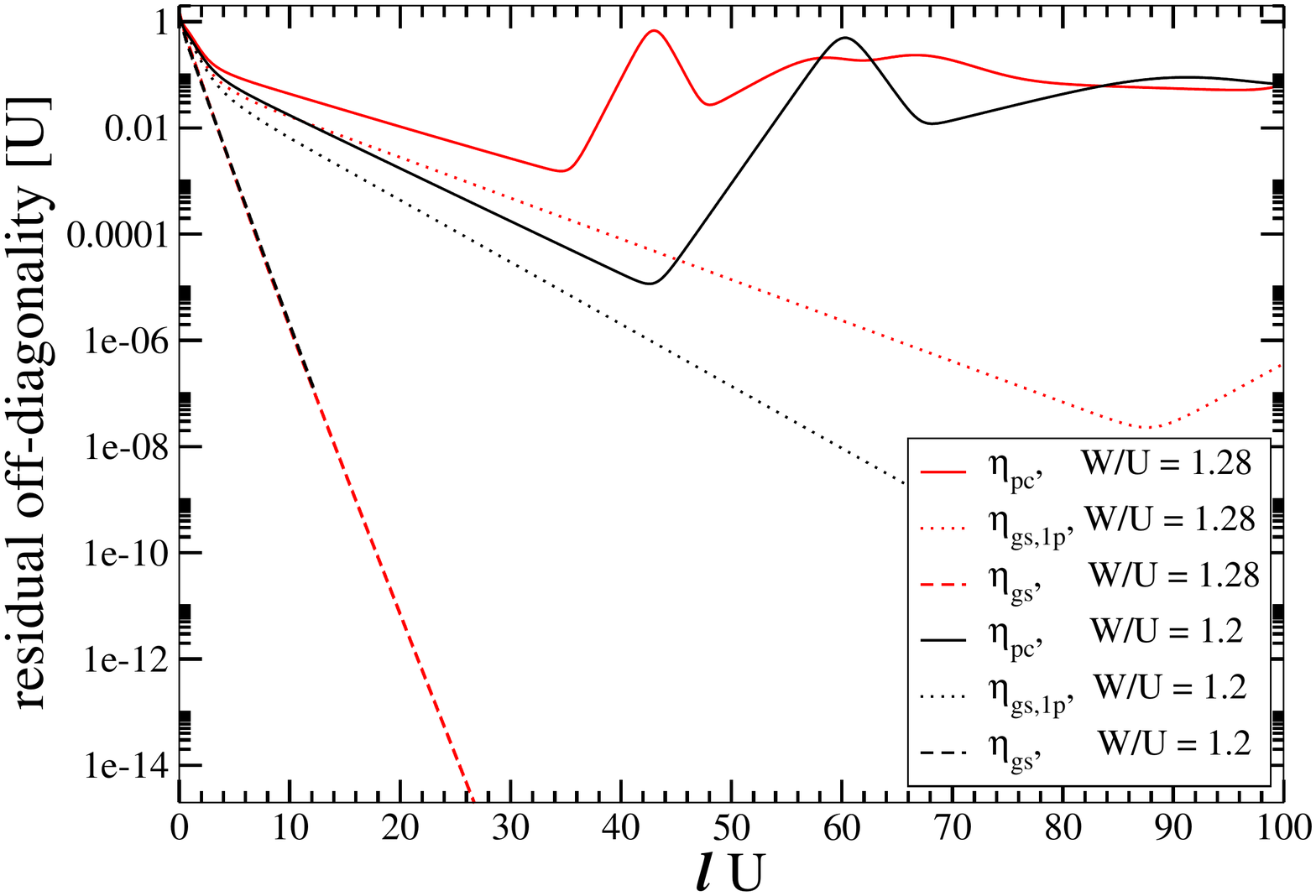}
\end{center}
\caption{(Color online) Behavior of the residual off-diagonality (ROD) in a 
double  plaquette calculation for various choices of the generator
for various initial ratios $W/U$.}
\label{fig_ROD_gen}
\end{figure}

In this section the influence of the choice of the generator is studied.
First, we consider the ROD  defined in Sect.\ \ref{chap:method} for the 
gs-generator \eqref{gs}, gs,1p-generator (\ref{gs1p}),
 and the pc-generator (\ref{pc}).
The results for the ROD obtained in a double plaquette calculation at 
half-filling are shown in Fig.\ \ref{fig_ROD_gen} as functions of the 
continuous flow  parameter $\ell$. 

For the gs-generator, the ROD converges for all values of the ratio $W/U$. In 
contrast, the ROD for the gs,1p-generator and even more
pronounced the ROD for the pc-generator show non-monotonic behavior 
for larger values of $W/U$. 

Non-monotonic behavior of the ROD suggests that the
intended transformation does not succeed.
There is no strict statement that a successful CUT has to
have a monotonic ROD. It is well possible that the ROD displays
local maxima which indicate that some energy eigenstates are 
re-ordered, see for instance Ref.\ \onlinecite{dusue04a}. 
If the CUT is performed
without approximation any unitary transformation is as good as any other.
But since we have to truncate many terms the upturn of the ROD
indicates a potential loss of accuracy. If the total norm of the off-diagonal
terms is large there is still a significant transformation to be done.
In the course of this transformation the truncation of terms may
introduce significant errors. In return, a quickly decreasing ROD
indicates that all coefficients to be eliminated decay fast and
significant truncation errors are less likely. But we like to stress
that the behavior of the ROD is only an indicator for possible
truncation errors which eventually may imply that the
intended mapping breaks down.

The faster convergence of the gs-generator is straightforward
to understand because  the gs-generator comprises only terms which create
DOs from the reference ensemble or which
annihilate them, see Eq.\ \eqref{gs} and 
Fig.\  \ref{fig:gs}. As long as there is a finite charge gap $\Delta_g$ these
processes are exponentially suppressed: $\propto \exp(-\Delta_g \ell)$.
For the gs,1p-generator the processes starting from one 
DO creating two additional DOs can be more difficult to suppress
if they decrease the total energy. This is possible if the DOs
disperse and a DO at high energies decays into three DOs at lower
energy. 

The pc-generator aims in addition at eliminating 
processes starting from two and more DOs so that there are even more
processes which may decrease the total energy while the number
of DOs increases. Hence we are not surprised to see that
the pc-generator induces a flow of the ROD which 
displays even more pronounced non-monotonic behavior.

From these observations the conclusion to always favor the
gs-generator suggests itself. But it is in fact a trade-off.
The gs-generator is quicker to implement and more robust in its
convergence, but it achieves less because it decouples only
the subspace without any DOs. For deriving only an extended
Heisenberg model this is completely sufficient and hence for this aim the
gs-generator is the generator of choice. But if one is additionally
interested in an explicit description of the dynamics of DOs
the other generators are advantageous as we will illustrate next.

\begin{figure}[htb]
\begin{center}
     \includegraphics[width=0.80\columnwidth]{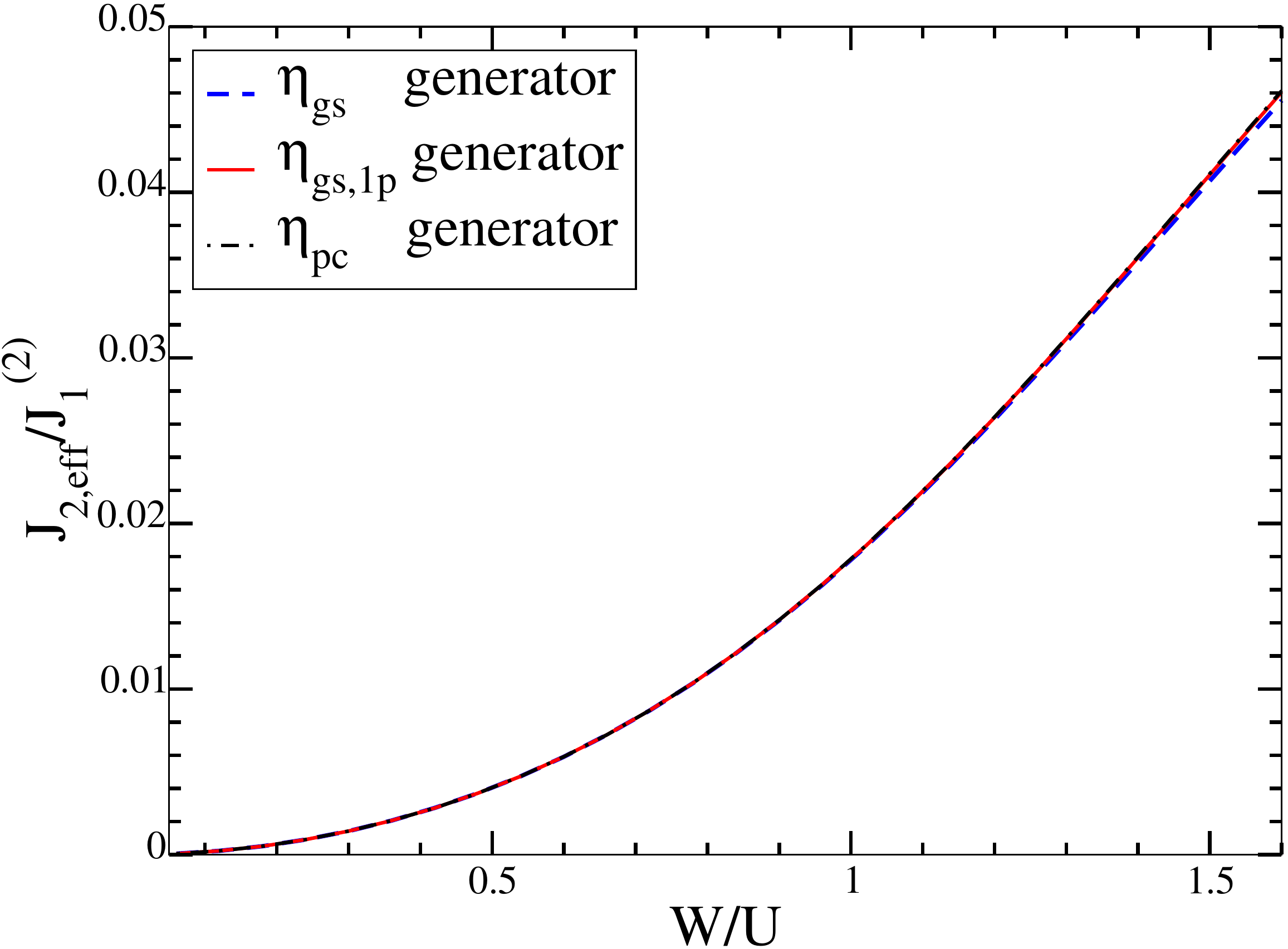}
\end{center}
\caption{(Color online) Behavior of the next nearest neighbor interaction $J_2$
 for different generators as function of $W/U$ determined in an upto4 
calculation. The results for the three generators almost coincide
even for large values of $W/U$.}
\label{fig_J2_gen}
\end{figure}

To see how the coupling constants are influenced by the choice of
generator a few exemplary results are shown.
One of the most important magnetic coupling constants besides the nearest 
neighbor coupling $J_1$ is the Heisenberg interaction between next-nearest 
neighbors $J_2$, i.e., diagonal over a plaquette. 
The behavior of this exchange coupling as function of $W/U$ for 
the three generators is depicted in Fig.\ \ref{fig_J2_gen}.

The curves for all three generators almost coincide. This underlines that  it 
is completely sufficient to use the gs or the gs,1p-generator for the 
determination of  this coupling. All processes contributing to this
coupling are included in the CUT induced by the gs-generator. 
This can be understood from the 
fact, that the magnetic coupling $J_2$ describes an interaction within the 
subspace of half-filled states, i.e., the subspace without any DO. 
This subspace is decoupled from the remainder of the Hilbert space
by all three generators. If no truncation errors occurred, all 
three generators would indeed yield precisely the same result, cf.\
Ref.\ \onlinecite{fisch10a}.

\begin{figure}[htb]
\begin{center}
     \includegraphics[width=0.80\columnwidth]{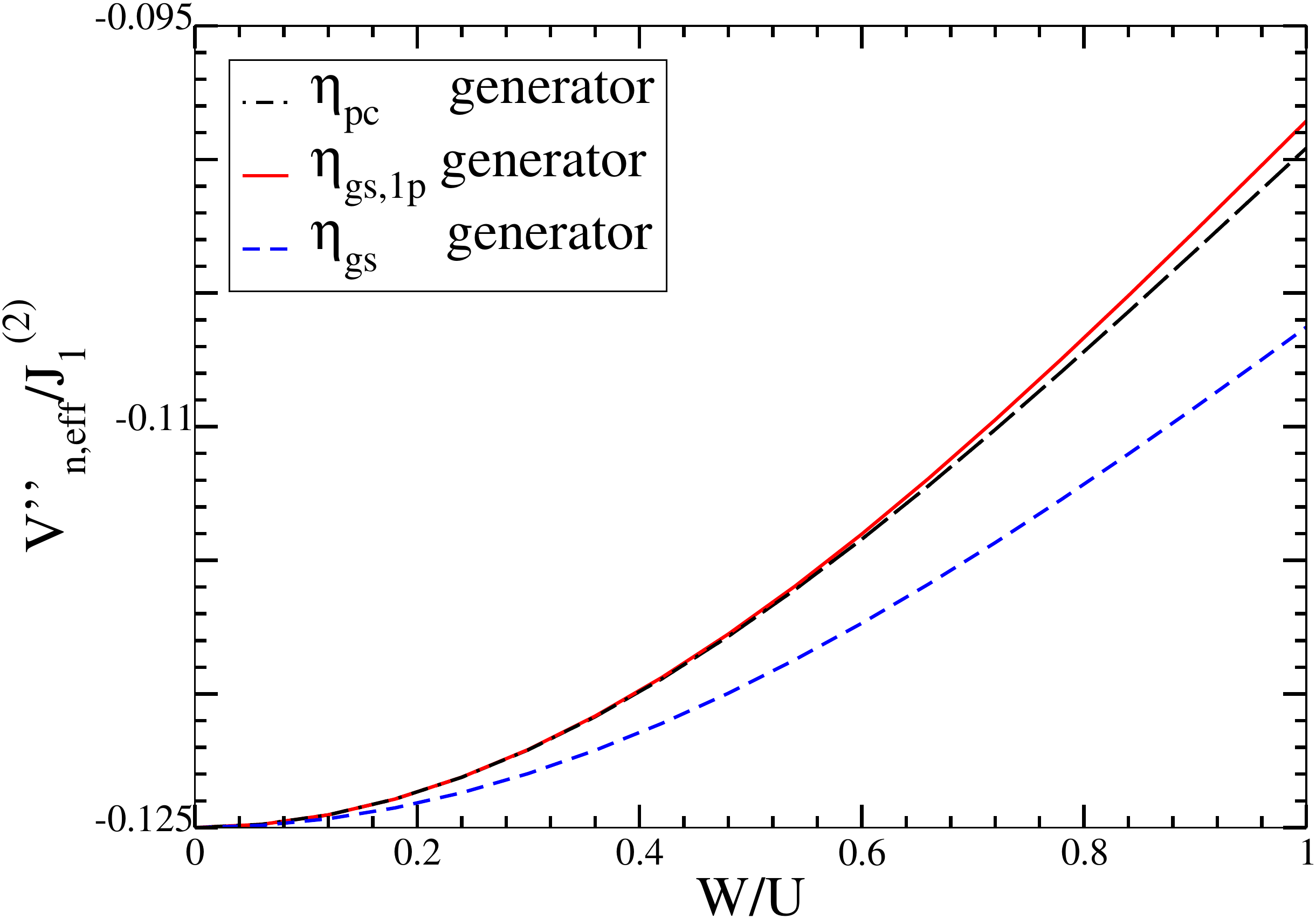}
\end{center}
\caption{(Color online) Density-density interaction between third nearest 
neighbors for different generators obtained in an upto4 calculation. The 
results for the $pc$- and the $gs,1p$-generator agree very well 
whereas the results of $gs$-generator show larger deviations.}
\label{fig_V_n_gen}
\end{figure}
In contrast to a pure spin-spin coupling the term 
\begin{align}
\hat{H}_{Vn}'' &= V''_n\sum_{\alpha, \beta}
\sum_{\langle\langle i,j\rangle\rangle}
\left\{\left(1-\hat{n}_{i,\alpha}^{\phantom\dagger}\right)
\hat{c}_{i,\bar{\alpha}}^\dagger\hat{c}_{j,\bar{\beta}}^{\phantom\dagger}
\left(1-\hat{n}_{j,\beta}^{\phantom\dagger}\right)\bar{n}_k\right.
\nonumber\\
& \left. +\hat{n}_{i,\alpha}^{\phantom\dagger}\hat{c}_{i,\bar{\alpha}}^\dagger
\hat{c}_{j,\bar{\beta}}^{\phantom\dagger} \hat{n}_{j,\beta}^{\phantom\dagger}
\bar{n}_k+\text{h.c.}
\right\}. 
\label{def:V_n_ss}
\end{align}
acts on two DOs. It describes the hopping of an electron from a singly 
occupied site to an empty site under the condition that site $k$ is occupied 
by a DO. It is a process which is not active on the subspace with zero
or only one DO. We do not expect that the results for the different
generators agree.
Indeed, the results for this coupling constant (Fig.\ \ref{fig_V_n_gen}) 
show rather  large deviations of the results obtained by the gs-generator 
from the results  obtained by  the other two generators. This illustrates
that the gs-generator induces a different unitary transformation than
the other two generators. Note that the deviations do not necessarily imply 
that the gs-result is less accurate because it results from the
representation of the Hamiltonian in a different basis.

In view of the above arguments, it is surprising that the results of
the gs,1p-generator and the pc-generator are so close to each other.
From their definitions we expect that the pc- and the gs,1p-results
agree very well for processes involving a single DO, but not necessarily
for processes involving two DOs.

In conclusion, the question which  generator is optimum cannot
be answered generally. It depends on the particular objective of
the intended investigation.

%%%%%%%%%%%%%%%%%%%%%%%%%%%%%%%%%%%%%%%%%%%%%%%%%%%%%%%%%%%%%%%%%%%%%%%%%%%%%%
%%%%%%%%%%%%%%%%%%%%%%%%%%%%%%%%%%%%%%%%%%%%%%%%%%%%%%%%%%%%%%%%%%%%%%%%%%%%%%
\section{apparent charge gap}
\label{chap:gap}
%%%%%%%%%%%%%%%%%%%%%%%%%%%%%%%%%%%%%%%%%%%%%%%%%%%%%%%%%%%%%%%%%%%%%%%%%%%%%%
%%%%%%%%%%%%%%%%%%%%%%%%%%%%%%%%%%%%%%

In the previous sections we have started to discuss the issue
for which conditions the intended mapping from the Hubbard model
to a generalized $t$-$J$ model is possible and justified.
Qualitatively it is obvious that $U$ must be large enough. But
quantitative indicators are needed. Here we aim at giving a
quantitative estimate for the parameter range in which the 
mapping is justified.

\subsection{General Considerations}

The basis of the transformation from the Hubbard model to the $t$-$J$ model is 
the elimination of charge fluctuations. These charge fluctuations correspond to
changes in the number of DOs. The corresponding processes are the ones that 
we consider  to be off-diagonal. To be able to eliminate such processes the 
subspaces with  differing numbers of DOs have to be separated in energy. 
In Sect.\ \ref{chap:method} we introduced the residual off-diagonality (ROD) 
as a  measure to decide if for a given $\ell$ the off-diagonal terms are small 
enough to be neglected.

\begin{figure}[htb]
\begin{center}
        \includegraphics[width=0.95\columnwidth]{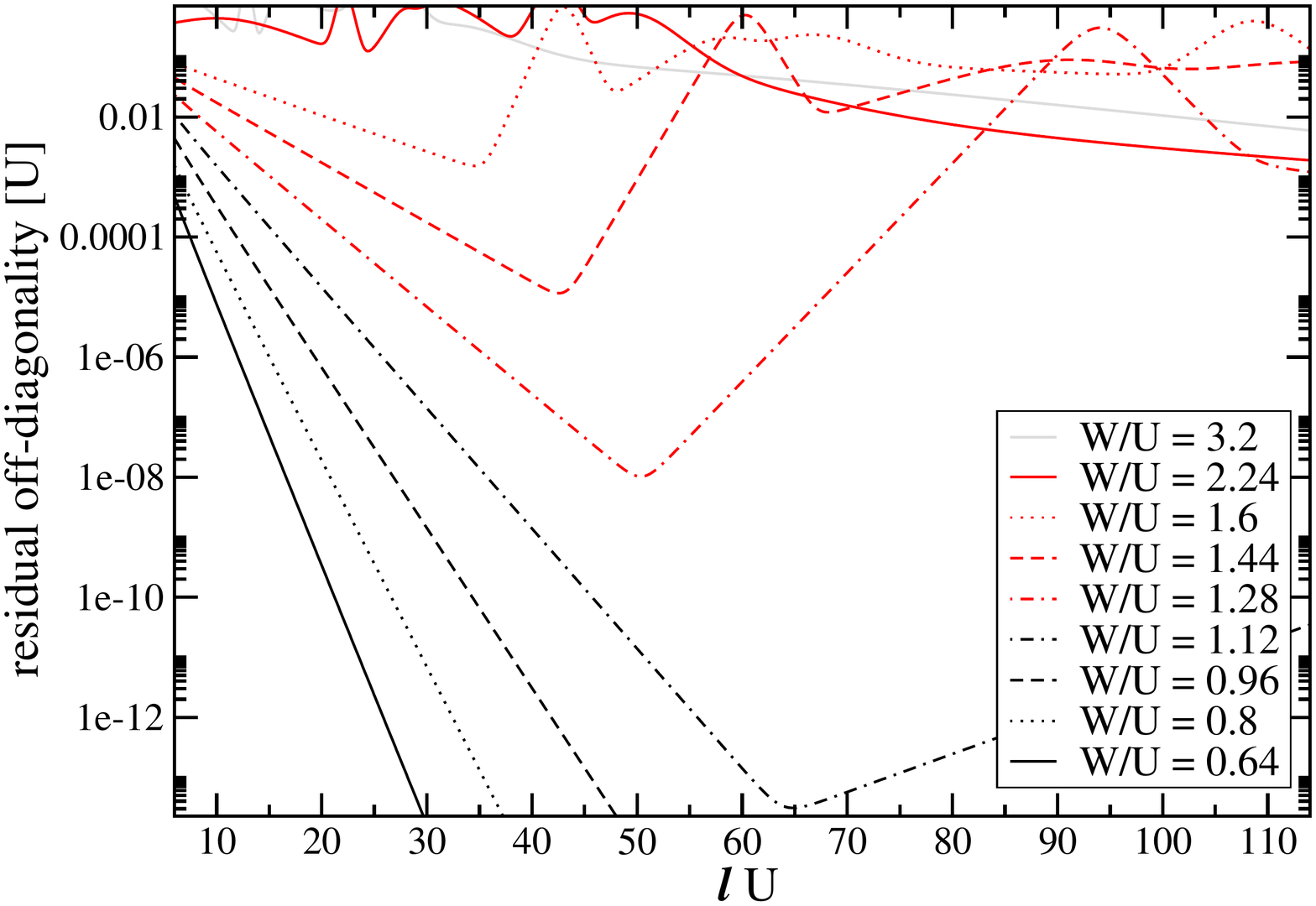}
\end{center}
\caption{(Color online) Behavior of the residual off-diagonality (ROD) for the 
pc-generator obtained in a double plaquette calculation for different values of
 $W/U$. The ROD measures to which extent charge fluctuations are eliminated 
as function of the flow parameter $\ell$.}
\label{fig_ROD_doppel}
\end{figure}
Figure \ref{fig_ROD_doppel} shows the behavior of the ROD of the pc-generator
for the double plaquette calculation at various values of $W/U$.
For small values of the flow parameter $\ell U< 2.24$ the ROD decreases 
exponentially. If the ratio $W/U$ is increased to $W/U=1.12$ the ROD evolves 
non-monotonically. For $\ell U<64$ the ROD falls below $10^{-13}$ to rise again
 for larger values of $\ell$. For even higher values of $W/U$ this increase 
sets in for smaller and smaller $\ell$.
  
We argued in the preceding section that a non-monotonic behavior
 of the ROD as observed in Fig.\ \ref{fig_ROD_doppel} is a first clue for a
possible breakdown of the mapping.
Although the non-monotonicity indicates problems of the mapping already for 
$W/U =1.12$ no sign of a possible breakdown can be seen in the coupling 
constants, see for instance Fig.\ \ref{fig_J2_gen}. 
This is due to the fact, that the dominant spin coupling 
constants are already converged to their value in the effective model for small
 values of $\ell$. Thus processes appearing only at large $\ell$ have no 
influence on these values.

To make progress in determining the range of validity of the mapping 
we have to study the separation of energy scales between the states 
without DOs and the states with DOs. To this end
 we investigate the energies of an added electron or an added hole
to the system, that means the density-of-states (DOS)
which comprises the LHB and the UHB for large $U$, see 
Figs.\ \ref{fig_DOS_half} and  \ref{fig_DOS_dop}. 
If $W$ is increased the bands approach each other and eventually touch so
that there is no energy separation anymore. States with differing number of DOs
have the same energy so that charge fluctuations can not be eliminated.
 In a paramagnetic 
description by dynamic mean-field theory (DMFT) the Mott-insulating phase
becomes unstable at this very point \cite{georg96,nishi04b,garci04,karsk05}.

Turning the argument around we use the gap $\Delta_g$ 
between the lower and the upper 
Hubbard band as quantitative indicator of the energy separation of 
subspaces with different number of DOs. If  $\Delta_g$ is finite
there is good physical reason to regard the mapping of the
Hubbard model onto the $t$-$J$ model as justified. If
 $\Delta_g$ vanishes the mapping has to break down.

There is one additional aspect to which we have to draw the
reader's attention. In infinite dimensions, where DMFT is exact,
one may suppress long-range magnetic order and consider the paramagnetic
phase which does not show an spin-spin correlation between different sites
so that the insulating paramagnetic phase behaves like the reference ensemble
\eqref{ro} for $\delta=0$. In particular, the charge gap does not
depend on the spin state.

But in finite dimensions, even without long-range magnetic order
one has to expect that the charge gap will generically depend on the
spin state. Hence there is no well-defined charge gap without
specifying the state of the spin degrees of freedom.
Thus we have to introduce the concept of the \emph{apparent} charge gap
$\Delta_g$ \cite{reisc04} which is designed to describe the energy separation
of subspaces with different number of DOs if an electron
is added to the disordered reference ensemble \eqref{ro}.
The apparent charge gap is not rigorously defined and
it cannot be measured precisely in experiment
because it does not capture bandtails of the  Hubbard bands 
which carry little spectral weight. But it is an
estimate for the energy separation between states
without DOs and states with DOs. Hence it provides an appropriate
estimate of the range of validity of the mapping from the
Hubbard model to the $t$-$J$ model.

We calculate $\Delta_g$ for the effective $t$-$J$ model derived before.
In the half-filled case the gap is calculated by estimating the lowest possible
 energy of an added DO, for details see below. Calculations for the 
half-filled case indicate a closure of the gap for $W/U\approx 0.9$ 
\cite{reisc04}.
In the doped case the calculation of $\Delta_g$  is divided into two steps as 
can be understood from the DOS sketched in  Fig.\ \ref{fig_DOS_dop}.

To calculate the apparent gap we first determine the lowest possible energy 
$\Delta_{\text{UHB}}$ of a DO in the upper Hubbard band. 
In a second step we calculate the maximum energy for 
the destruction of a DO $\Delta_{\text{LHB}}$, that means for
adding an electron to an empty site. Hence in the doped case we
use
\begin{align}
\Delta_g = \Delta_{\text{UHB}}-\Delta_{\text{LHB}},
\label{dg-doped}
\end{align}
while in the undoped case we have
\begin{align}
\Delta_g = 2\Delta_{\text{UHB}}.
\label{dg-undoped}
\end{align}
Note that the seemingly discontinuous definition of
$\Delta_g$ as function of $\Delta_{\text{UHB}}$ stems from 
the discontinuous evolution of the Fermi level which jumps
upon hole doping from the middle between the Hubbard bands to 
the edge of the lower Hubbard band.

\subsection{Calculation of $\Delta_g$}

The apparent charge gap is calculated for the effective $t$-$J$ model.  
A full diagonalization of the Hamiltonian is not feasible and it would
not provide what we need, namely the charge gap above the
disordered reference ensemble \eqref{ro}. Hence we  apply a 
Lanczos approach in terms of operators. The Lanczos approach is appropriate
because we only aim at extremum eigenvalues. 
Since we have to deal with operators 
acting on the reference ensemble \cite{reisc04}, 
which is a mixed state, the Liouville 
formulation of this method has to be used \cite{mori65,fulde93,viswa94}. The 
evolution of an operator $\hat{A}$ is given by the Liouville superoperator 
\begin{align}
\mathcal{L}\,\hat{A}= \left[\hat{H}_{\text{eff}},\,\hat{A}\right]\,.
\end{align}
The effect of this superoperator applied to the creation operator of a DO 
consists of moving the DO and changing the spin background. 
With this operator a basis of operators $\left\{\hat v_0,...\hat v_n \right\}$
 describing the DO with momentum $k$ and the effect on its surrounding spins is
 built recursively. In the first part of the calculation the minimal energy of 
a DO with momentum $k$ is calculated. The calculation starts with the vector 
\begin{align}
\hat{v}_0 = \frac{1}{\sqrt{N}}\sum_{\vec{r}}e^{i\vec{k}\vec{r}}
\hat{n}^{\phantom\dagger}_{\vec{r},\downarrow}\,
\hat{c}^\dagger_{\vec{r},\uparrow}\,.
\label{vo}
\end{align}
where $N$ denotes the number of lattice sites and the vector $\vec{r}$ 
determines the actual position of the DO. The action of this operator is to put
 an $\uparrow$ electron on a site which is already occupied by a
$\downarrow$  electron so that a doubly occupied site is created.
From this starting vector the basis is built recursively by
\begin{align}
\hat{v}_{i+1} = \mathcal{L}\hat{v}_i - a_{i}\hat{v}_i - b_i^2\hat{v}_{i-1}
\end{align}
according to the rules of the Lanczos tridiagonalization.
The scalar product of the Liouville formulation \cite{fulde93} is defined as
\begin{align}
\left(\hat{A}|\hat{B}\right) = \text{Tr}\left(\hat{A}^\dagger
\hat{B}^{\phantom\dagger}\hat{\rho}_0^{\phantom\dagger}\right)\,.
\end{align}
with the statistical operator $\hat{\rho}_0$ of the reference ensemble 
(\ref{ro}). 
The prefactors $a_i$ are given through the projection onto $\hat v_i$
\begin{align}
a_i &= \frac{\left(\hat{v}_i|\mathcal{L}\hat{v}_i\right)}{\left(\hat{v}_i|
\hat{v}_i\right)}
\end{align}
and the $b_i$ are given by
\begin{align}
b_i^2 = \frac{\left(\hat{v}_i|\hat{v}_i\right)}{\left(\hat{v}_{i-1}|
\hat{v}_{i-1}\right)} \,.
\end{align}
In this  operator basis the Liouville superoperator takes 
tridiagonal form with the coefficients $a_i$ on the diagonal
the coefficients $b_i$ on the secondary diagonals.

If infinitely many iterations were performed the dispersion
 of a DO relative to the disordered spin background would be given as the 
lowest energy in the subspace spanned by the calculated operator basis. In 
real calculations only a few iterations are feasible due to the humongous
number of terms in the effective Hamiltonian. Starting with the vector 
$\hat{v}_0$ in (\ref{vo}) consisting of one single operator, the commutation 
leads to increasingly complicated terms whose appropriate superposition
describes $\hat{v}_j$. The 
effort grows exponentially with the number of iterations. Thus we 
restrict ourselves to a finite basis  $\left\{\hat v_0 \ldots \hat 
v_n\right\}$. 
The lowest  energy calculated in the  subspace spanned by 
$\left\{\hat v_1 \ldots \hat v_n\right\}$ yields an upper bound 
$\Delta_\text{UHB}$ to the real dispersion of the DO. Note that 
in the following  we denote by $\Delta_\text{UHB}$ this upper bound
in order to keep the notation simple.

In the half-filled case the apparent charge gap 
results from Eq.\ \eqref{dg-undoped}.
For finite doping the particle-hole symmetry is lost, the Fermi energy
jumps to the LHB, and the value 
$\Delta_{\text{LHB}}$ has to be determined. 
To this end, we start from the modified  vector 
\begin{align}
\hat{u}_0 = \frac{1}{\sqrt{N}}\sum_{\vec{r}}e^{i\vec{k}\vec{r}} 
\hat{c}^\dagger_{\vec{r},\uparrow}
\left(1-\hat{n}^{\phantom\dagger}_{\vec{r},\downarrow}\right).
\end{align}
This operator destroys a DO by placing a single electron on an empty site. 
The value for $\Delta_{\text{LHB}}$ we obtain from the calculation
in a finite subspace $\left\{\hat u_0 \ldots \hat u_n\right\}$ 
is a lower bound to the true maximum energy.
 Finally the apparent  charge gap is given by \eqref{dg-doped}
in dependence on the doping level $\delta$.
As argued before the mapping to the $t$-$J$ model is justified
as along as $\Delta_g\geq0$.

\begin{figure}[htb]
\begin{center}
        \includegraphics[width=0.95\columnwidth]{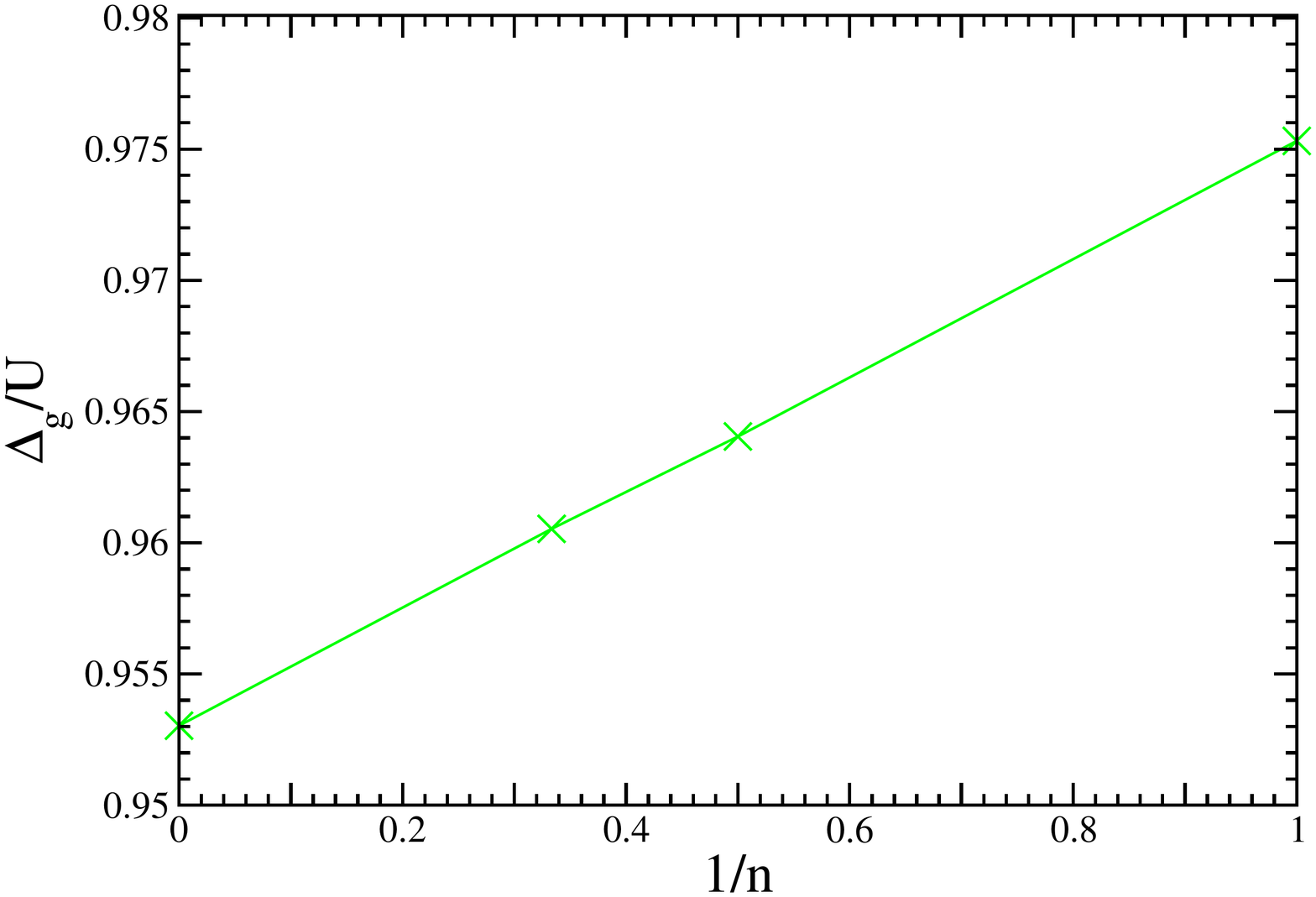}
\end{center}
\caption{(Color online) Linear extrapolation in $\frac{1}{n}$ of
 the apparent charge gap 
at half-filling for $W/U= 0.5$
obtained by $n$ iterations.}
\label{fig_gap_durchn}
\end{figure}

Extremum values of $E_k$ occur at the high symmetry points of the 
Brillouin zone. Thus we avoid costly calculation of the whole dispersion
and focus on the momenta $\vec{k} = \left(0,0\right)$ and 
$\vec{k} = \left(\pi,\pi\right)$ where the lattice constant is set to unity. 
The calculations rely on the the nearest neighbor effective model. 
Previous calculations in the half-filled case \cite{reisc04} showed that 
there is no significant change in the results obtained for different truncation
 schemes because the main uncertainty results from  the 
limited number  of iterations in the Lanczos tridiagonalization.
 The truncation of the effective model used plays 
only a minor role. Since only a few iterations were feasible we 
additionally perform an extrapolation. 
The results for the gaps are extrapolated 
in $1/n$ with $n$ denoting the number of 
iterations. By extrapolating to $n=\infty$ we obtain an estimate for 
$\Delta_g$; for more details we refer to Ref.\ \onlinecite{reisc04}.
Figure \ref{fig_gap_durchn} displays the extrapolation  for the
 half-filled case.
For the doped case, $\Delta_\text{LHB}$ and $\Delta_\text{UHB}$
are extrapolated separately in $1/n$.

%%%%%%%%%%%%%%%%%%%%%%%%%%%%%%%%%%%%%%%%%%%%%%%%%%%%%%%%%%%%%%%%%%%%%%%%%%%%%%
%%%%%%%%%%%%%%%%%%%%%%%%%%%%%%%%%%%%%%
\subsection{Results for $\Delta_g$}
\label{chap:res_gap}
%%%%%%%%%%%%%%%%%%%%%%%%%%%%%%%%%%%%%%%%%%%%%%%%%%%%%%%%%%%%%%%%%%%%%%%%%%%%%%%
%%%%%%%%%%%%%%%%%%%%%%%%%%%%%%%%%%%%%

The apparent charge gap is computed for the effective $t$-$J$ model 
derived by a CUT with NN truncation  using the pc-generator or the 
gs,1p-generator. The gs-generator is not used in this 
context because the resulting effective model mixes a single DO
with the subspaces of two and more DOs.
The gap is calculated for various doping levels as function of $W$. 
Thus the value $W/U$ up to which 
the mapping is justified is estimated from $\Delta_g(W/U) = 0$.
The minimum $\Delta_\text{UHB}$ of the dispersion of a DO is found 
for a vanishing momentum. The maximum energy $\Delta_\text{LHB}$
for the destruction of a hole is 
found for a momentum $\vec{k} = \left(\pi,\pi\right)$.

\begin{figure}[htb]
	\centering
	\includegraphics[width=1\columnwidth]{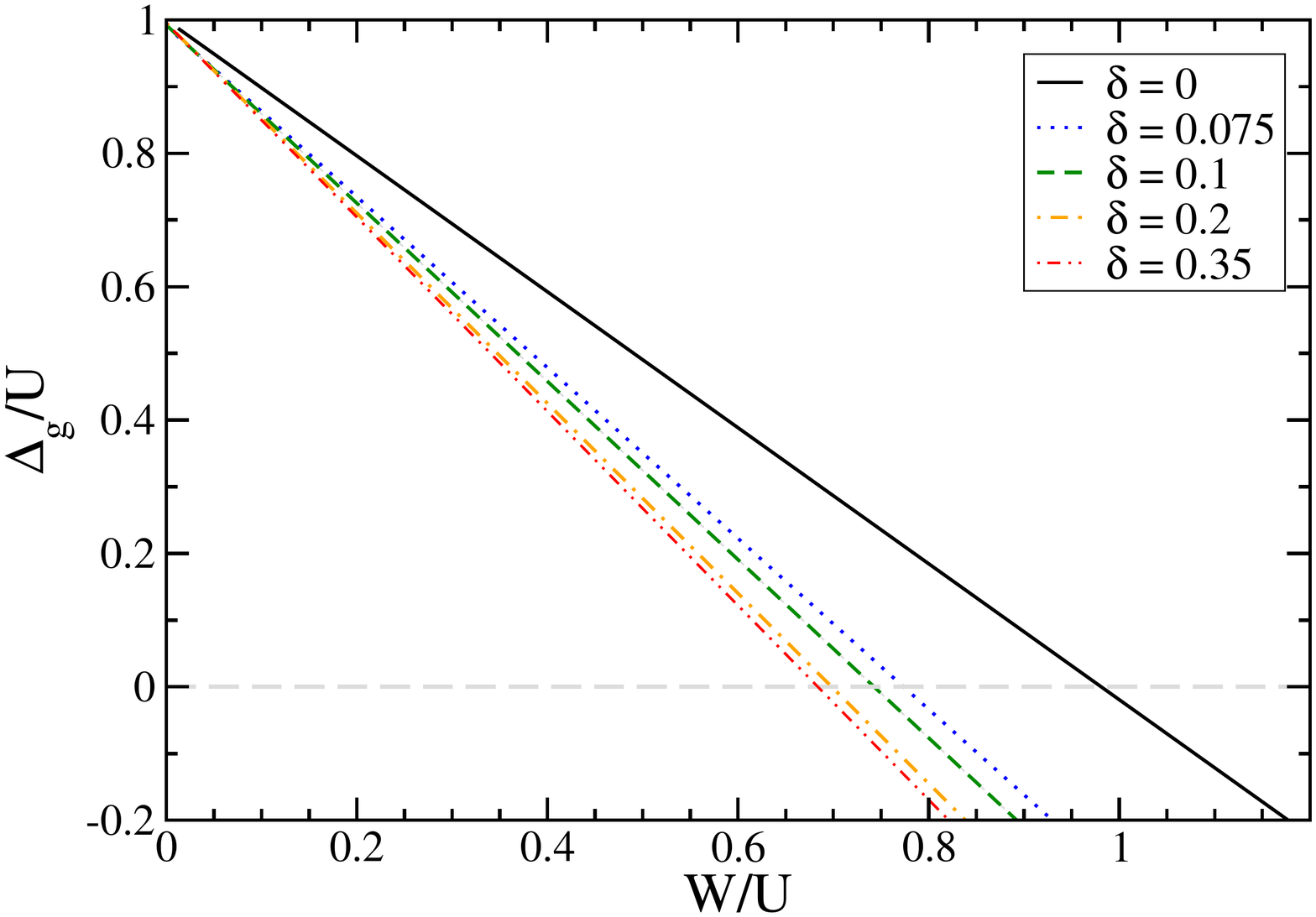}
	\caption{(Color online) The extrapolated apparent charge gap 
as function of  $W/U$ for various $\delta$.}
	\label{fig:1durchn}
\end{figure}
The results for the extrapolated $\Delta_g$ are displayed in 
Fig.\ \ref{fig:1durchn} for various values of the doping $\delta$. 
For  vanishing bandwidth $W=0$ the 
apparent gap is given by the  Hubbard repulsion $U$. Thus the curves of
$\Delta_{\text{app}}/U$ start at unity. Then the gap decreases
 almost linearly until $\Delta_{\text{app}} = 0$ is reached. Negative values of
 the gap  indicate the breakdown of the mapping. 
The linear decrease of the gap has also been observed
for the  half-filled case \cite{gebha97}. 
The decrease leads to a closure of the charge gap for
 $W/U = 1$. 
For the Bethe lattice with $z\rightarrow\infty$ a closure of the 
gap was found for $W/U = 0.89$ \cite{nishi04b} which agrees well
with our estimate in view of the different
lattices and techniques. Other numerical evaluations of DMFT 
for the Bethe lattice yield a closure of the insulating gap at
  $W/U\approx0.84$ \cite{garci04} or at $W/U\approx0.83$ \cite{karsk05}.

\begin{figure}[htb]
	\centering
	\includegraphics[width=0.98\columnwidth]{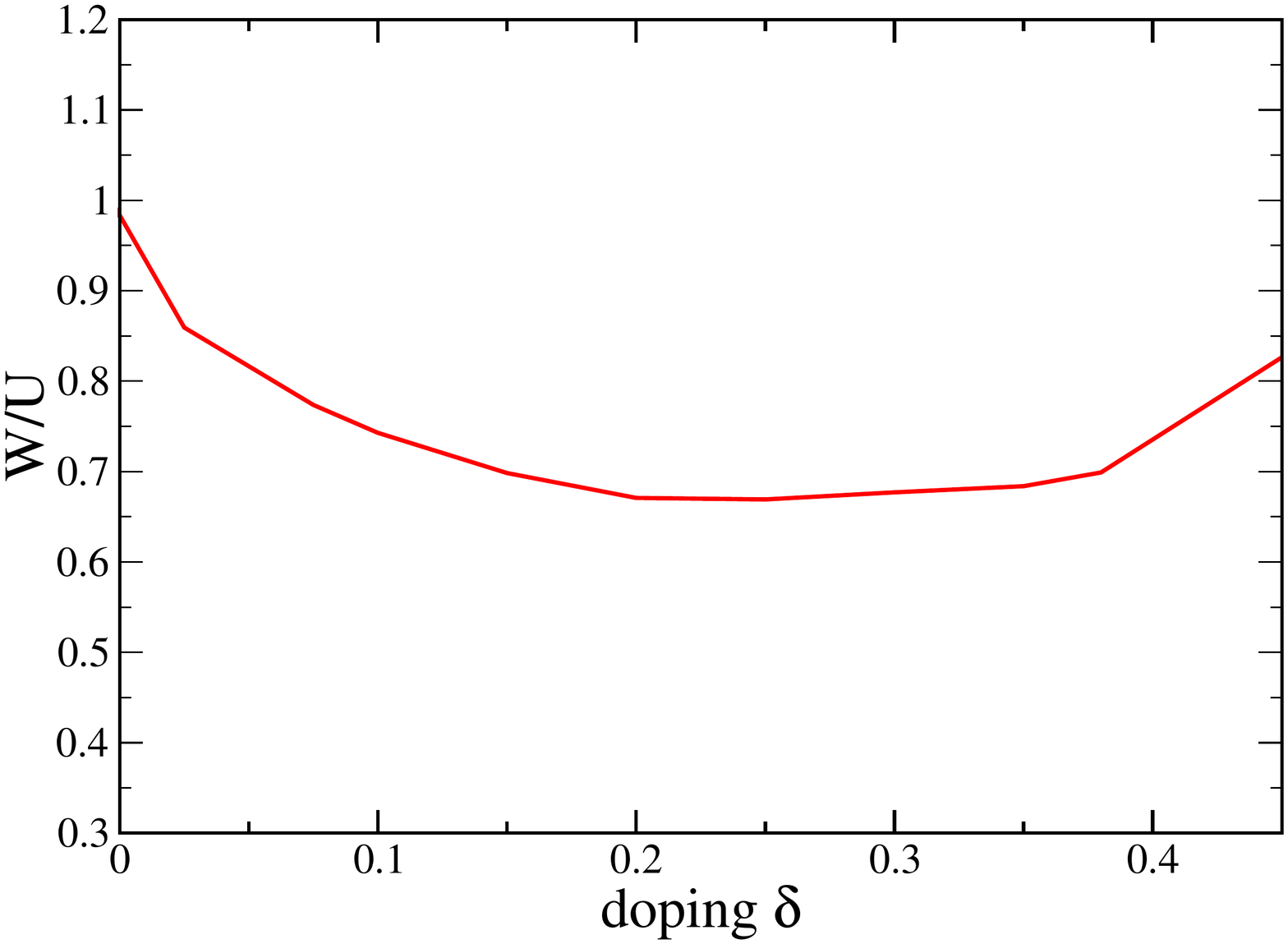}
	\caption{(Color online) Limiting values for $W/U$ up to which a mapping
 of the Hubbard model to the effective $t$-$J$ model is justified. It is
derived from the zeros $\Delta_g(W/U)=0$ as function of doping.}
	\label{fig:appli}
	\end{figure}
Our results indicate that the apparent charge gap $\Delta_g$ for the square 
lattice closes at $W/U \approx 0.98$ for $\delta=0$. 
Upon doping $\Delta_g$ vanishes even faster upon increasing bandwidth
so that the range of applicability of the mapping `Hubbard model $\to$
$t$-$J$ model' is reduced.
From the values of $W/U$ where $\Delta_g$ beomes zero we estimate
this range of applicability. The result is shown in Fig.\ 
\ref{fig:appli} which represents one of the central results of this
work. Our approach provides the first systematic derivation of this diagram of 
applicability as function of doping.

The range of applicability decreases from $W/U \approx 0.98$ for $\delta=0$ to 
$W/U \approx 0.67$ for $\delta = 0.25$. Then the range of 
applicability increases again slightly to $W/U = 0.73$ for $\delta = 0.4$.
The plateau in $W/U(\delta)$ and the moderate increase are
rather unexpected, cf.\ Ref.\ \cite{milli91}. We do not
have an obvious explanation for it.
In contrast, the decrease of the range of applicability
for $\delta \le 0.3$ meets the qualitative expectation
since a doped system has a higher mobility of charges so that the
energy separation of sectors of differing number of DOs becomes
smeared out.

The relative constant limiting value for $W/U$ below which
the use of a generalized $t$-$J$ model is justified provides
interesting information on the applicability of $t$-$J$ models
for doped systems. The use of $t$-$J$ models is very widespread
in theoretical studies for the high-$T_c$ superconductors
based on cuprates. Commonly used parameters are $W/U \approx 0.7$ and 
$\delta <0.3$ \cite{ogata08}. Our results indicate that the
use of $t$-$J$ models is indeed justified. But caution is required
 in the doping range  $0.18\lessapprox \delta\lessapprox 0.25$
where $W/U \approx 0.7$ is at about the limit of applicability.
Thus our results shed light on the important question of the applicability
of a commonly used model. It is remarkable that the issue of
how to justify this model for significant levels of dopings
has attracted so little attention so far.

%%%%%%%%%%%%%%%%%%%%%%%%%%%%%%%%%%%%%%%
%%%%%%%%%%%%%%%%%%%%%%%%%%%%%%%%%%%%%%%
\section{results for the relevant coupling constants}
\label{chap:results_coup}

In the preceding section we comprehensively discussed the applicability of the
derivation of a generalized $t$-$J$ model. The result of this
discussion is summarized in the estimated 
range of applicability shown in  Fig.\ \ref{fig:appli}.
In the present section, we provide the coupling constants
which ensue from the CUT of the Hubbard model to the $t$-$J$ model.
Results are given  in the range $W/U \le 1.0$ because the mapping
definitely breaks down beyond.

All results shown  are derived from upto4 calculations using the 
pc-generator. Additionally we performed random double plaquette 
calculations with the gs,1p-generator to check  if there are changes in the 
coupling constants when higher truncation schemes are applied. 
No significant differences are found. Thus the upto4 truncation appears to be 
sufficient to determine the coupling constants.
The results for the half-filled case ($\delta =0$) shown in the following 
figures agree perfectly with the results obtained by Reischl et al.\ 
\cite{reisc04}.

Although the mapping generates a large number of terms, only few of them
are really relevant in the final effective model. Most others only have 
very small prefactors.

\subsection{Chemical Potential}

\begin{figure}[htb]
	\centering
	\includegraphics[width=0.98\columnwidth]{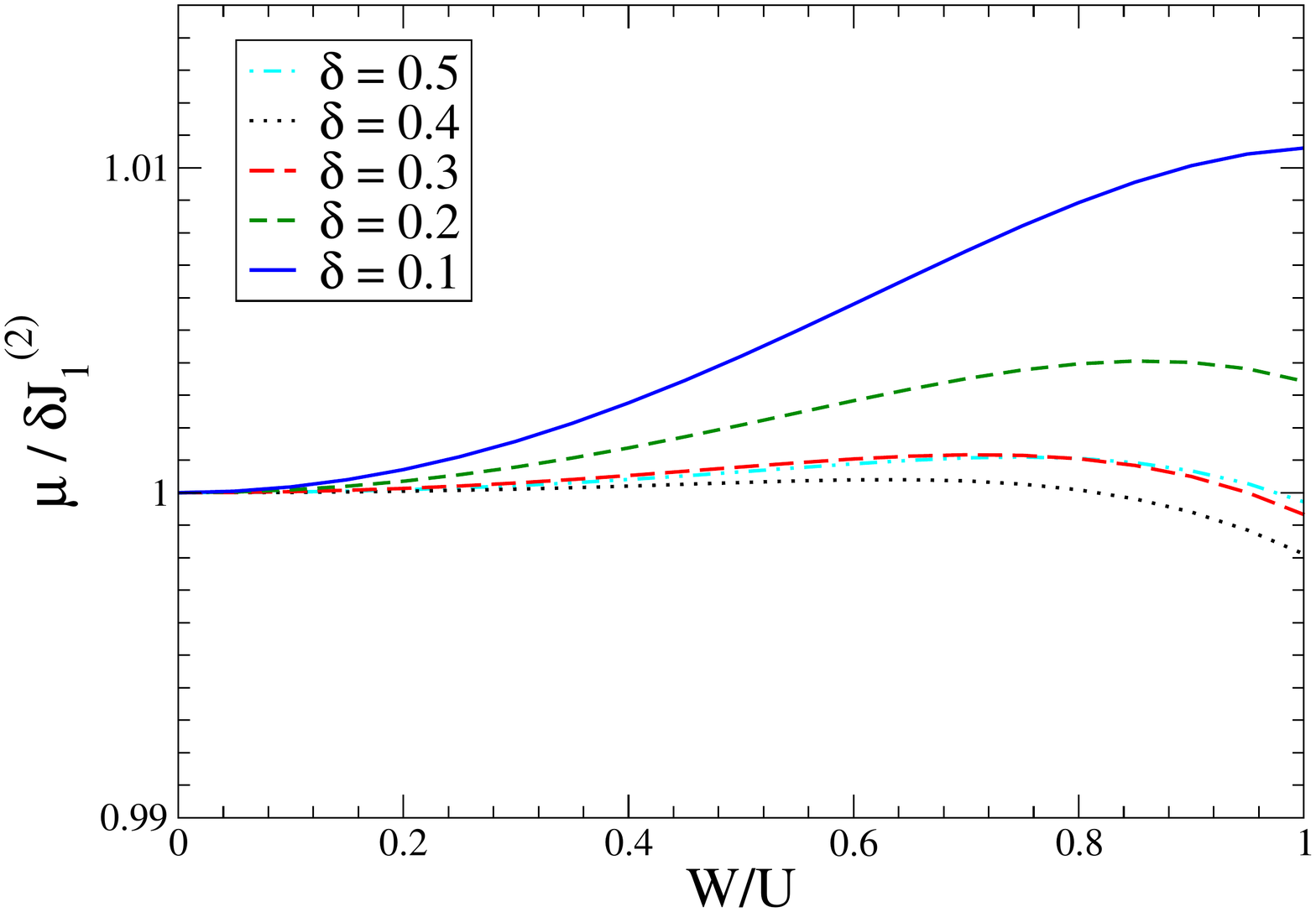}
	\caption{(Color online) Chemical potential relative to 
	  $\delta J_1^{(2)}$for various doping concentrations $\delta$.}
	\label{fig:mu_n}
\end{figure}

First, we consider the chemical potential as defined in 
(\ref{eq:mu}). In leading quadratic order of $\frac{t}{U}$ it is 
proportional to $\delta$. Therefore $\mu$ will be 
shown in units of $\delta J_1^{(2)}$ as defined in (\ref{J_1}).
The ratio $\mu/\delta$ shows almost no dependence on 
$\delta$, see Fig.\ \ref{fig:mu_n}.
As function of $W/U$ the chemical potential stays rather constant 
and even for $W/U = 1$ the deviations are small. 
The dependence on $W/U$ is greater for smaller values of $\delta$
concentration.
\noindent

\subsection{Spin Terms}
The dominant terms of the effective model are the Heisenberg-type spin 
interactions.  
\begin{align}
H_{\text{Heisenb.} |i-j|}= 
\sum_{i,j}J_{|i-j|}\vec{S}_i\vec{S}_j\,.
\label{Heise}
\end{align}
The largest contribution of this type is the Heisenberg exchange  $J_1$ 
between  nearest neighbors. 
All results are shown relative to the leading perturbative result 
$J_1^{(2)} = \frac{4t^2}{U}$.

\begin{figure}[htb]
\begin{minipage}[hbt]{0.5\columnwidth}
	\centering
	\hspace*{-0.35cm}
	\includegraphics[width=1.2\columnwidth]{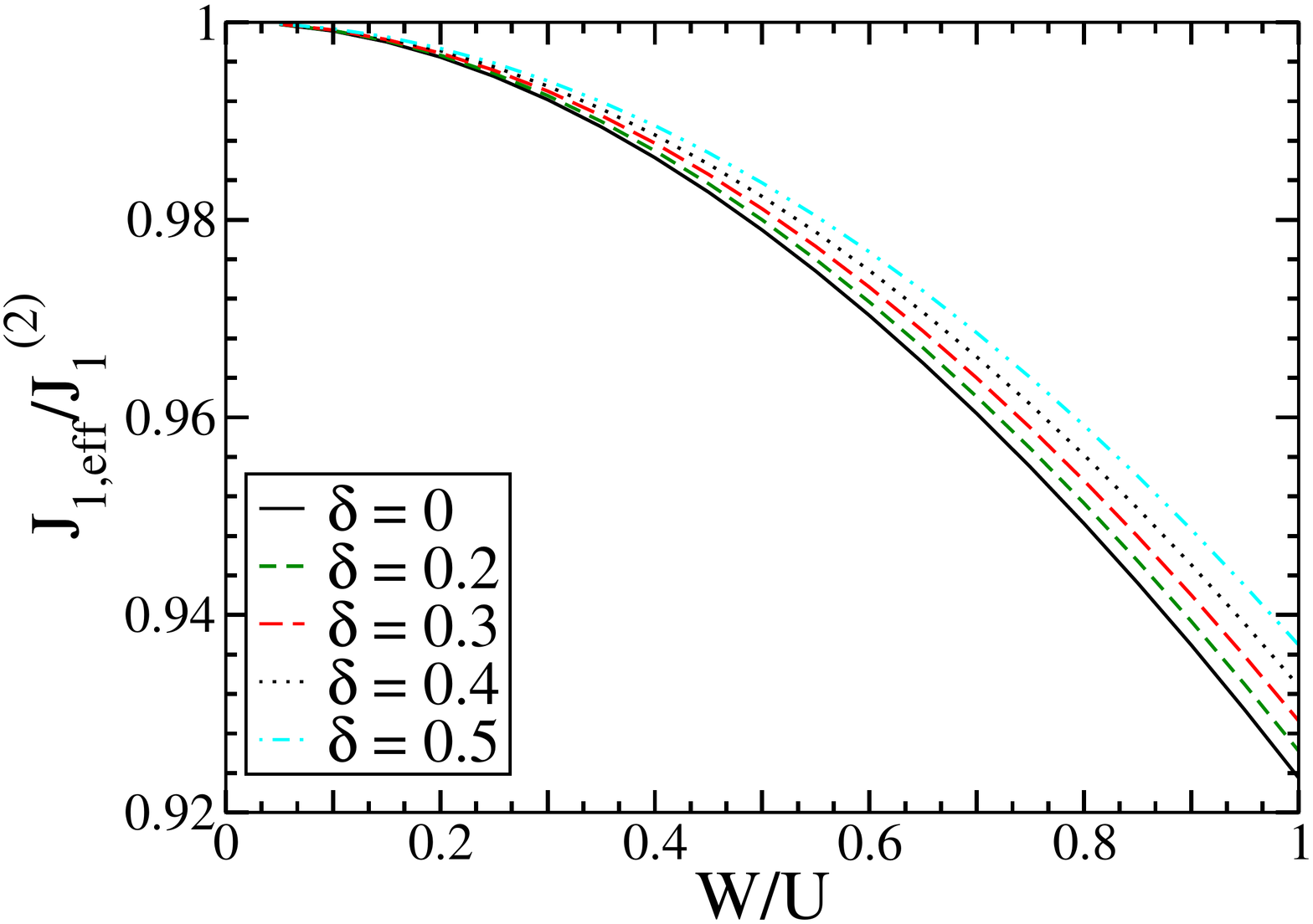}
			\end{minipage}
\centering
\hspace*{-0.2cm}
\begin{minipage}[hbt]{0.5\columnwidth}
	\centering
	\includegraphics[width=1.2\columnwidth]{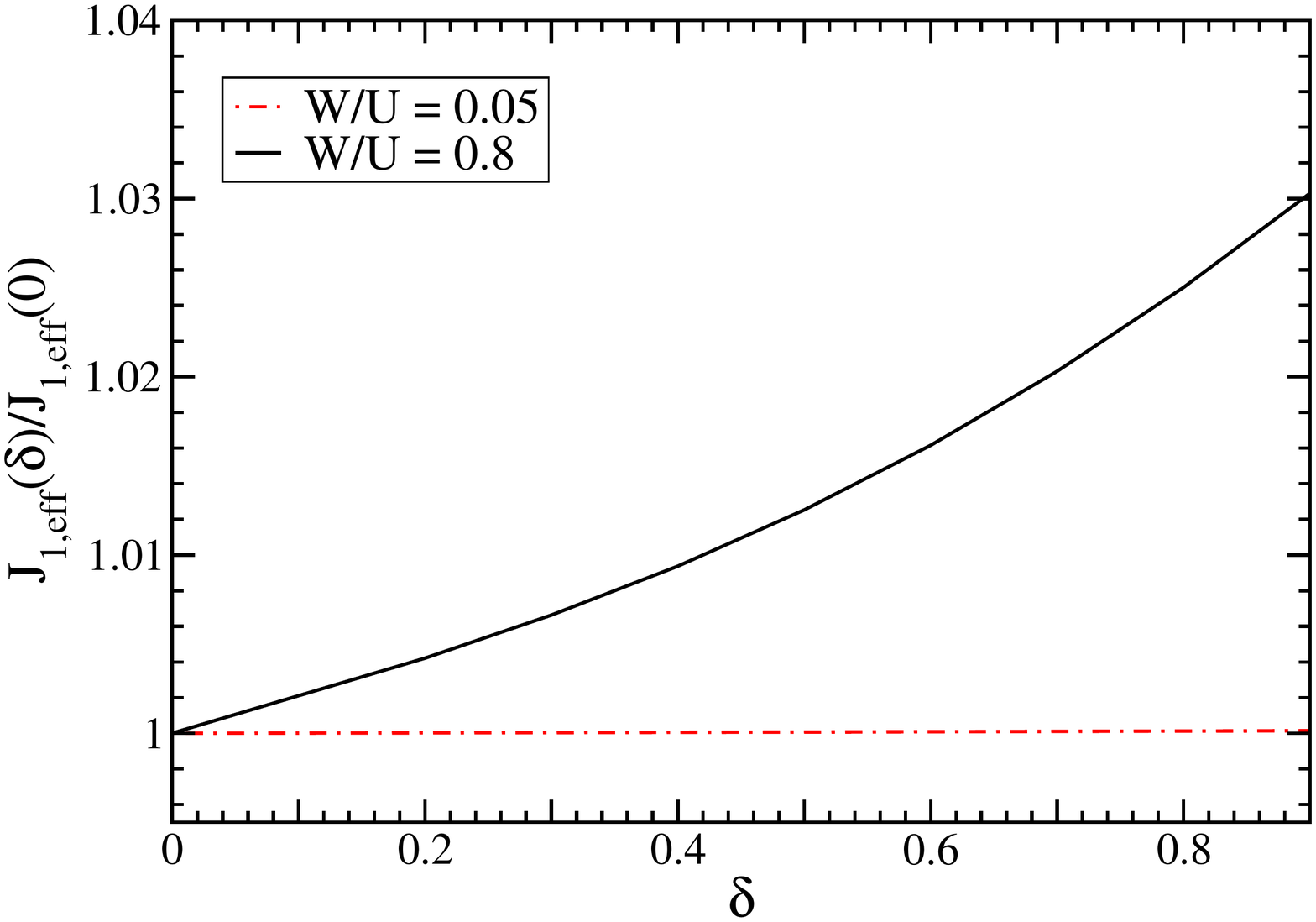}
 \end{minipage}
\caption{(Color online) Dependence of $J_1$ on $W/U$ for various values of 
$\delta$ (left panel). The dependence on $\delta$
for two ratios $W/U$ is depicted in the right panel. The 
undoped value for $W/U = 0.05$ is found to be 
$J_1(0) \approx 1.5621\cdot 10^{-4}\,U$. For $W/U = 0.8$ $J_1(0)$ takes the 
value $0.0379\,U$.}
\label{fig:J_1_dop_n}
\end{figure}

The left panel of Fig.\ \ref{fig:J_1_dop_n} shows the dependence of  
$J_1$ on the ratio $W/U$. 
Starting from $J_1^{(2)}$ for $W/U=0$ the coupling constant takes slightly 
smaller values for larger $W/U$. Additionally, the doping dependence of 
$J_1$ for various values of $W/U$ is shown relative to its value for 
the undoped system in the right panel of Fig.\ \ref{fig:J_1_dop_n}.
$J_1$ increases with  $\delta$. The doping has a 
greater influence for larger values of $W/U$, but the effect remains rather 
small. Even for $W/U = 0.8$ the doping causes a change in $J_1$ of only  about 
$3\%$.
\noindent

\begin{figure}[htb]
\centering
	\includegraphics[width=1\columnwidth]{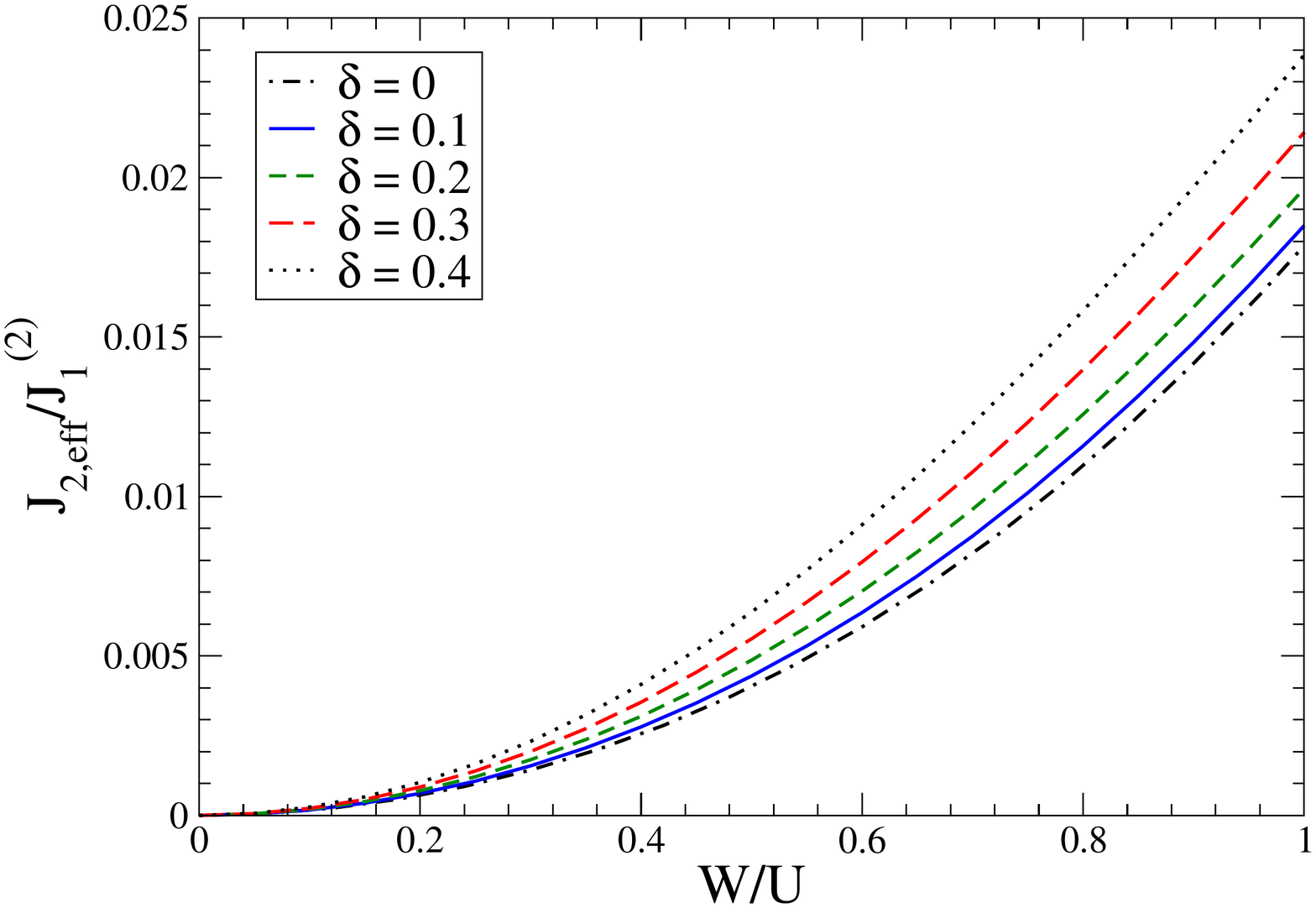}
	\caption{Effective $J_2$ for various doping concentrations as 
function of $W/U$.}
		\label{fig:J_2_dop2}
		\end{figure}

The Heisenberg exchange between next-nearest (diagonal) neighbors $J_2$ 
as well as the exchange $J_3$  between neighbors at a linear distance of 
two lattice spacings  are much smaller than $J_1$.
Both terms appear in fourth order of $\frac{t}{U}$.
Even for $W/U = 1$ $J_2$ is smaller than $0.03J_1$, see Fig.\ 
\ref{fig:J_2_dop2}.
Surprisingly, $J_2$ shows a slightly more significant (relative) 
dependence on the dopant 
concentration than $J_1$, see Fig.\ \ref{fig:J_2_dop2}. 
Yet, in view of the small absolute values of $J_2$, this doping
dependence can be neglected.

\begin{figure}[htb]
\begin{minipage}[hbt]{0.5\columnwidth}
	\centering
	\hspace*{-0.35cm}
	\includegraphics[width=1.2\columnwidth]{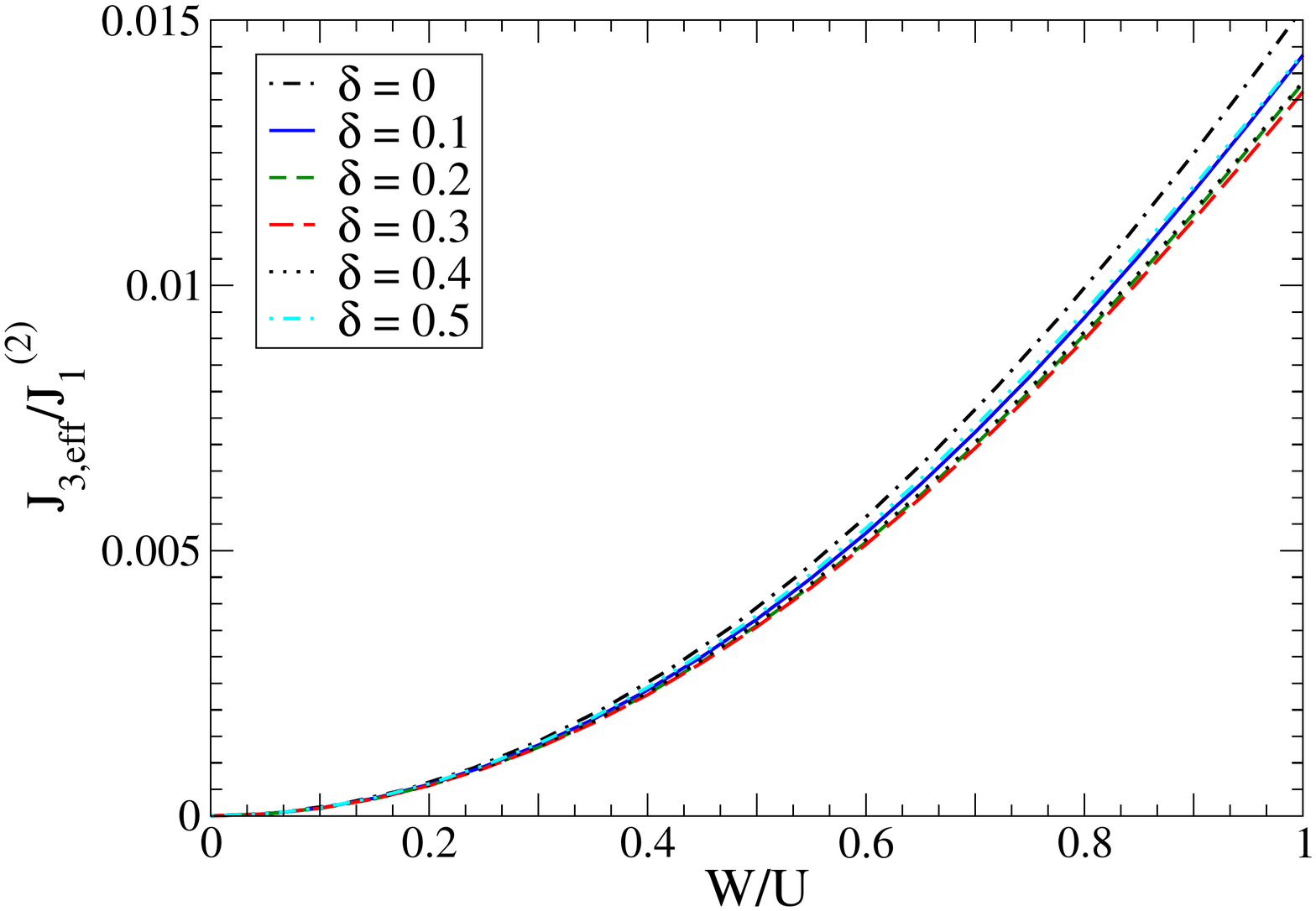}
			\end{minipage}
\centering
\hspace*{-0.2cm}
\begin{minipage}[hbt]{0.5\columnwidth}
	\centering
	\includegraphics[width=1.2\columnwidth]{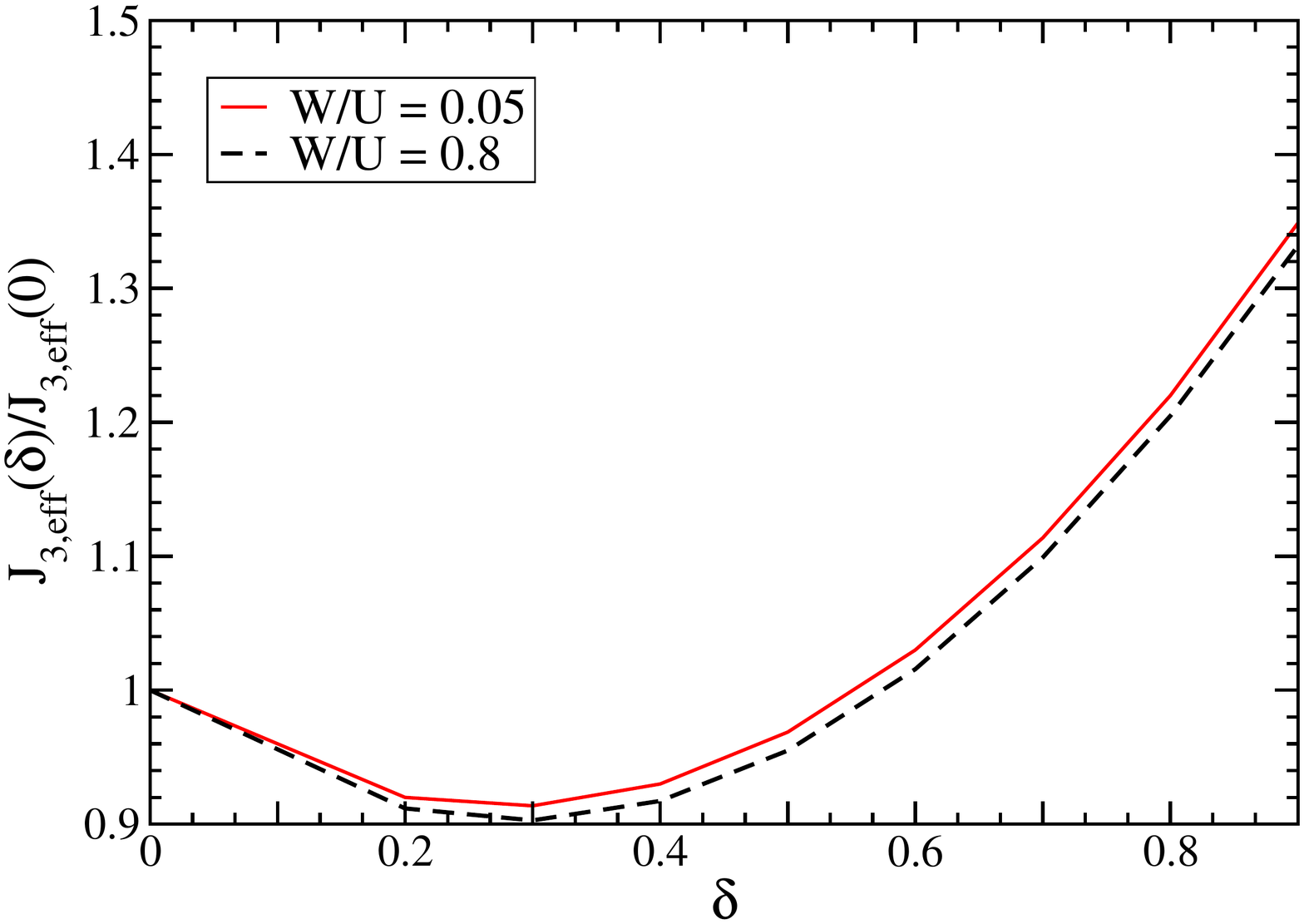}
 \end{minipage}
\caption{(Color online)  Dependence of $J_3$ on $W/U$ for various values of 
$\delta$ (left panel). The dependence on $\delta$
for two ratios $W/U$ is depicted in the right panel. The undoped values are 
$J_3(0) \approx 6.1039\cdot10^{-9}\,U$ for 
$W/U = 0.05$ and $J_3(0) \approx 3.9802\cdot10^{-4}\,U$ for $W/U = 0.8$.}
\label{fig:J_3_n}
\end{figure}

As can be seen in the right panel of Fig.\ \ref{fig:J_3_n} the coupling $J_3$ 
shows a counter-intuitive behavior. First it decreases upon doping
but increases again beyond $\delta \approx 0.3$.

\begin{figure}[htb]
	\centering
	\includegraphics[width=0.2\columnwidth]{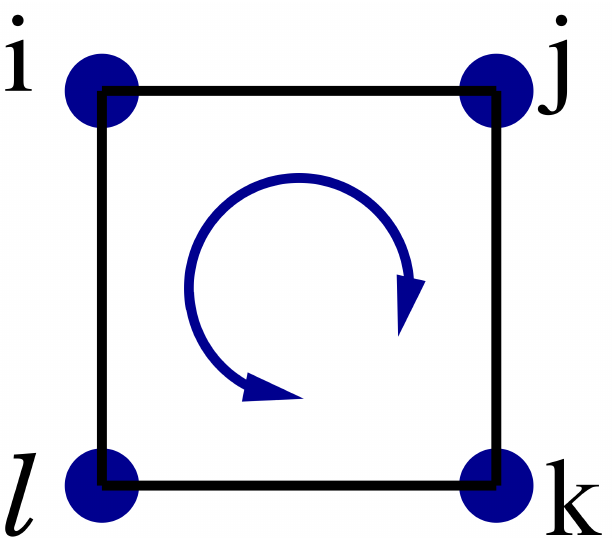}
	\caption{(Color online) Positions of spins interacting via the ring 
exchange $H_\Box$.}
	\label{rin}
\end{figure}

Besides the 2-spin terms in \eqref{Heise} the two dimensional square lattice 
also allows for 4-spin interactions. The leading contribution is given by
 \begin{align}
 \hat{H}_{\Box} &= J_\Box \sum_{<i,j,k,l>} 
\left[\left(\vec{S}_i\vec{S}_j\right)\left(\vec{S}_k\vec{S}_l\right) + 
\left(\vec{S}_i\vec{S}_l\right)\left(\vec{S}_j\vec{S}_k\right) -\right.
\nonumber\\ 
& \left.\left(\vec{S}_i\vec{S}_k\right)
\left(\vec{S}_j\vec{S}_l\right)\right]
\label{def:J_ring}
 \end{align}
which we sloppily call ring exchange although the complete
ring exchange comprises also nearest-neighbor and diagonal 
2-spin couplings \cite{brehm99}. We do so since these
2-spin terms are accounted for by $J_1$, $J_2$, and $J_3$ in \eqref{Heise}.
The term \eqref{def:J_ring} describes the interaction of the four spins on a 
plaquette, see Fig.\ \ref{rin}, and it occurs first in order 
$\left(\frac{t}{U}\right)^4$\,  \cite{takah77}. Its importance is
discussed at length in the literature, see for instance.
Refs.\ 
\onlinecite{schmi90b,brehm99,mulle02a,katan02a,reisc04,schmi05b} 
and references therein.
The magnetic excitations in planar cuprates may not be understood without 
considering ring exchange \cite{mulle04n,loren99c,notbo07}.

\begin{figure}[htb]
\begin{minipage}[hbt]{0.5\columnwidth}
	\centering
	\hspace*{-0.35cm}
		\includegraphics[width=1.2\columnwidth]{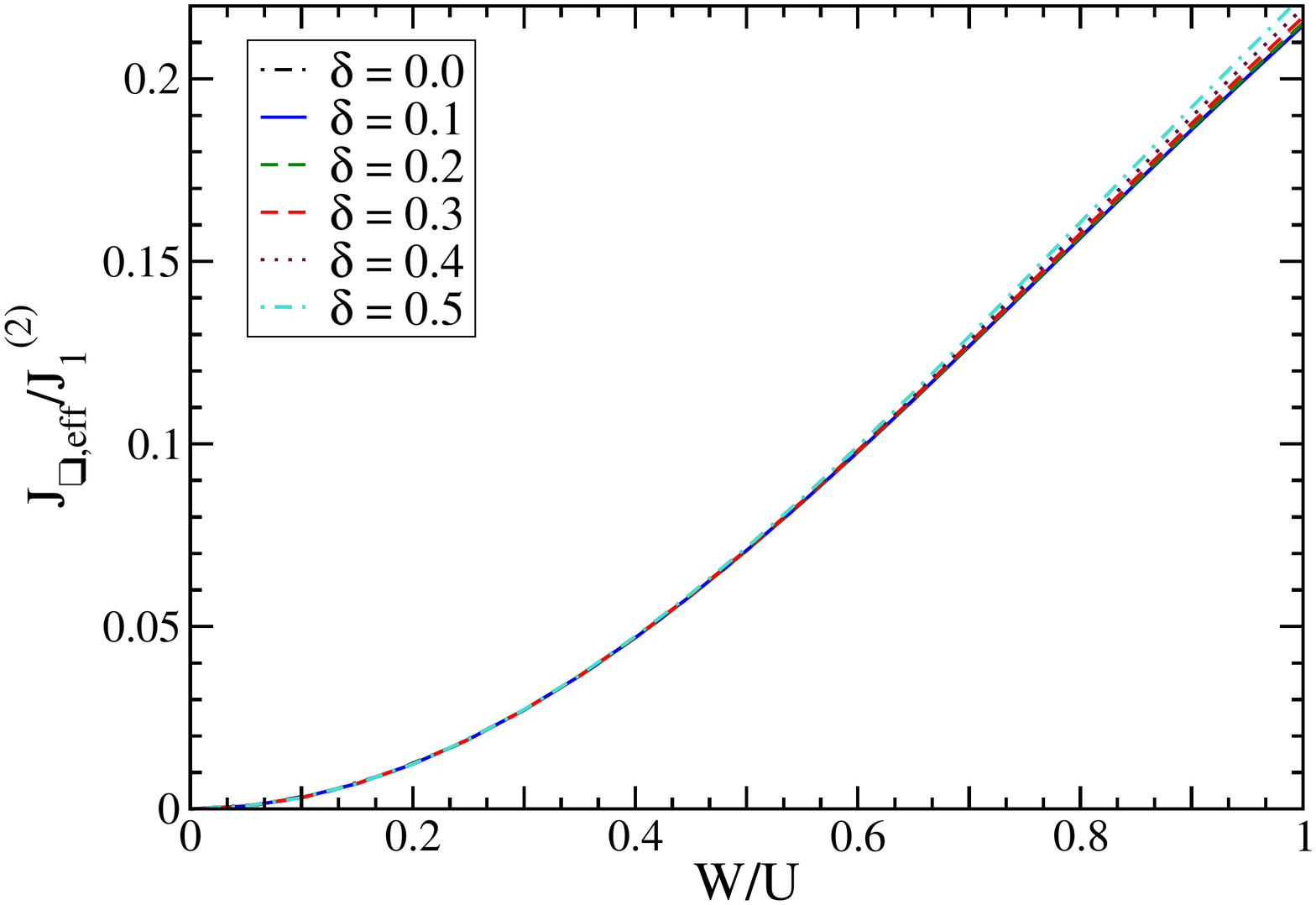}
			\end{minipage}
\centering
\hspace*{-0.2cm}
\begin{minipage}[hbt]{0.5\columnwidth}
	\centering
	\includegraphics[width=1.2\columnwidth]{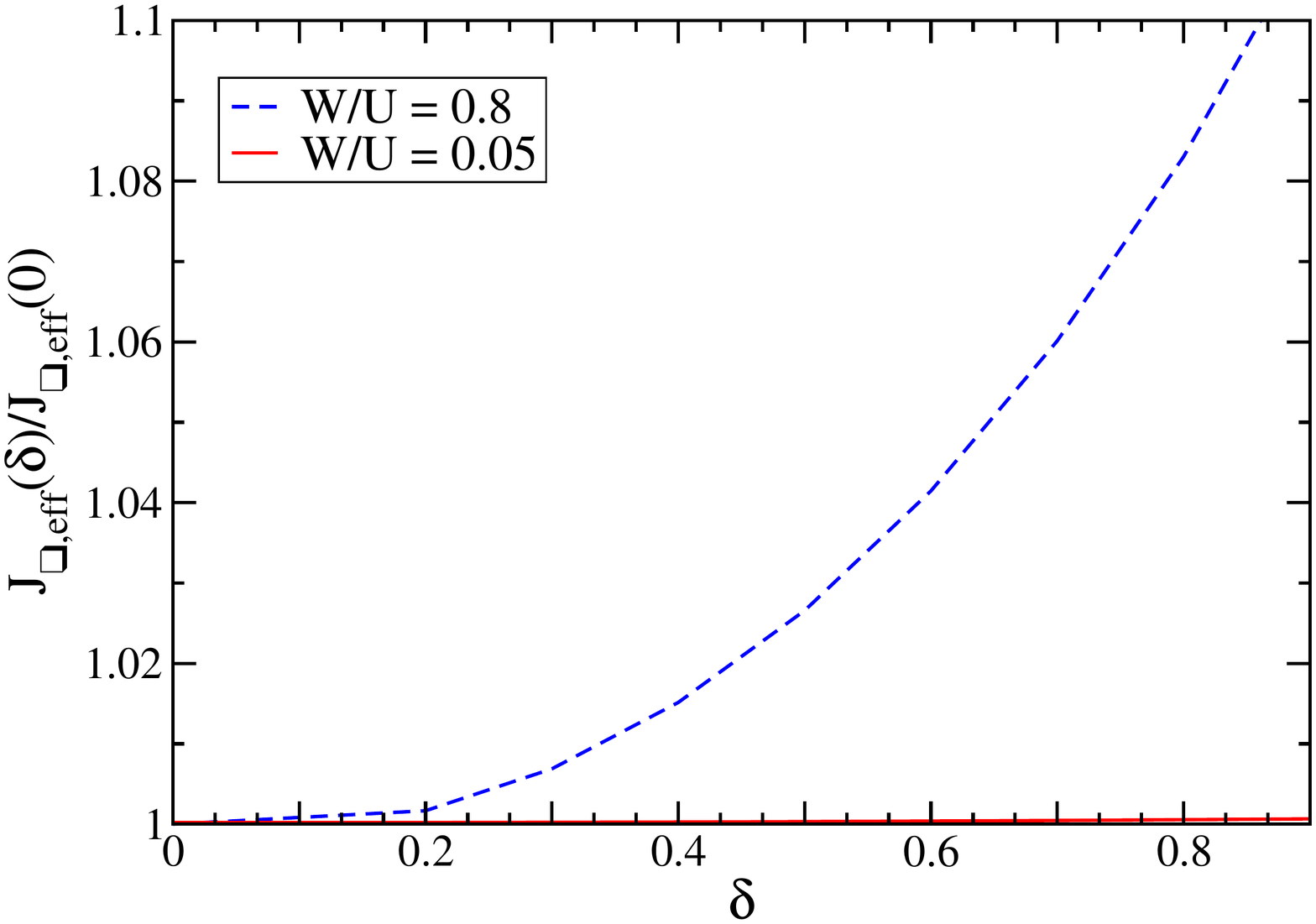}
 \end{minipage}
\caption{(Color online) Effective ring exchange term $J_\Box$ for 
various values of  $\delta$ (left panel) and its doping 
dependence of $J_\Box$ (right panel) relative to the undoped values 
$J_\Box \approx 2.4390\cdot10^(-7)\,U$ for $W/U = 0.05$ and 
$J_\Box(0) \approx 0.0125\,U$ for $W/U = 0.8$.}
\label{fig:J_ring_d}
\end{figure}

Compared to other quartic exchange couplings such as
$J_2$ or $J_3$ the ring exchange is much more
important, see its values in Fig.\ \ref{fig:J_ring_d}.
The ring exchange  takes values of up to $20\%$ of $J_1$. 
Hence this term must not be neglected in an effective model.

The ring exchange shown in Fig.\ \ref{fig:J_ring_d} 
displays nearly no dependence on the  doping $\delta$. 
Even for doping as large as  $\delta = 0.8$ the change in 
the coefficient is less than $1.12$ percent. 
Thus while ring exchange is an important process its doping dependence 
can safely be omitted.

\begin{figure}[htb]
	\centering
	\includegraphics[width=1\columnwidth]{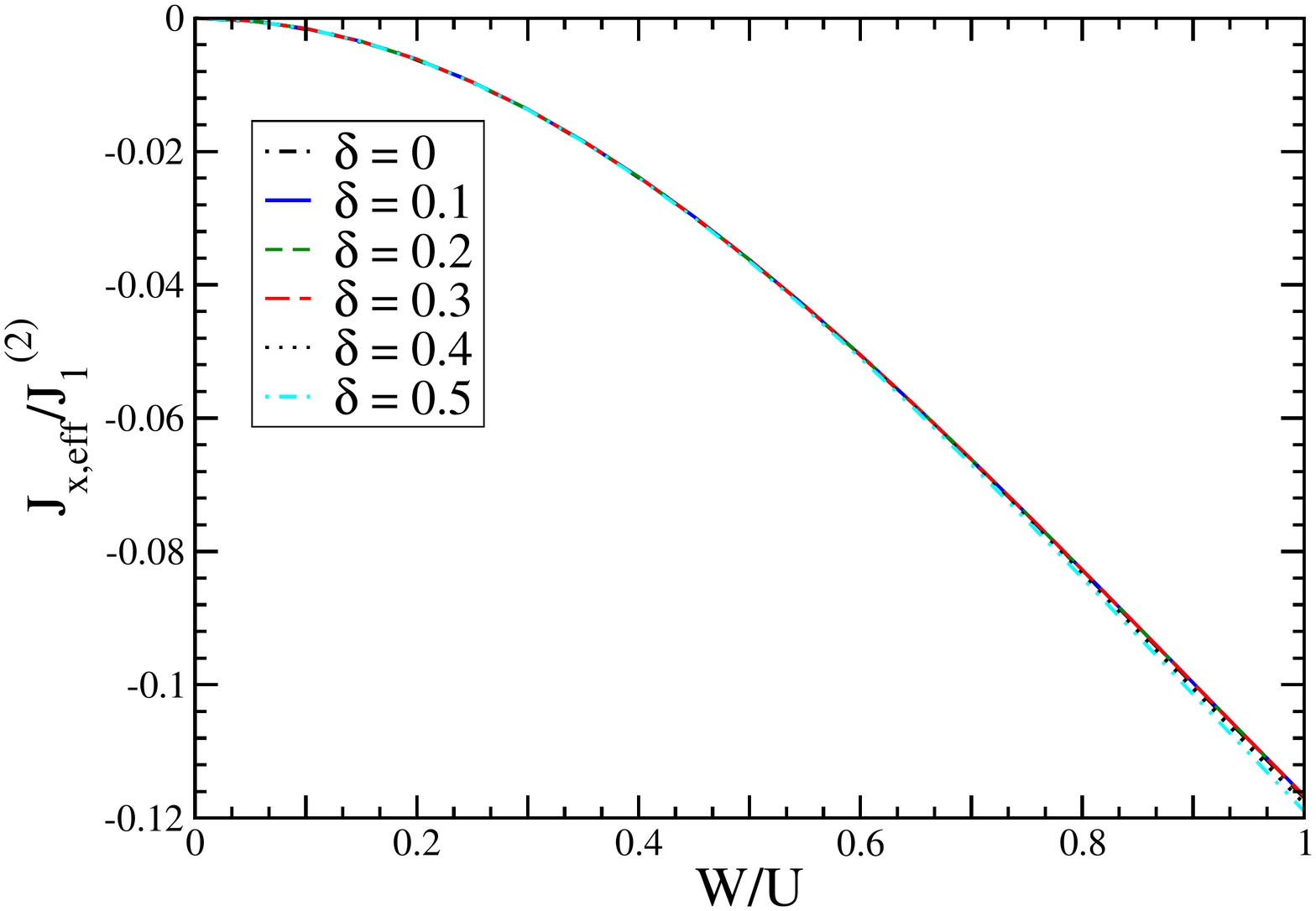}
	\caption{(Color online) Effective $J_{\times}$ as defined in Eq.\ 
\ref{def:J_cross} for various doping concentrations $\delta$.}
	\label{kreuz}
\end{figure}

The second 4-spin term is the cross exchange
\begin{align}
\hat{H}_\times = J_\times \sum_{<i,j,k,l>}\left(\vec{S}_i\vec{S}_k\right)
\left(\vec{S}_j\vec{S}_l\right).
\label{def:J_cross}
\end{align}
In this term the spins are located  on the same sites as for the ring exchange,
but the inner products are taken of the diagonal spins. 
The corresponding coupling constant is shown 
in Fig.\ \ref{kreuz}. It takes smaller values than $J_\Box$ and hardly shows
any doping dependence.
\noindent

\subsection{Interaction of Double Occupancies}

\begin{figure}[htb]
	\centering
	\includegraphics[width=1\columnwidth]{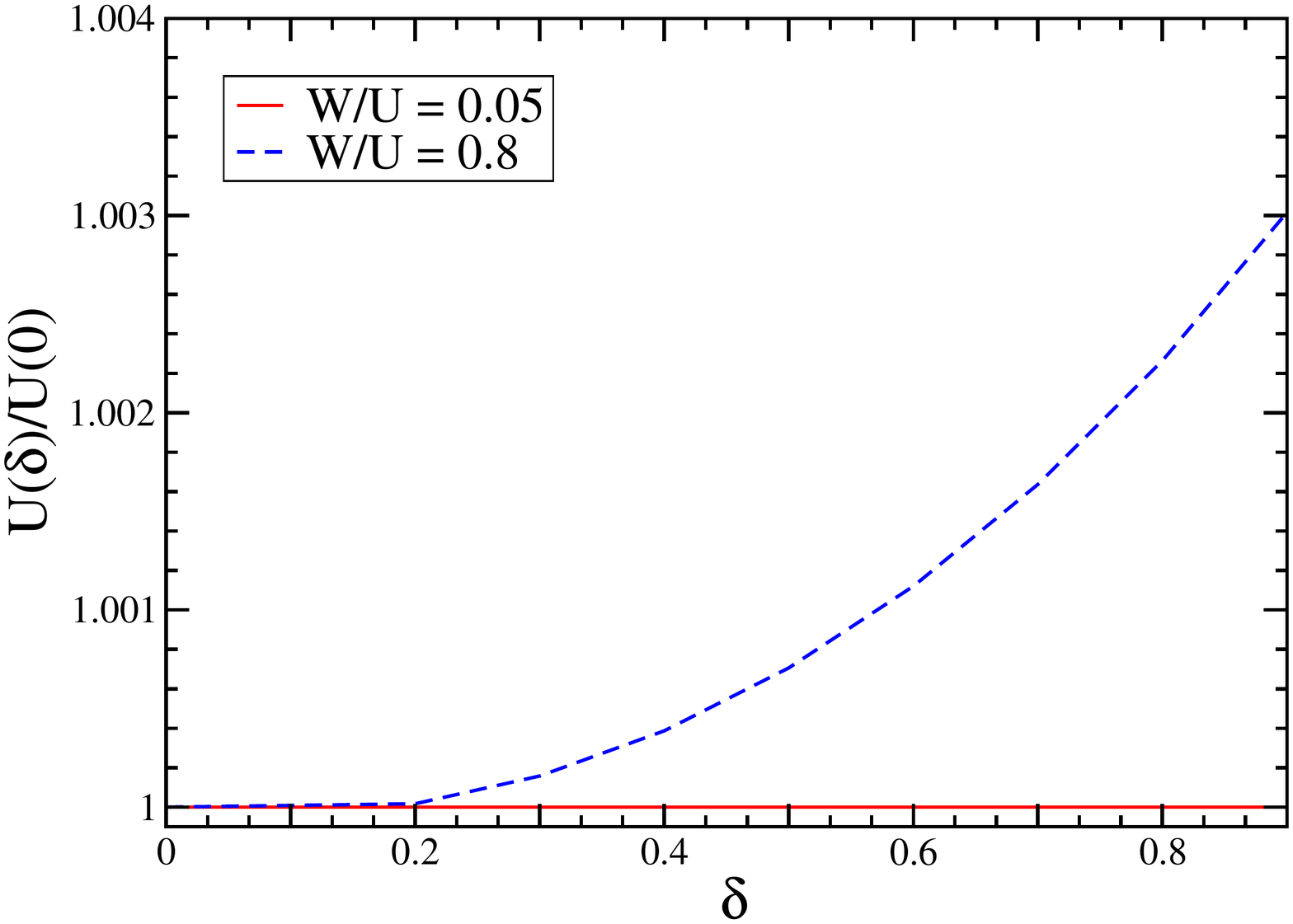}
	\caption{(Color online) Hubbard repulsion for the doped case relative 
to the value in the half-filled case as function of $\delta$ for various
values of $W/U$. The doped values are given relative to the undoped values of 
$U(0) \approx 1.0003\,U$ for $W/U = 0.05$ and $U(0) \approx 1.0787\,U$ for 
$W/U = 0.8$. }
	\label{fig:U_alle}
\end{figure}

The effective generalized $t$-$J$ model also contains interactions between DOs.
First, we consider the Hubbard repulsion $U$ which determines the energy costs 
for the creation of a single DO. So strictly speaking it does
not represent a true interaction. Since the  deviations of
 the doped values of $U$ from the ones in the half-filled case 
are small we directly  show the doped values relative to the half-filled ones
in Fig.\ \ref{fig:U_alle}. This coupling constant shows nearly no dependence 
on the doping $\delta$. 
Hence the influence of doping on $U$ may be neglected.

\begin{figure}[htb]
	\includegraphics[width=1\columnwidth]{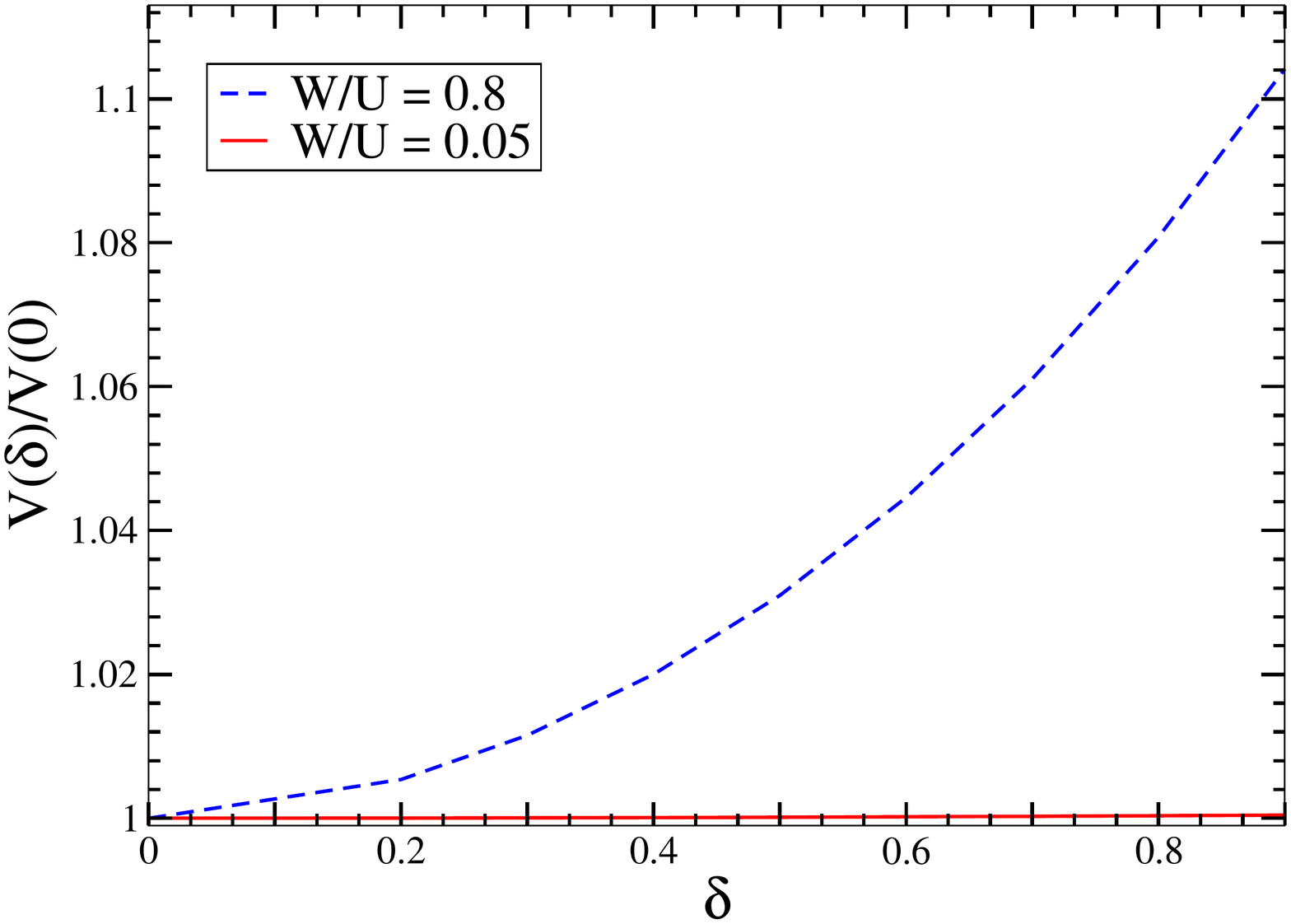}
	\caption{(Color online) Effective density-density 
interaction as function of doping for various values of $W/U$.
Results are given relative to the values at zero doping
$V(0) \approx -3.9056\cdot10^{-5}\,U$ for 
$W/U = 0.05$ and $V(0) \approx -9.2525\cdot10^{-3}\,U$ for $W/U = 0.8$.}
	\label{fig:V_alle}
	\end{figure}

The following interaction terms are active only in the presence of at 
least two DOs. Thus these terms have to be seen as genuine 2-DO 
interactions.
Among them density-density interactions of various distances appear.
The density-density interaction between nearest neighbors reads
\begin{align}
\hat{H}_V = V \sum_{<i,j>}\bar{n}_{i,\delta}\bar{n}_{j,\delta},
\end{align}
where $\bar{n}_\delta$ denotes the operator counting the number of electrons on
 a site compared to the average filling, cf.\ Tab.\ \ref{table:list}. 
At half-filling this
 term only contributes if site $i$ and site $j$ are either empty or truly 
doubly occupied. Figure \ref{fig:V_alle} only shows an increase in the
 coupling constant of about  $1\%$ 
under the influence of doping for $W/U = 0.8$.
\noindent

\begin{figure}[htb]
	\centering
	\includegraphics[width=1\columnwidth]{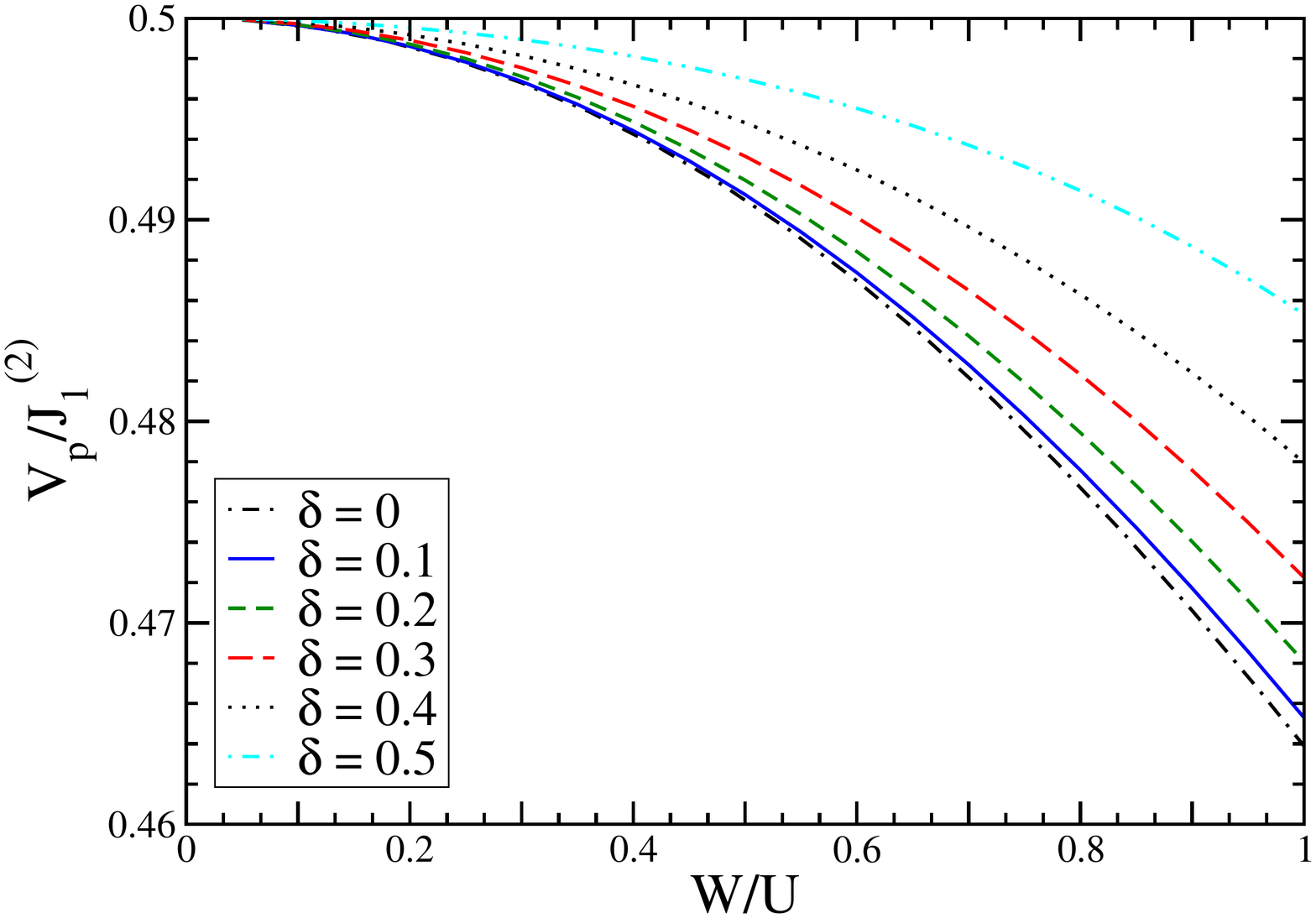}
	\caption{(Color online) Effective pair interaction $V_p$ as 
defined in Eq. \ref{eq:V_p} for various $\delta$ as a function of $W/U$.}
	\label{fig:V_p_t}
\end{figure}

A second type of interaction is correlated hopping.
 The most important term of this kind is the hopping of two electrons to a 
nearest neighbor site which is initially empty, see Eq.\ \ref{eq:V_p}. 
Since the empty state also corresponds to a DO, the effect of the
 term is to exchange the two DOs. The results for various
doping levels are depicted in Fig.\ \ref{fig:V_p_t}.

\begin{figure}[htb]
	\centering
	\includegraphics[width=1\columnwidth]{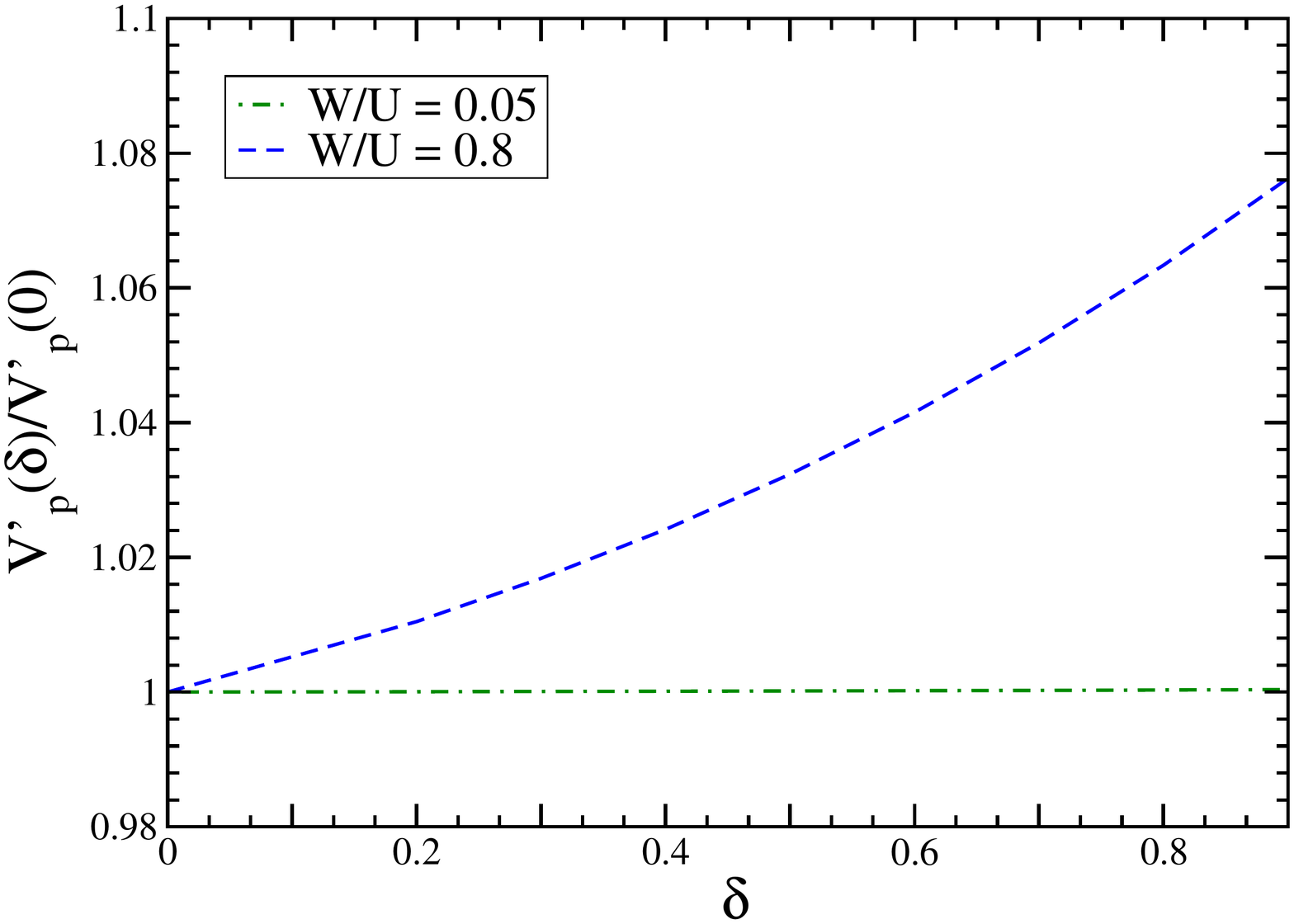}
  \caption{(Color online) Effective pair interaction $V_p'$ on $W/U$ 
for various $\delta$. In the undoped case 
$V'_p$ takes the value $V_p'(0) \approx 3.9035\cdot10^{-5}\,U$ for 
$W/U = 0.05$ and $V_p'(0) \approx 8.5167\cdot10^{-3}\,U$ for $W/U = 0.8$.}
	\label{fig:V_p_alle}
\end{figure}

Besides the nearest neighbor pair interaction $V_p$ there are also pair 
interaction terms between three spins. One of these terms is the interaction 
of three spins on a plaquette which reads
\begin{align}
\hat{H}'_{\text{pair}} &= V'_{p} \sum_{\sigma}\sum_{<i,j,k>}
\left[\hat{c}_{k,\sigma}^\dagger\hat{c}_{k,\bar{\sigma}}^\dagger
\hat{c}_{i,\bar{\sigma}}^{\phantom\dagger}\hat{n}_{i,\sigma}^{\phantom\dagger}
\hat{c}_{j,\sigma}^{\phantom\dagger}
(1-\hat{n}_{j,\bar{\sigma}}^{\phantom\dagger}) +\right.
\nonumber\\ 
&\left.\hat{c}_{k,\sigma}^\dagger\hat{c}_{k,\bar{\sigma}}^\dagger
\hat{c}_{i,\bar{\sigma}}(1-\hat{n}_{i,\sigma}^{\phantom\dagger})
\hat{c}_{j,\sigma}^{\phantom\dagger}\hat{n}_{j,\bar{\sigma}} + 
\text{h.c.}\right].
\label{def:V_p_s}
\end{align} 
The sites $i$ and $j$ are supposed to be  diagonal neighbors  with a 
common adjacent site $k$. One possible process consists of the hopping of an 
electron from a singly occupied site $j$ to an empty site $k$. Simultaneously,
an electron from the doubly occupied site $i$ hops to site $k$ forming a DO on
 this site. The corresponding effective coupling constant $ V'_{p}$
is depicted in Fig.\  \ref{fig:V_p_alle} as function of doping.

Since this correlated hopping imposes an additional constraint on the state of 
site  $k$ it is half as large as the nearest neighbor term $V_p$. 
Even for large values of $W/U$ the coupling constant is 
increased only by $8\%$ for large doping.
\noindent

\begin{figure}[htb]
	\centering
	\includegraphics[width=0.6\columnwidth]{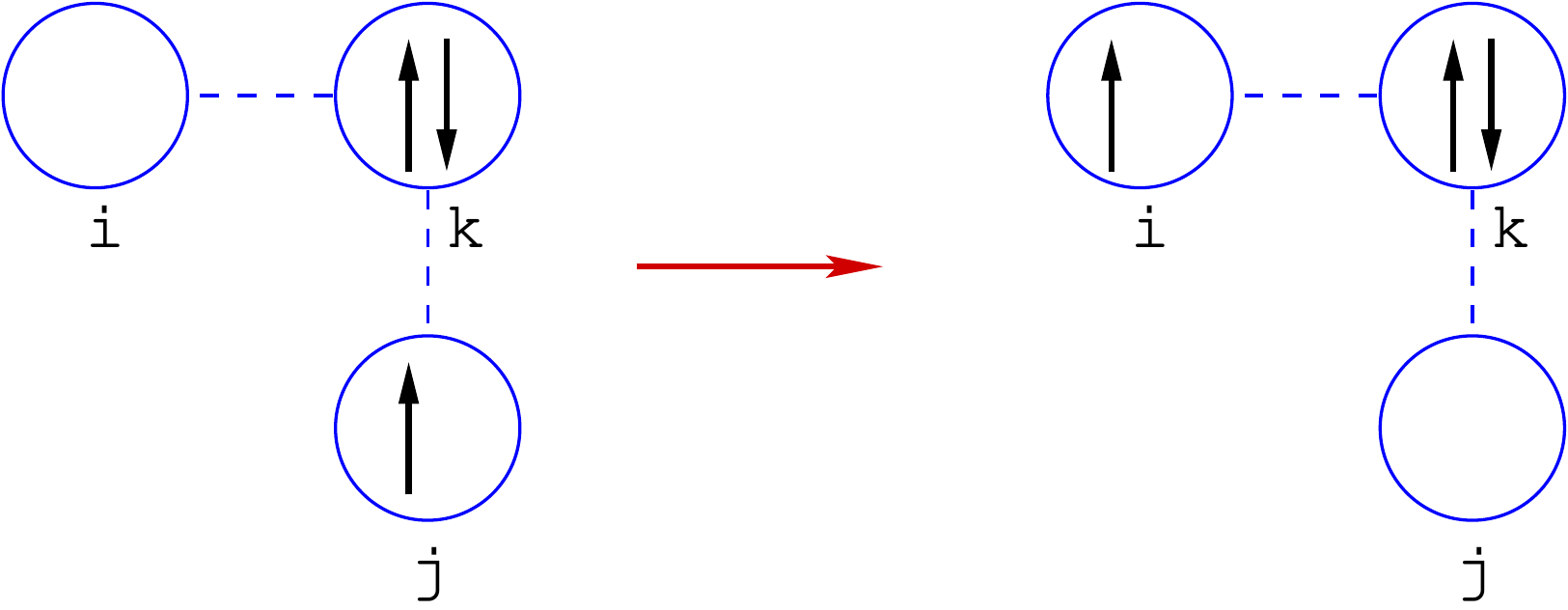}
  \caption{(Color online) Example for processes comprised in 
$\hat{H}_{V'_n}$}
	\label{sk}
\end{figure}

The last class of terms considered are correlated hopping terms 
such as
\begin{align}
\hat{H}_{V'_n} &= V'_n\sum_{\alpha, \beta}\sum_{<i,j,k>}
\left\{\left(1-\hat{n}_{i,\alpha}^{\phantom\dagger}\right)
\hat{c}_{i,\bar{\alpha}}^\dagger
\hat{c}_{j,\bar{\beta}}^{\phantom\dagger}
\left(1-\hat{n}_{j,\beta}^{\phantom\dagger}\right)\bar{n}_k+\right.
\nonumber\\
&\left.\hat{n}_{i,\alpha}^{\phantom\dagger}\hat{c}_{i,\bar{\alpha}}^\dagger
\hat{c}_{j,\bar{\beta}}^{\phantom\dagger} 
\hat{n}_{j,\beta}^{\phantom\dagger}\bar{n}_k+\text{h.c.}
\right\}.
\label{def:V_n_s}
\end{align}
One of the processes described by $\hat{H}_{V'_n}$ is the hopping of an 
electron from a singly occupied site $j$ to an empty site $i$ 
under the condition that site $k$ is occupied by a DO, see Fig.\ \ref{sk}.
Sites $i$ and $j$ are diagonal neighbors on a plaquette
and $k$ joint adjacent neighbor. 

\noindent
Due to the constraint that site $k$ has to be occupied by a DO and site $i$ 
has to be empty in the beginning, these processes rely on the presence of two 
DOs which justifies to view them as true interactions. The number of DOs is not
 changed by this process.
Processes such as $\hat{H}_{V'_n}$ appear in second order of $\frac{t}{U}$. 

\begin{figure}[htb]
\begin{minipage}[hbt]{0.5\columnwidth}
	\centering
	\hspace*{-0.35cm}
	\includegraphics[width=1.2\columnwidth]{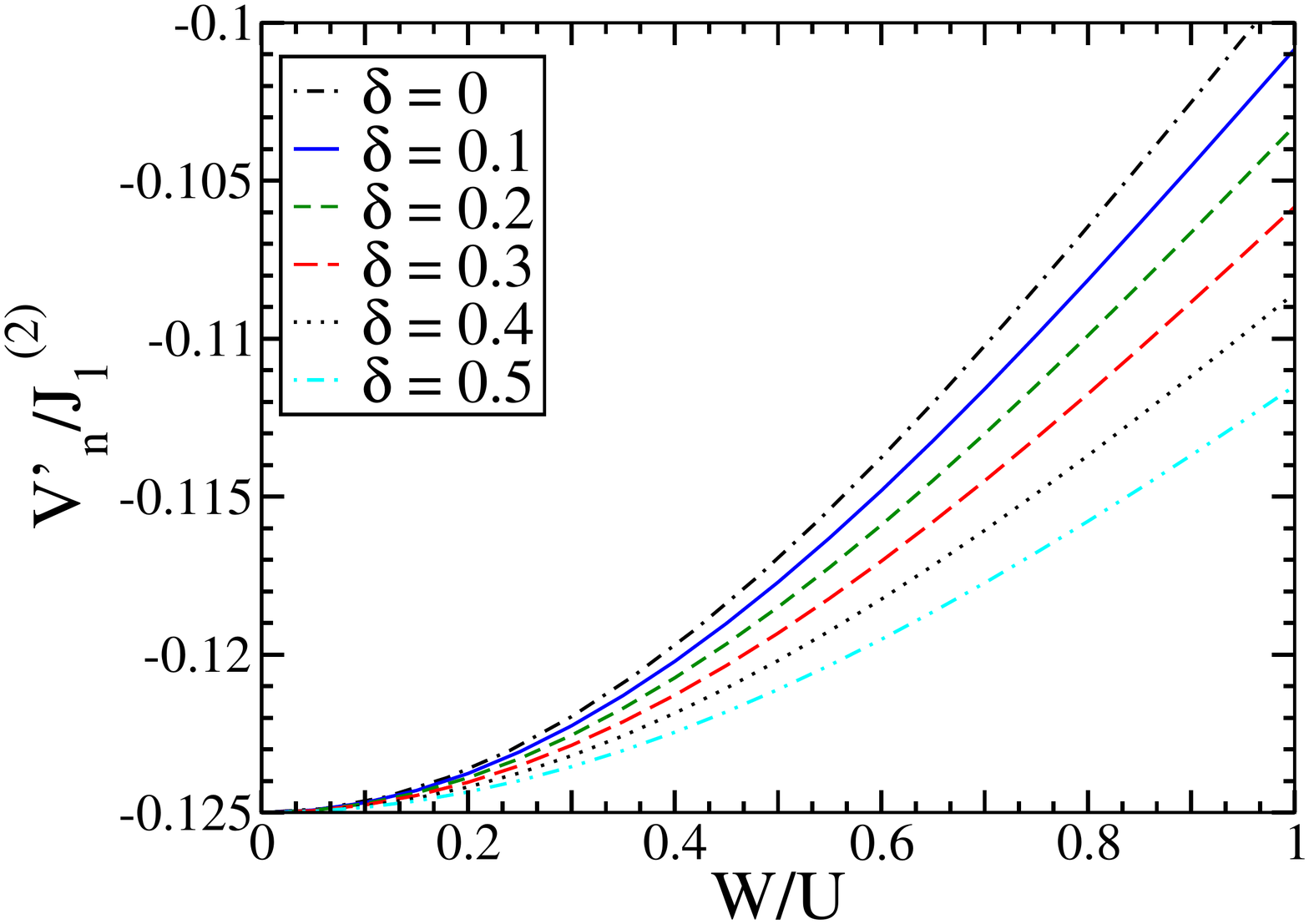}
			\end{minipage}
\centering
\hspace*{-0.2cm}
\begin{minipage}[hbt]{0.5\columnwidth}
	\centering
	\includegraphics[width=1.2\columnwidth]{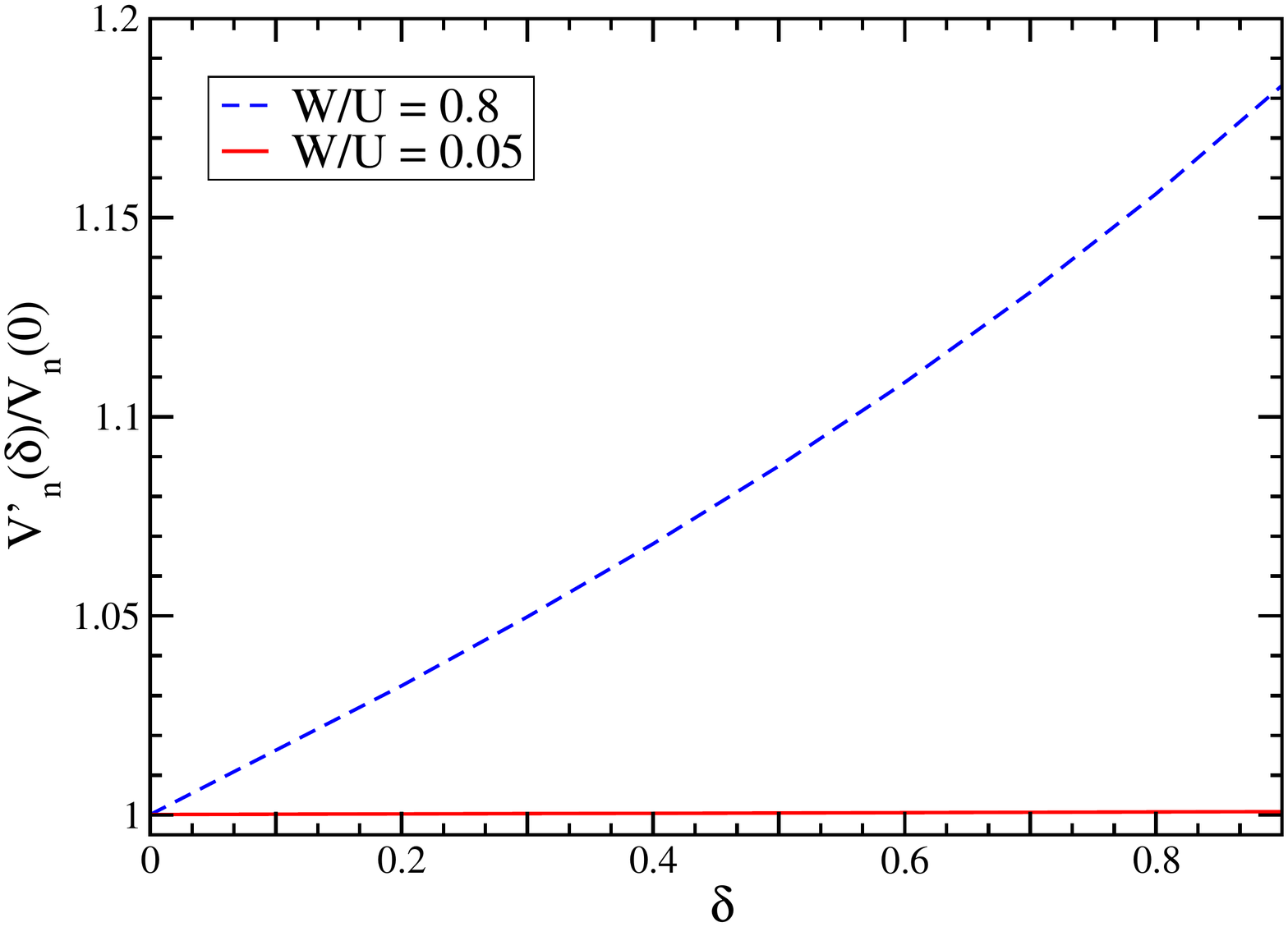}
 \end{minipage}
\caption{(Color online) Effective correlated hopping process $V'_n$ over 
diagonal neighbors for various $\delta$ (left panel) 
and as function of $\delta$ for the values
$W/U = 0.05$ and $W/U = 0.8$ (right panel). The undoped values are 
$V'_n \approx -1.9518\cdot10^{-5}\,U$ ($W/U = 0.05$) and 
$V'_n \approx -4.2584\cdot10^{-3}\,U$ ($W/U = 0.8$).}
\label{fig:V_n_str_n}
\end{figure}

The corresponding coupling constant  is shown in the left
panel of Fig.\ \ref{fig:V_n_str_n} as function of $W/U$.
In the right panel of Fig.\ \ref{fig:V_n_str_n}, the value for the coupling 
constant $V'_n$ in the doped case is shown relative to its value in the 
half-filled case. 
For large values of $W/U$, the coupling $V'_n$ shows a noticeable dependence 
on  $\delta$. Note that besides the correlated hopping defined in
\eqref{def:V_n_s} and illustrated in Fig.\ \ref{sk},
there is correlated hopping  between three sites  located on three
sites in a row. The corresponding coupling constant $V''_n$ shows 
the same behavior as $V'_n$ so that we do not show it here for
brevity.

For too large values of $W/U$, i.e., $W/U \gg 1$, 
the curves for the coupling constants are not 
smooth anymore (not shown here) \cite{reisc04}. 
This observation is explained by
the breakdown of the  mapping as it is indicated by the behavior of the 
apparent charge gap (see Sect.\ \ref{chap:gap}).

\subsection{Hopping Terms}

\begin{figure}[htb]
\begin{minipage}[hbt]{0.5\columnwidth}
	\centering
	\hspace*{-0.35cm}
	\includegraphics[width=1.2\columnwidth]{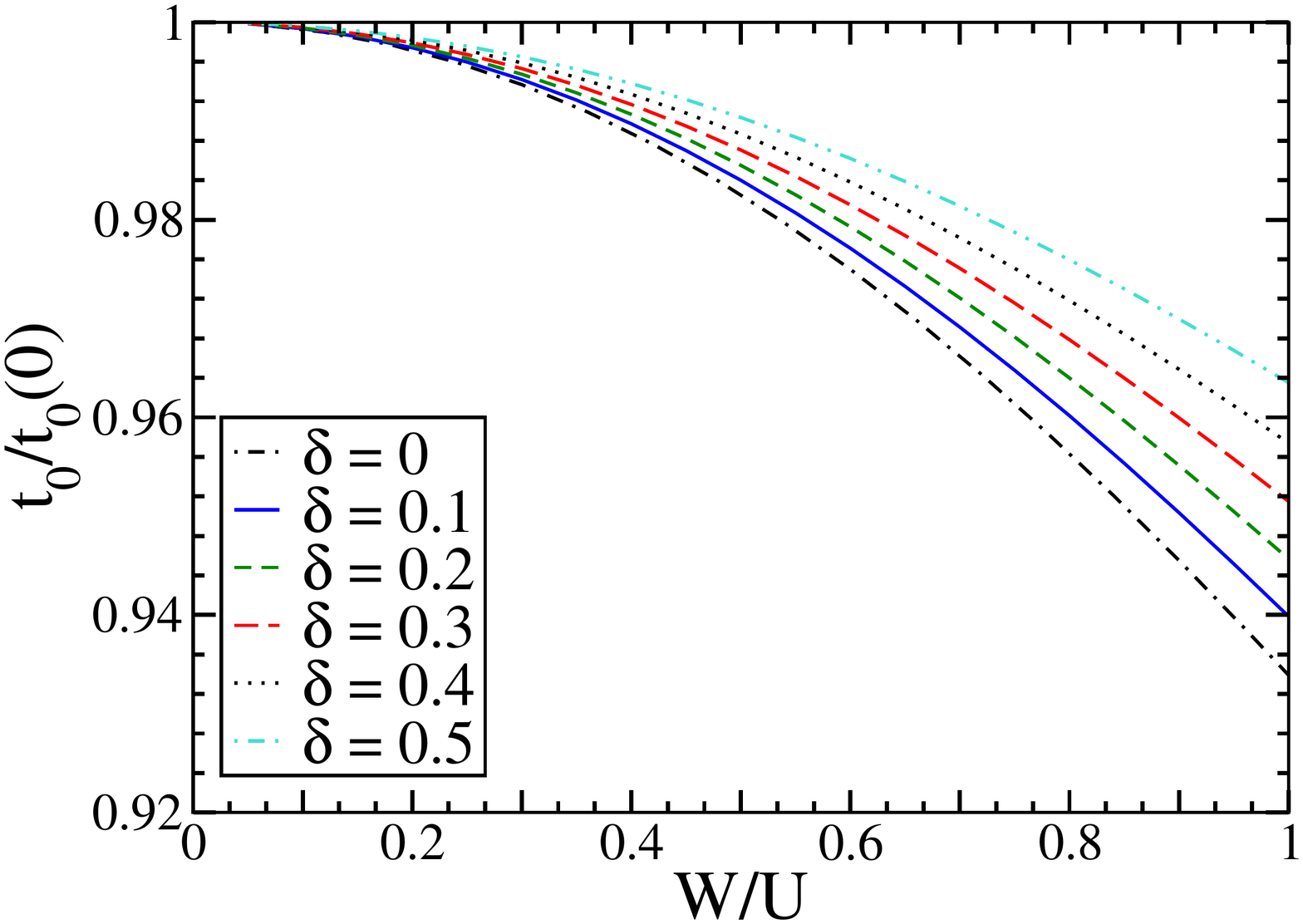}
			\end{minipage}
\centering
\hspace*{-0.2cm}
\begin{minipage}[hbt]{0.5\columnwidth}
		\centering
	\includegraphics[width=1.2\columnwidth]{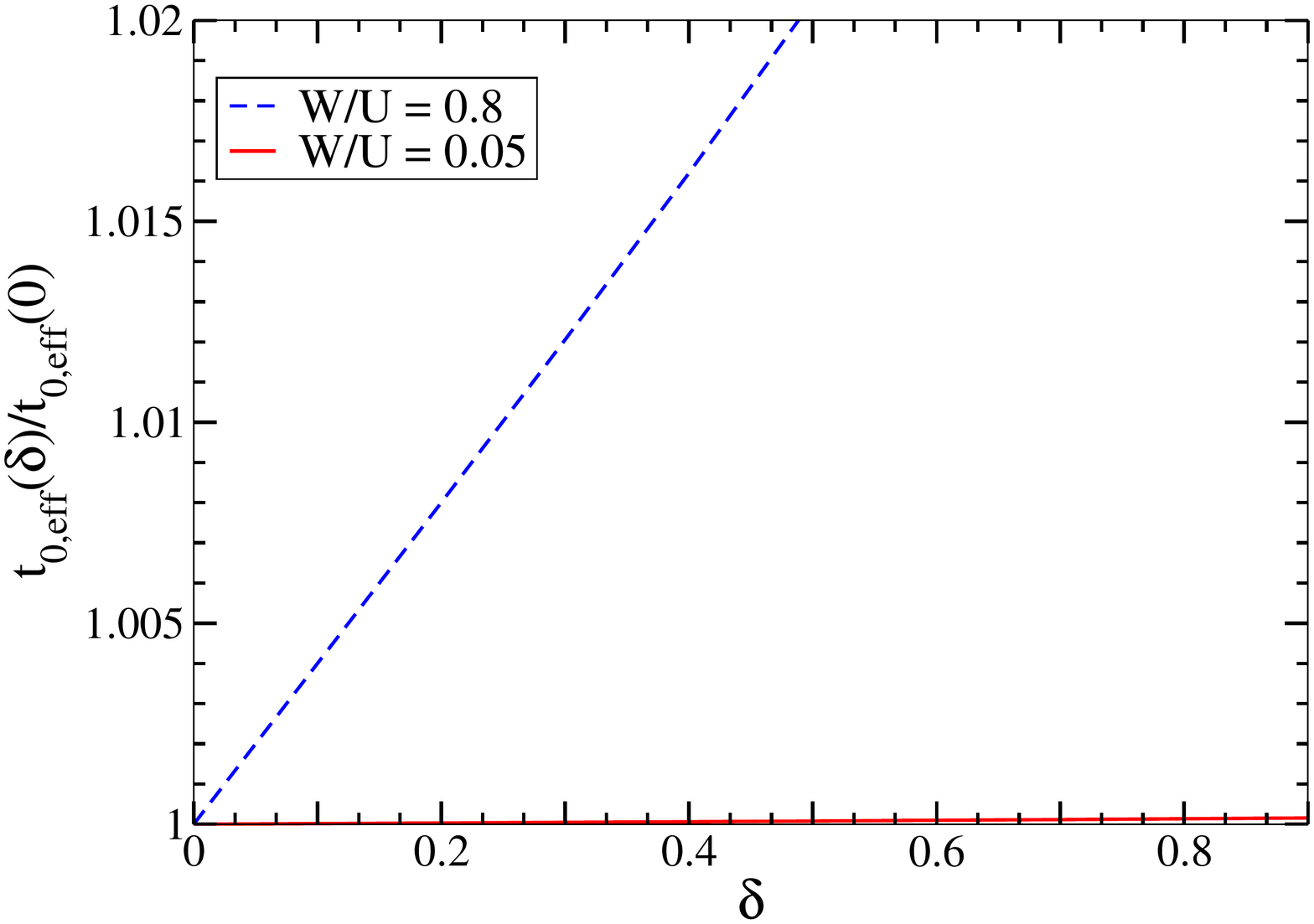}
 \end{minipage} 
\caption{(Color online) Effective NN hopping without changing the number of DOs
 compared to its initial, bare value (left panel). Its doping dependence 
relative to the values at half-filling
$t_0(0) \approx 6.2490\cdot10^{-3}\,U$ for $W/U = 0.05$ and 
$t_0(0) \approx 0.0956\,U$ for $W/U = 0.8$ is depicted in the right panel.}
\label{fig:t_0_n}
\end{figure}

The first term to be considered is $\hat{T}_0$ which was introduced before in 
(\ref{TO}). The initial 
$\hat{T}_0$ represents hopping processes by one lattice spacing without a 
change in the number of DOs. The corresponding coupling constant $t_0$ is 
shown in the left panel of Fig.\ \ref{fig:t_0_n} relative to its 
unrenormalized  value.
The deviation of the coupling $t_0$ from its bare 
value is proportional to $\left(\frac{t}{U}\right)^2$.
It is increased upon increased doping. 
To examine the doping dependence the renormalized value of $t_0$
 in the doped case is compared to the one in the 
half-filled case in the right panel of Fig.\ \ref{fig:t_0_n}. 
Under the influence of doping the hopping parameter is increased linearly. But 
even for $W/U=0.8$ the parameter is changed only by  a few percent.

\begin{figure}[htb]
\begin{minipage}[hbt]{0.5\columnwidth}
	\centering
	\hspace*{-0.35cm}
	\includegraphics[width=1.2\columnwidth]{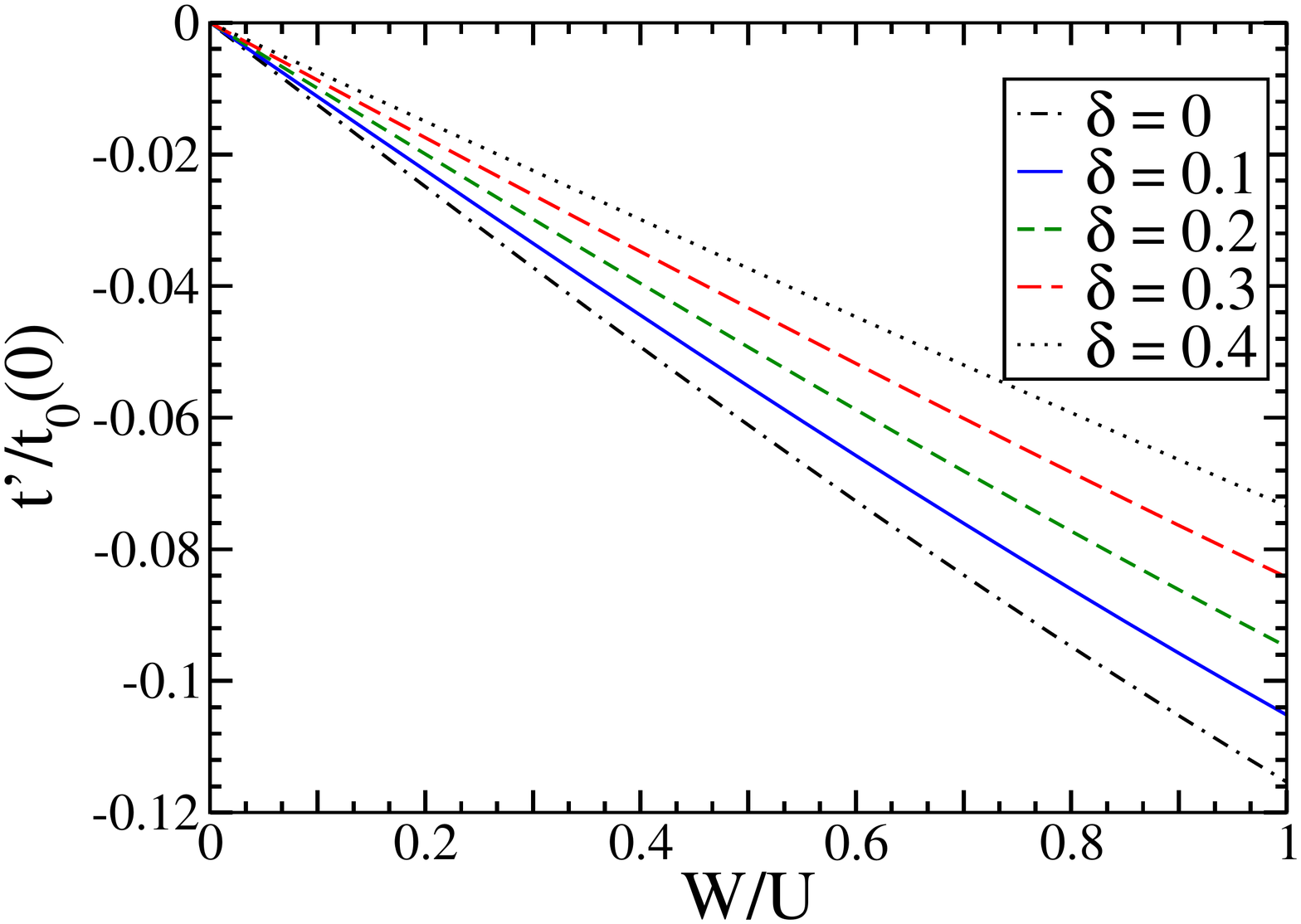}
		\end{minipage}
\centering
\hspace*{-0.2cm}
\begin{minipage}[hbt]{0.5\columnwidth}
		\centering
	\includegraphics[width=1.2\columnwidth]{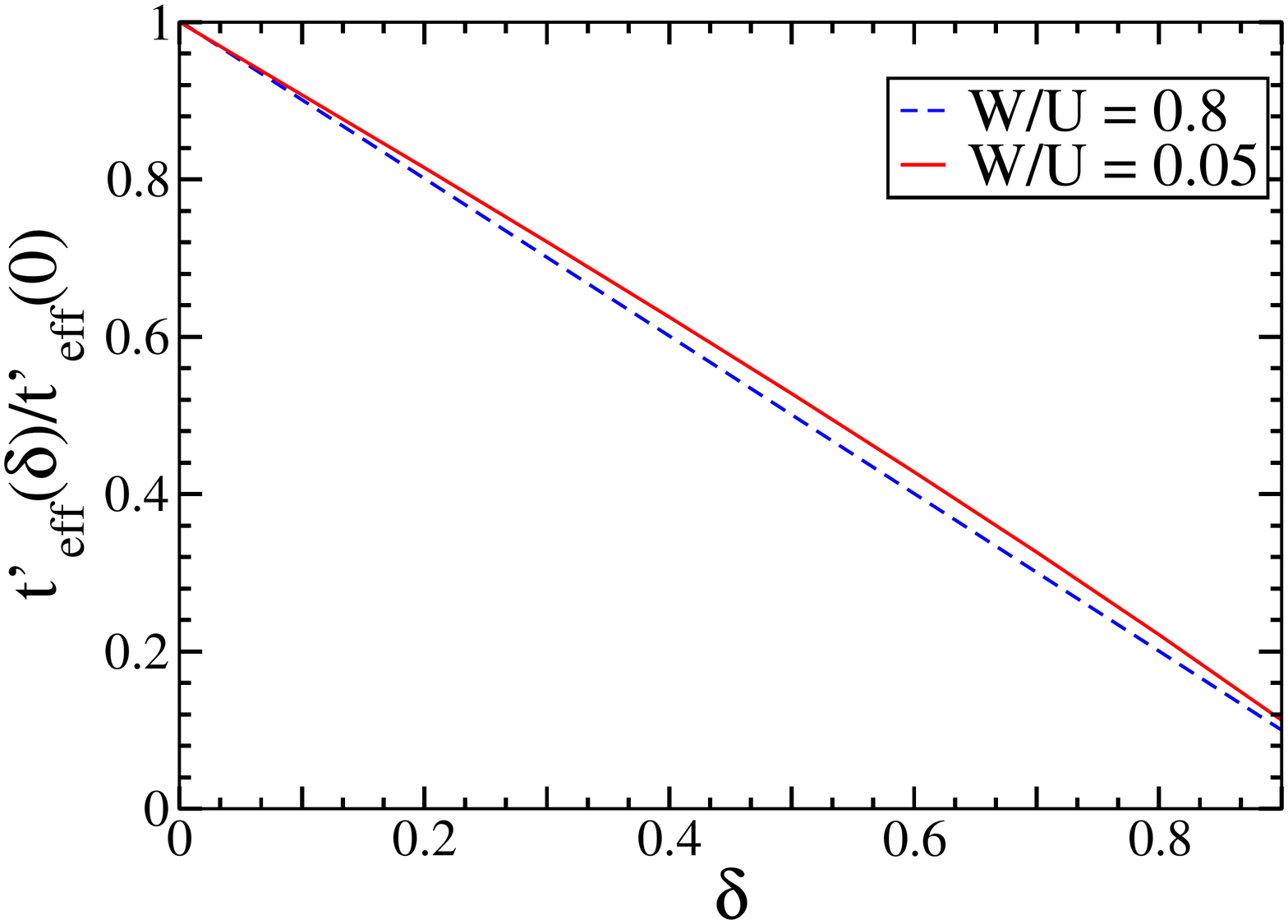}
 \end{minipage}
\caption{(Color online) Effective
NNN Hopping  $t'$ between diagonal sites as function
of $W/U$ for various values of  $\delta$ 
(left panel) and its  doping dependence for $W/U=0.05$ and $W/U=0.8$ (right 
panel). In the undoped case $t'$ takes the values
$t'(0) \approx -3.9053\cdot10^{-5}\,U$ ($W/U = 0.05$) and 
$t'(0) \approx -9.4785\cdot10^{-3}\,U$ for $W/U = 0.8$.}
\label{fig:t_str_n}
\end{figure}

Hopping  also occurs between diagonal sites on a plaquette, for 
instance in 
\begin{align}
\hat{T}'_0 = t'\sum_{\sigma}\sum_{<<i,j>>}
\left[\left(1-\hat{n}_{i,\sigma}^{\phantom\dagger}\right)
\hat{c}_{i,\bar{\sigma}}^\dagger\hat{c}_{j,\bar{\sigma}}^{\phantom\dagger}
(1-\hat{n}_{j,\sigma}^{\phantom\dagger})+\right.
\nonumber\\
\left.\hat{n}_{i,\sigma}^{\phantom\dagger}
\hat{c}_{i,\bar{\sigma}}^\dagger\hat{c}_{j,\bar{\sigma}}^{\phantom\dagger}
\hat{n}_{j,\sigma}^{\phantom\dagger}+\text{h.c.}\right]\,.
\label{def:t_s}
\end{align}
Here the double bracket under the sum indicates next-nearest neighbors (NNN) 
on the square  lattice.
The same process also appears between third nearest neighbors with
 a distance of two lattice sites. The corresponding coupling constant is 
denoted by $t''$. Since $t''$ and $t'$ show very similar behavior we only show 
the results for $t'$ in Fig.\ \ref{fig:t_str_n}.

The hopping $t'$ decreases linearly for increasing ratio $W/U$
with slopes depending on the doping, see left panel of Fig.\ \ref{fig:t_str_n}.
Relative to its values at half-filling the decrease as function
of doping hardly depends on $W/U$, see right panel of Fig.\ \ref{fig:t_str_n}.
It is remarkable that the constant is decreased to almost 0 for
 $\delta \rightarrow 1$. This is actually the \emph{only} 
significant dependence on doping that we found.
But one has to keep in mind that the absolute values of $t'$ are small.
Note that the sign of $t_0$ is positive whereas $t'$ and $t''$ are negative.

\begin{figure}[htb]
\begin{minipage}[hbt]{0.5\columnwidth}
	\centering
	\hspace*{-0.35cm}
	\includegraphics[width=1.2\columnwidth]{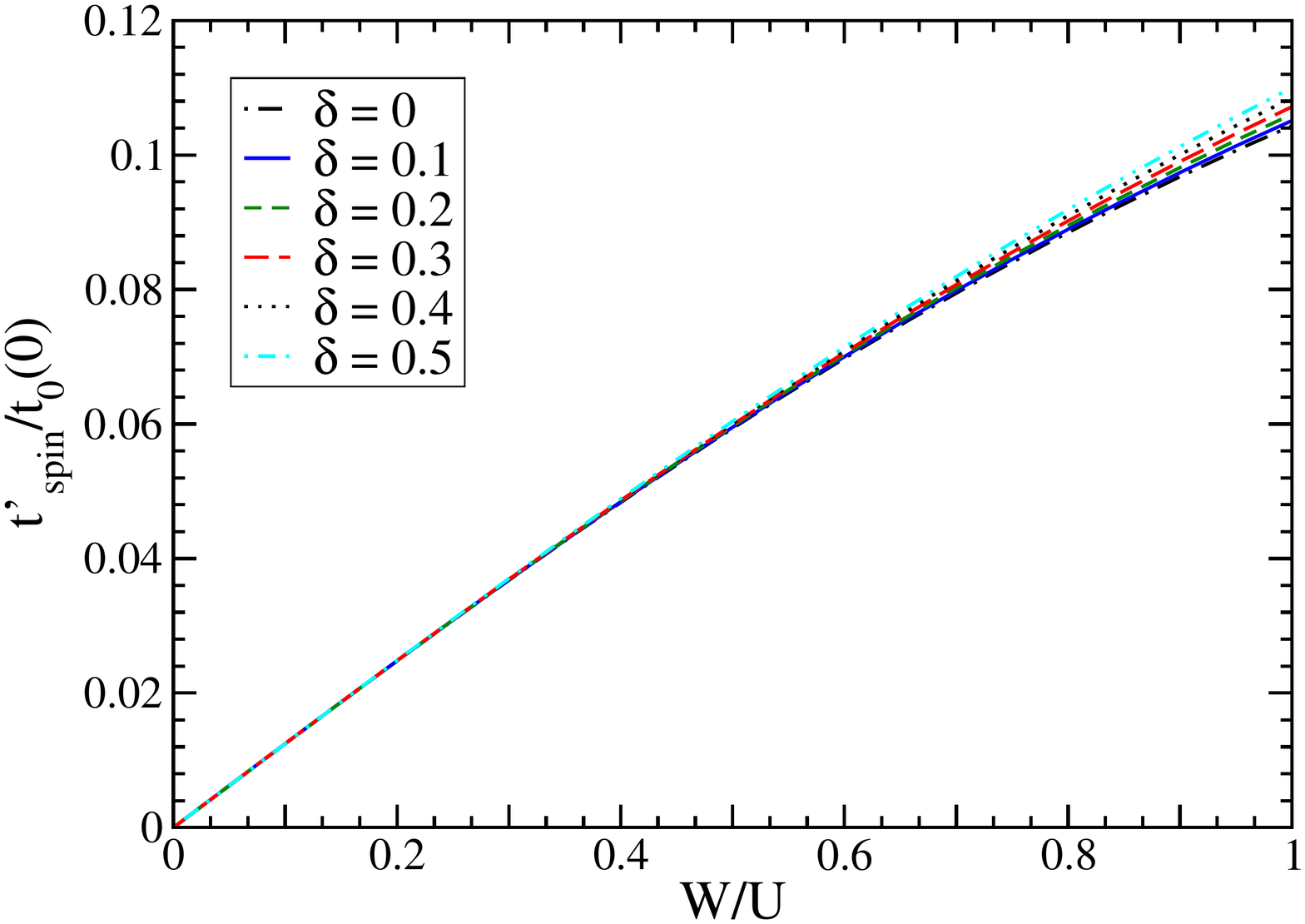}
			\end{minipage}
\centering
\hspace*{-0.2cm}
\begin{minipage}[hbt]{0.5\columnwidth}
		\centering
	\includegraphics[width=1.2\columnwidth]{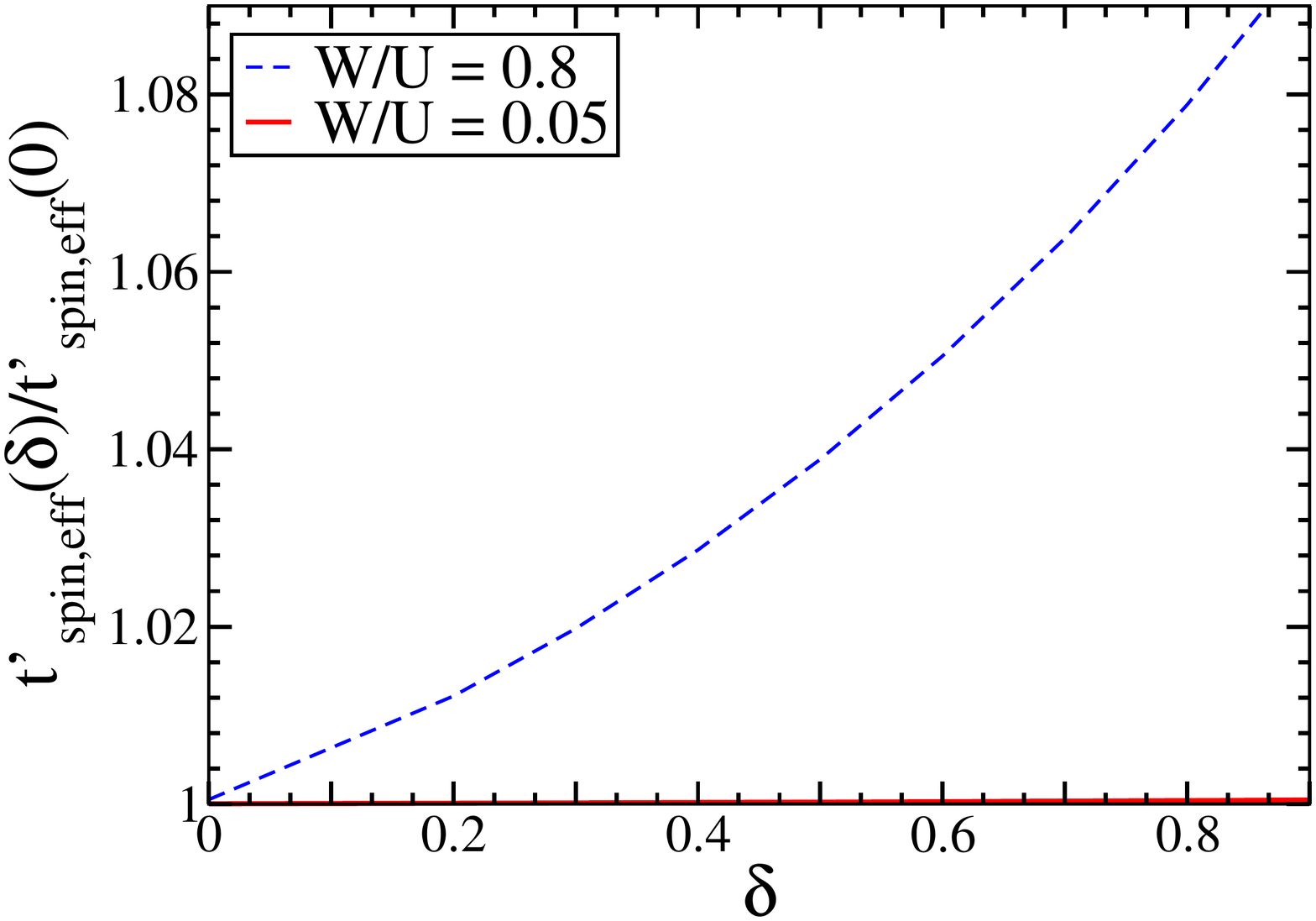}
 \end{minipage}
\caption{(Color online) Effective spin dependent hopping 
$t'_{\text{spin}}$ for various values of $\delta$ (left panel) 
and its dependence on $\delta$ for two
values of $W/U$, namely
$W/U=0.05$ ($t'_{\text{spin}}(0) \approx 3.9042\cdot10^{-5}\,U$) and 
$W/U=0.8$ ($t'_{\text{spin}}(0) \approx 8.8496\cdot10^{-3}\,U$) (right panel).}
\label{fig:t_spin_diag_n}
\end{figure}

An interesting coupling generated in second order of $\frac{t}{U}$ is the spin 
dependent hopping described by
\begin{align}
\hat{T}'_{\text{spin}} &= t'_{\text{spin}}\sum_{\alpha\beta}
\sum_{<i,k,j>}\left[\left[\left(1-\hat{n}_{i,\alpha}^{\phantom\dagger}\right)
\hat{c}_{i,\bar{\alpha}}^\dagger\vec{\sigma}_{\bar{\alpha},\bar{\beta}}
\hat{c}_{j,\bar{\beta}}^{\phantom\dagger}
(1-\hat{n}_{j,\beta}^{\phantom\dagger})\right.\right.
\nonumber\\
 & \left.\left.+\hat{n}_{i,\alpha}^{\phantom\dagger}
\hat{c}_{i,\bar{\alpha}}^\dagger\vec{\sigma}_{\bar{\alpha}\bar{\beta}}
\hat{c}_{j,\bar{\beta}}^{\phantom\dagger}\hat{n}_{j,\beta}^{\phantom\dagger}+
\text{h.c.}\right]\vec{S}_k\right]
\label{def:t_spin_s}\hspace*{1cm}
\end{align}
where the sum runs over two diagonal neighbors $i$ and $j$ which have a common 
nearest neighbor $k$. 
The size of the corresponding hopping parameter $t'_{\text{spin}}$, see
Fig.\ \ref{fig:t_spin_diag_n}, is comparable to the spin independent parameter
$t'$. This shows that the  induced NNN spin dependent hopping processes are 
as important as the spin  independent ones. This was first observed by
Reischl et al. \cite{reisc04} at half-filling.

Upon doping the spin dependent hopping term is increased. 
But even for larger values of $W/U$ the value of $t'_{\text{spin}}$
 is increased  only by a few percent. 
The analogous process also appears between third nearest neighbors. The 
corresponding
coupling constant behaves similar to $t'_{\text{spin}}$ so that we do not 
display it here. Both processes concern three sites. Thus it does not  
matter significantly 
whether the sites are aligned linearly or in a right angle on a  plaquette.

Compared to the spin independent hopping $t'$ the spin 
dependent hoppings $t'_{\text{spin}}$ and $t''_{\text{spin}}$ are not 
negligible. But all of them are fairly small compared
to the bare NN hoping $t_0$.
Thus one can either stick to a pure $t$-$J$ model or include more extended 
hopping terms. But if one opts for including hopping over two
lattice spacings one should incorporate  spin independent hopping
 $t'$ as well as the spin dependent hoppings  $t'_{\text{spin}}$ and 
$t''_{\text{spin}}$.
The doping dependence of the spin dependent hopping elements may be neglected. 
In contrast, the hopping element $t'$ shows a rather strong dependence 
on the doping concentration $\delta$.

%%%%%%%%%%%%%%%%%%%%%%%%%%%%%%%%%%%%%%%
%%%%%%%%%%%%%%%%%%%%%%%%%%%%%%%%%%%%%%%

\section{Summary}
\label{chap:summ}

We  presented a systematically controlled mapping of a fermionic Hubbard 
model to a generalized $t$-$J$ model. The conceptual foundation
of this mapping was analyzed carefully. In particular, we developed
a scheme for this mapping which covers also the interesting case
of substantial doping. Remarkably, this issue had so far attracted
only little attention.

In the derivation of the generalized $t$-$J$ model we eliminate the
 charge fluctuations by self-similar continuous unitary transformations. 
Processes that change the number of double occupancies are rotated away.
Thereby, we obtain the effective coupling constants as function of
the doping $\delta$ and of the ratio $W/U$ 
where $W$ is the bandwidth and $U$ the local interaction.
Note that the generalized $t$-$J$ model comprises 
the magnetic degrees of freedom as well as
the kinetics and the interactions of double occupancies.

We extended the concept of the apparent charge gap $\Delta_g$ \cite{reisc04}
from half-filling to the doped system. This gap is not the true
physical gap but it measures the energy separation of
subspaces with differing number of double occupancies \emph{irrespective}
of the spin state. We argue that as long as $\Delta_g$ is finite
the mapping to a $t$-$J$ model is justified. A vanishing $\Delta_g$ 
indicates the breakdown of this mapping. By estimating
the parameter where $\Delta_g(W/U,\delta)=0$ holds we
derived a diagram of applicability of the $t$-$J$ model
shown in Fig.\ \ref{fig:appli}. As expected the applicability is reduced 
upon doping $\delta$. But it levels at intermediate values of doping
so that the commonly assumed parameters for the description of
 high-$T_c$  cuprates lie within the range of applicability. 
To our knowledge, no such result was derived before.

Furthermore we find that the coupling constants of the effective model 
show hardly any doping dependence. The only coupling which exhibits a 
significant dependence on $\delta$  is the hopping parameter $t'$ 
describing hopping between diagonal neighbors. 
Relative to its value at half-filling $t'$ exhibits a strong doping 
dependence. But the absolute value of this hopping element remains small. Thus 
within a wide range of doping the $t$-$J$ model with constant coupling 
constants is appropriate. Besides the usually considered terms,
the 4-spin ring exchange on each plaquette should be included.

Technically we used recently developed types of infinitesimal generators 
for the continuous unitary transformation \cite{fisch10a}.  
They only decouple certain  subspaces of the Hilbert space which simplifies 
and accelerates the  calculations.
So far, the pc-generator was used which leads to a particle number conserving
effective model; the particles are the double occupancies
\cite{reisc04,yang10}. We extended the gs-generator introduced
previously for the ground state of
matrices \cite{dawso08} and of many-body systems \cite{fisch10a}
to mixed reference ensembles. The gs-generator is particularly suited
to obtain the purely magnetic Heisenberg model since it
efficiently decouples the subspace of the reference ensemble from the remainder
of the Hilbert space. If, however, the dynamics of
the double occupancies matters as well, the gs,1p-generator turned out
to be a good compromise between efficiency and sufficient decoupling.
This generator decouples the reference ensemble and the states
with one double occupancy from the rest of the Hilbert space.
We found that the couplings derived from a faster gs,1p-calculation
agree very well with the results from a slower pc-calculation.
We expect that these generator or modifications of them
will continue to play an important role in the systematic derivation
of effective models.

%Outlook
The present analysis for the square lattice can certainly be 
extended to other types of lattices such as the triangular
lattice which has already been analysed by perturbative
CUTs \cite{yang10}, the honeycomb lattice \cite{meng10}, or
more sophisticated lattices such as the kagom\'e lattice and so on.
In this way, the effects of subleading magnetic exchange
processes such as ring exchange can be analysed quantitatively.

Another route to extend the present calculation
is to also transform the observables, for instance the
standard fermionic creation operator. At half-filling,
one will then be able to compute the spectral weight in the upper Hubbard band,
that means in the subspace with one double occupancy.
But there should be also weight in the subspaces with
three and more double occupancies. To our knowledge, no estimate
whatsoever exists for the weight in such trans-Hubbard bands.

More generally, the systematic derivation of effective
models in other contexts can be tackled by adapting 
the ideas of the present work. For instance, the reliable downfolding of 
interacting fermionic models with many bands to models with a minimum
number of bands and Hubbard-type of interactions
is a long standing field of
research \cite{herbs78a,herbs78b, gunna90}
which continues to attract much attention, see for instance
Refs.\ \onlinecite{aryas04,cano07,miyak08,cano10}.
We think that continuous unitary transformations 
provide an promising approch to make
systematic and controlled progress in this field.

Hence we expect that the systematic derivation
of effective models by means of continuous unitary
transformations will continue to evolve into a 
field with widespread applications.

\begin{acknowledgments}
We thank A.A. Reischl, C. Raas, K.P. Schmidt, E. Koch, 
N. Lorscheid and S. Schmitt for fruitful 
discussions. We gratefully acknowledge support by the Studienstiftung des 
deutschen Volkes.
\end{acknowledgments}

 %\bibliographystyle{apsrev}
 %\bibliography{liter-shv3}

\end{document}